\documentclass[fleqn,usenatbib]{mnras}

\usepackage{newtxtext,newtxmath}
\usepackage[T1]{fontenc}
\usepackage{ae,aecompl}

\newtoggle{referee}            %
\toggletrue{referee}         %
\iftoggle{referee}{%           %
    \usepackage{color,soul}    %
}{%                            %
}                              %
\usepackage{graphicx}	% Including figure files
\usepackage{amsmath}	% Advanced maths commands
\usepackage{amssymb}	% Extra maths symbols
\usepackage{wasysym}  % Extra astronomy symbols
\usepackage{bm}       % Bold maths
\usepackage{etoolbox}
\makeatletter % Workaround to only color the year for cite commands
  \patchcmd{\NAT@citex}
    {\@citea\NAT@hyper@{%
      \NAT@nmfmt{\NAT@nm}%
      \hyper@natlinkbreak{\NAT@aysep\NAT@spacechar}{\@citeb\@extra@b@citeb}%
      \NAT@date}}
    {\@citea\NAT@nmfmt{\NAT@nm}%
    \NAT@aysep\NAT@spacechar\NAT@hyper@{\NAT@date}}{}{}

  \patchcmd{\NAT@citex}
    {\@citea\NAT@hyper@{%
      \NAT@nmfmt{\NAT@nm}%
      \hyper@natlinkbreak{\NAT@spacechar\NAT@@open\if*#1*\else#1\NAT@spacechar\fi}%
        {\@citeb\@extra@b@citeb}%
      \NAT@date}}
    {\@citea\NAT@nmfmt{\NAT@nm}%
    \NAT@spacechar\NAT@@open\if*#1*\else#1\NAT@spacechar\fi\NAT@hyper@{\NAT@date}}
    {}{}
\makeatother

\newcommand\Msun{\text{M}_{\astrosun}} % requires the wasysym package
\newcommand\Zsun{\text{Z}_{\astrosun}} % requires the wasysym package
\newcommand\colt{\textsc{colt}} % code name
\newcommand\HI{{H\,\textsc{i}}} % neutral hydrogen
\newcommand\HII{{H\,\textsc{ii}}} % ionized hydrogen
\title[Ly$\alpha$ escape from high-$z$ galaxies]{The physics of Lyman $\balpha$ escape from high-redshift galaxies}

\author[A.\ Smith et al.]{Aaron~Smith,$^{1,2}$\thanks{E-mail: \href{mailto:arsmith@mit.edu}{arsmith@mit.edu}}\thanks{Einstein Fellow.}
  Xiangcheng~Ma,$^{3,4}$
  Volker~Bromm,$^{2}$
  Steven~L.~Finkelstein,$^{2}$ \newauthor Philip~F.~Hopkins,$^{4}$
  Claude-Andr\'{e} Faucher-Gigu\`{e}re$^{5}$
  and
  Du\v{s}an Kere\v{s}$^6$
  \\
  $^{1}$Department of Physics,
  Massachusetts Institute of Technology, Cambridge, MA 02139, USA \\
  $^{2}$Department of Astronomy, The University of Texas at Austin, Austin, TX 78712, USA \\
  $^{3}$Department of Astronomy and Theoretical Astrophysics Center,
  University of California Berkeley, Berkeley, CA 94720, USA \\
  $^{4}$TAPIR, MC 350-17, California Institute of Technology, Pasadena, CA 91125, USA \\
  $^{5}$Department of Physics and Astronomy and CIERA, Northwestern University, Evanston, IL 60647, USA \\
  $^{6}$Department of Physics,
  University of California at San Diego,
  La Jolla, CA 92093, USA
}

\date{Accepted XXX. Received YYY; in original form ZZZ}

\pubyear{2018}

\begin{document}
\label{firstpage}
\pagerange{\pageref{firstpage}--\pageref{lastpage}}
\maketitle

\begin{abstract}
Lyman $\alpha$ (Ly$\alpha$) photons from ionizing sources and cooling radiation undergo a complex resonant scattering process that generates unique spectral signatures in high-redshift galaxies. We present a detailed Ly$\alpha$ radiative transfer study of a cosmological zoom-in simulation from the Feedback In Realistic Environments (FIRE) project. We focus on the time, spatial, and angular properties of the Ly$\alpha$ emission over a redshift range of $z = 5$--$7$, after escaping the galaxy and being transmitted through the intergalactic medium (IGM). Over this epoch, our target galaxy has an average stellar mass of $M_\star \approx 5 \times 10^8\,\Msun$. We find that many of the interesting features of the Ly$\alpha$ line can be understood in terms of the galaxy's star formation history. The time variability, spatial morphology, and anisotropy of Ly$\alpha$ properties are consistent with current observations. For example, the rest frame equivalent width has a $\text{EW}_{\text{Ly}\alpha,0} > 20\,\text{\AA}$ duty cycle of $62\%$ with a non-negligible number of sightlines with $> 100\,\text{\AA}$, associated with outflowing regions of a starburst with greater coincident UV continuum absorption, as these conditions generate redder, narrower (or single peaked) line profiles. The lowest equivalent widths correspond to cosmological filaments, which have little impact on UV continuum photons but efficiently trap Ly$\alpha$ and produce bluer, broader lines with less transmission through the IGM. We also show that in dense self-shielding, low-metallicity filaments and satellites Ly$\alpha$ radiation pressure can be dynamically important. Finally, despite a significant reduction in surface brightness with increasing redshift, Ly$\alpha$ detections and spectroscopy of high-$z$ galaxies with the upcoming \textit{James Webb Space Telescope} is feasible.
\end{abstract}

% Select between one and six entries from the list of approved keywords. Don't make up new ones.

%%%%%%%%%%%%%%%%%%%%%%%%%%%%%%%%%%%%%%%%%%%%%%%%%%

%%%%%%%%%%%%%%%%% BODY OF PAPER %%%%%%%%%%%%%%%%%%

\section{Introduction}
\label{sec:intro}
Lyman~$\alpha$ (Ly$\alpha$) emission from stellar populations and active galactic nuclei is a powerful diagnostic of high-$z$ objects, enabling the effective identification and spectroscopic confirmation of these sources \citep{Bromm2011,Finkelstein_2016}. However, due to the complex physics of the Ly$\alpha$ radiative transfer process, it has proven difficult to interpret the results from theoretical models and observational data \citep{Dijkstra_PASA_2014}. For example, the fraction of Ly$\alpha$ emitters (LAEs) among the galaxy population has been found to first increase with redshift, and then drop dramatically at $z > 6$ \citep{Stark_2011,Schenker_2014}. Intriguingly, there is evidence that the redshift evolution is less dramatic than previously thought \citep{Pentericci_2018}, and the main difference between LAEs and non-LAEs is that the latter are significantly dustier \citep{De_Barros_2017}. The Ly$\alpha$ escape fractions from low mass $z < 6$ galaxies can therefore be quite high, especially given the tendency of supernova explosions and radiation pressure to drive galactic outflows, generating a redshifted Ly$\alpha$ line \citep{Stark_2017}. At lower redshifts, where more data is available, LAEs with high equivalent widths exhibit bluer UV slopes and are younger, less massive, and have lower star formation rates \citep[e.g.][]{Hayes_2015,Trainor_2016}. Given the increasing quantity and quality of high-$z$ data, we anticipate additional progress in using LAE visibility and quasar absorption spectra as probes of reionization \citep{Furlanetto_2006,Dayal_2011,Jensen_2013,Mason_2018}, Ly$\alpha$ haloes to study the circumgalactic medium \citep[CGM;][]{Steidel_2011,Hayes_2013,Momose_2016,Leclercq_2017,Erb_2018}, and spatially/spectrally resolved properties to better understand the formation and evolution of galaxy populations.

On the theoretical side, accurate Monte Carlo radiative transfer (MCRT) simulations allow us to study the resonant scattering of Ly$\alpha$ photons within the interstellar medium (ISM) of galaxies and their subsequent transmission through the intergalactic medium (IGM). Substantial progress has been made in understanding Ly$\alpha$ escape in homogeneous media \citep{Harrington_1973,Neufeld_1990}, a clumpy multiphase ISM \citep{Neufeld_1991,Hansen_Oh_2006,Dijkstra_2012,Laursen_2013,Duval_2014,Gronke_2014,Gronke_2016}, expanding shell environments \citep{Dijkstra_2006,Verhamme_2006,Gronke_2015}, and idealized anisotropic environments \citep{Zheng_Wallace_2014,Behrens_2014,Smith_2015}. These models have been used to interpret observed spectra with varying success \citep[e.g.][]{Verhamme_2008,Orlitova_2018}. On the other hand, MCRT has been applied to galaxy simulations in post-processing from cosmological initial conditions \citep{Tasitsiomi_2006,Laursen_AMR_2009,Faucher-Giguere_2010,Zheng_2010,Barnes_2011,Yajima_2012,Smith_2015,Trebitsch_2016} and isolated disk galaxy configurations \citep{Verhamme_2012,Behrens_Braun_2014}. However, the analysis and interpretation of MCRT results from hydrodynamical simulations is further complicated by the inability to isolate physical effects by adjusting parameters one at a time. Furthermore, successful models will eventually need to simultaneously and self-consistently resolve the sub-structure of the ISM and the large-scale structure of the increasingly neutral IGM in the high-$z$ Universe. Still, the field is maturing as multiple groups are incorporating Ly$\alpha$ MCRT into state-of-the-art adaptive resolution simulation pipelines, fostering high level theoretical and observational synergies.

Ideally, by matching Ly$\alpha$ simulations and observations we can extract additional missing information about high-$z$ galaxies. This requires more robust models for individual LAEs and the emergent statistical properties of galaxy populations. The main goal of this paper is to explore the physics of Ly$\alpha$ escape from individual high-$z$ galaxies as comprehensively as possible. Such realistic radiative transfer modeling can provide more accurate predictions for future observations and more meaningful constraints from currently available data. For example, \citet{Laursen_2018} have recently demonstrated that the UltraVISTA survey, with a narrowband filter tuned to detect Ly$\alpha$ emission at $z = 8.8$, only has a $\sim 10\%$ probability of detecting any LAEs at all once the survey has finished, which is a significantly smaller success rate than earlier predictions. This is attributed to more realistic (i.e. simulation-based) models and a lower than expected survey depth. Additionally, understanding Ly$\alpha$ escape is highly relevant for 21-cm cosmology \citep{Pritchard2012}. In fact, the recent tentative detection by the Experiment to Detect the Global Epoch of Reionization Signature (EDGES) of a global absorption trough centred at $z \approx 17$, presumably an imprint of the first luminous sources at cosmic dawn provides a case in point \citep{Bowman_2018,Madau_2018}. There, the physics of Ly$\alpha$ photon escape from high-$z$ sources is crucial to understand the local radiation background needed to achieve strong coupling of the spin and kinetic temperatures.

In this paper, we present a detailed Ly$\alpha$ radiative transfer study of a star-forming galaxy over the redshift range of $z = 5$--$7$, selected to be sufficiently bright to have detectable Ly$\alpha$ with the upcoming \textit{James Webb Space Telescope} (\textit{JWST}) but not too massive to be very dusty and less important for reionization. In Section~\ref{sec:methods}, we describe the high resolution cosmological simulation and various radiative transfer methodologies, including for ionizing, Ly$\alpha$, and UV continuum photons. In Section~\ref{sec:results}, we discuss the time, spatial, and angular resolved properties of the Ly$\alpha$ photons immediately after emission, after scattering through the ISM and CGM, and finally after transmission through the IGM. We also isolate properties for the recombination and collisional excitation mechanisms. Our simulations provide insights on the Ly$\alpha$ escape fraction, angular anisotropy, red-to-blue flux ratio, half-light radius, halo scale height, frequency moments, peak offset, full-width at half-maximum, and equivalent width of the emergent spectra. Beyond this, we also characterize and illustrate quantities related to optically-thin lines and the surface brightness, line flux, moment maps, azimuthal variations, and angular power spectra. In Section~\ref{sec:observability}, we consider the observability of our target galaxy to better inform LAE survey strategies. In Section~\ref{sec:insights}, we perform a statistical analysis to construct a model predicting the escape fraction of Ly$\alpha$ photons from high-$z$ galaxies based on the local emission environment. In Section~\ref{sec:Lya_radiation_pressure}, we calculate the role of Ly$\alpha$ radiation pressure from the post-processed cosmological simulations. Finally, in Section~\ref{sec:conc}, we briefly discuss the implications of our work and the desirable requirements of future Ly$\alpha$ radiative transfer studies.

\section{Methods}
\label{sec:methods}
We now describe the cosmological simulation and post-processing methods to accurately calculate Ly$\alpha$ observables. In particular, we provide the context for the results of our study in Section~\ref{sec:results}, regarding the spectral properties and evolution of high-redshift Ly$\alpha$ emitting galaxies.

\begin{figure*}
  \centering
  \includegraphics[width=\textwidth]{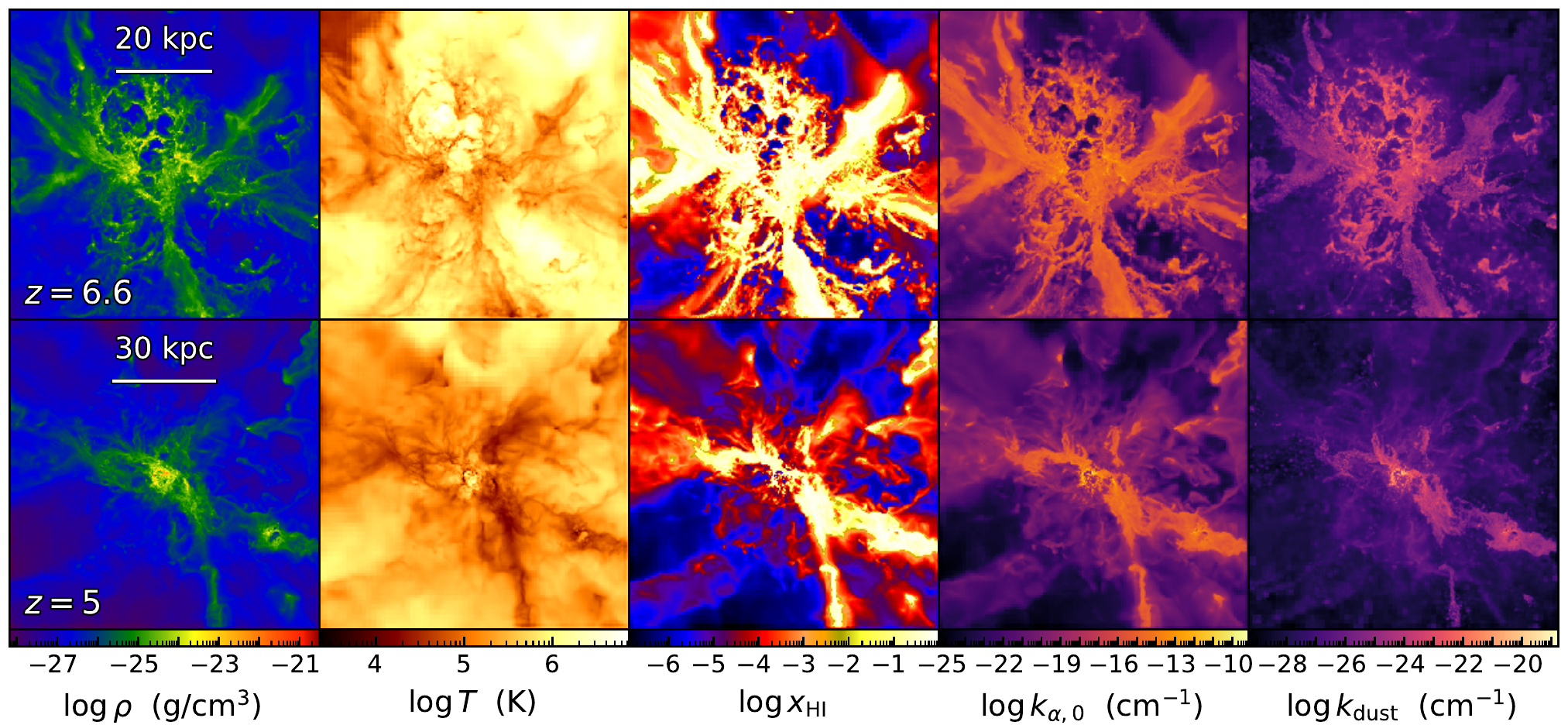}
  \caption{Projected images of the target galaxy at $z = 6.6$ and $z = 5$, illustrating the redshift evolution of the line-of-sight mass-weighted density $\rho$, temperature $T$, neutral hydrogen fraction $x_\text{\HI} \equiv n_\text{\HI} / n_\text{H}$, Ly$\alpha$ absorption coefficient at line centre $k_{\alpha,0}$, and dust absorption coefficient $k_\text{dust}$. See Section~\ref{sec:Lya_RT} for a description of the absorption coefficients employed in the Ly$\alpha$ radiative transfer calculations. The density, temperature, and ionization states are extremely anisotropic and undergo significant morphological changes with redshift.}
  \label{fig:spatial}
\end{figure*}

\subsection{Simulation setup}
\label{sec:sims}
In this paper we examine a single cosmological zoom-in simulation from the suite of galaxies presented in \citet{Ma_2018}, specifically the galaxy labeled \verb"z5m11c", which was selected based on being a typical low-mass LAE at $z = 5$. The simulation is a part of the Feedback In Realistic Environments project\footnote{See the FIRE project web site at: \href{http://fire.northwestern.edu}{http://fire.northwestern.edu}.} \citep[FIRE-2;][]{Hopkins_FIRE2_2017}, which is an updated version of the first implementation \citep[FIRE-1;][]{Hopkins_FIRE_2014}. The simulation employs the meshless finite-mass (MFM) method in the \textsc{gizmo} hydrodynamics code \citep{Hopkins_GIZMO_2015}, with initial conditions and halo selection criteria as described in \citet{Ma_2018}. The multi-scale zoom-in process ensures zero contamination from low-resolution particles within $2 R_\text{vir}$ at $z = 5$. For the \verb"z5m11c" model the initial particle masses are $m_\text{gas} = 7126.5\,\Msun$ for gas and $m_\text{DM} =  3.9 \times 10^4\,\Msun$ for dark matter, while the minimum Plummer-equivalent force softening lengths of gas, star, and high-resolution dark matter particles are $\epsilon_\text{gas} = 0.42\,\text{pc}$, $\epsilon_\star =  2.1\,\text{pc}$, and $\epsilon_\text{DM} = 42\,\text{pc}$, respectively. The softening lengths are fixed for stars and dark matter but adaptive for gas, with the minimum value in comoving units at $z > 9$ and physical units thereafter.

\begin{figure}
  \centering
  \includegraphics[width=\columnwidth]{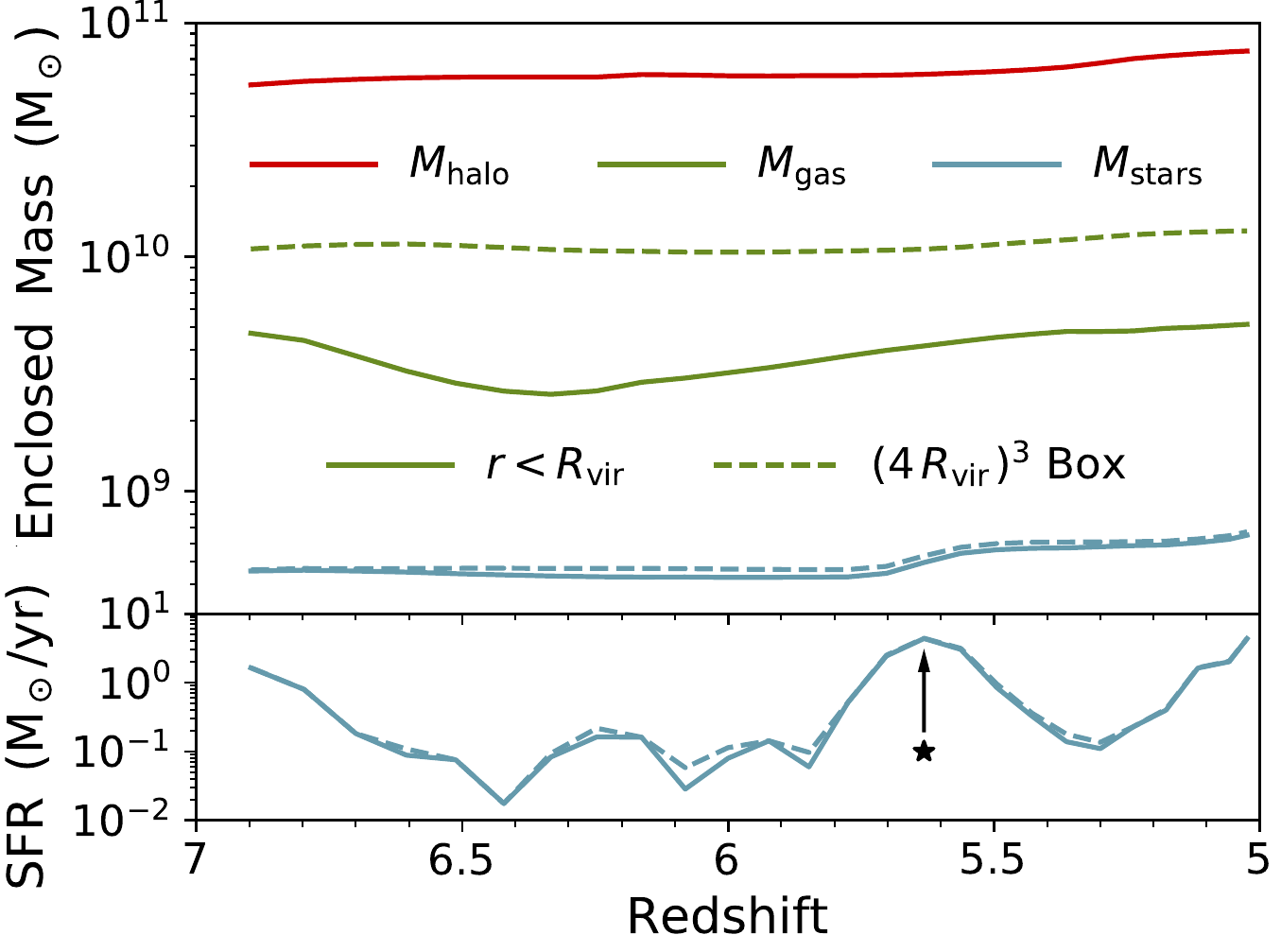}
  \caption{The redshift evolution of the target galaxy's halo mass $M_\text{halo}$, along with the total stellar mass $M_\star$ and gas mass $M_\text{gas}$, within the the virial radius $R_\text{vir}$. The total gas and stellar content within the $(4\,R_\text{vir})^3$ MCRT simulation domain remains fairly constant, although at $z \approx 5.63$ the galaxy undergoes significant starburst activity.}
  \label{fig:Mass}
\end{figure}

In Fig.~\ref{fig:spatial} we show the spatial distribution of the gas based on mass-weighted projections of hydrodynamical quantities related to Ly$\alpha$ radiative transfer, including the density, temperature, ionization state, and dust content. Furthermore, in Fig.~\ref{fig:Mass} we show the redshift evolution of the halo mass $M_\text{halo}$ of the target galaxy, along with the total stellar mass $M_\star$ and gas mass $M_\text{gas}$ within the virial radius $R_\text{vir}$ and $(4\,R_\text{vir})^3$ MCRT simulation domain. As a summary reference we provide several time-averaged quantities in Table~\ref{tab:time_avg_int}, both over the entire redshift range and further divided into intervals to indicate any evolution. The time-averaged masses are $\langle M_\text{halo} \rangle \approx 6.2 \times 10^{10}\,\Msun$, $\langle M_\text{gas} \rangle \approx 1.1 \times 10^{10}\,\Msun$, and $\langle M_\star \rangle \approx 5.2 \times 10^8\,\Msun$. The corresponding virial radius for the central halo is $\langle R_\text{vir} \rangle \approx 19\,\text{kpc}$, with a maximum circular velocity of $\langle V_\text{max} \rangle \approx 127\,\text{km\,s}^{-1}$ reached at a radius of $\langle R_\text{max} \rangle \approx 11\,\text{kpc}$. We use the Binary Population and Spectral Synthesis (BPASS) models \citep[version 2.0;][]{Eldridge_2008,Stanway_2016} to compute the spectral energy distribution (SED) of each star particle based on its age and metallicity. The BPASS models include binary stellar populations, taking into account effects from mass transfer, common envelope phases, binary mergers, and quasi-homogeneous evolution at low metallicities. The inclusion of binaries is often invoked to reproduce the nebular line properties in $z \approx 2$--$3$ galaxies \citep{Steidel_2016}. In this paper, we adopt a \citet{Kroupa_2002} initial mass function (IMF) from $0.1$--$100\,\Msun$, with IMF slopes of $-1.30$ from $0.1$--$0.5\,\Msun$ and $-2.35$ from $0.5$--$100\,\Msun$. Under these conditions, we calculate an intrinsic ionizing ($>13.6\,\text{eV}$) photon emission rate of $\langle \dot{N}_\text{ion} \rangle \approx 1.4 \times 10^{53}\,\text{s}^{-1}$ and a luminosity of $\langle L_\text{ion} \rangle \approx 4.9 \times 10^{42}\,\text{erg\,s}^{-1}$.

\subsection{Intrinsic emission}
The dominant source of Ly$\alpha$ photons is recombination radiation due to nebular emission. In a study by \citet{Ma_2015}, the approximate prescription for on-the-fly photoionization in the FIRE simulations was found to be fairly consistent with more accurate post-processing calculations. However, to ensure the reliability of the line emissivities and scattering opacities necessary for Ly$\alpha$ radiative transfer, we must first perform Lyman continuum~(LyC) radiative transfer to update the local ionization states.\footnote{Even though the ionization states change, we keep the gas temperature the same as the original simulations. This leads to some under-heated \HII\ regions, but these cells would only make a small change to the total Ly$\alpha$ emissivity as the recombination coefficient depends weakly on temperature. In the halo, the hydrodynamical simulations already fairly capture both photoheating by the ionizing background and shock-heating of accreted gas.} To retain the high resolution of the simulation we deposited the unstructured mesh data onto an adaptive octree grid structure with a refinement criterion of no more than two particles per leaf cell. We then employed an octree version of the Monte Carlo code described in \citet{Ma_2015,Ma_2016} based on the \textsc{sedona} code \citep{Kasen_2006}, which iteratively solves the ionization states assuming ionization equilibrium. The code accounts for photoionization from each star particle and a uniform, redshift-dependent meta-galactic ionizing background and collisional ionization \citep{Faucher-Giguere_2009}. Each of these mechanisms contribute to the resolved Ly$\alpha$ emissivity in the simulation. Therefore, the Ly$\alpha$ luminosity due to recombination is
\begin{equation} \label{eq:L_alpha_rec}
  L_\alpha^\text{rec} = h\nu_\alpha \int P_\text{B}(T) \alpha_\text{B}(T) n_e n_p \text{d}V \, ,
\end{equation}
where $h \nu_\alpha = 10.2$\,eV, the Ly$\alpha$ conversion probability per recombination event is $P_\text{B}(T) \approx 0.68$, the case~B recombination coefficient is $\alpha_\text{B}$, and the number densities $n_e$ and $n_p$ are for free electrons and protons, respectively \citep{Cantalupo_2008,Dijkstra_PASA_2014}. At these redshifts we also model the radiation arising from collisions between free electrons and neutral hydrogen atoms, which populate excited states and produce additional Ly$\alpha$ `cooling' emission. The Ly$\alpha$ luminosity due to collisional excitation is
\begin{equation} \label{eq:L_alpha_col}
  L_\alpha^\text{col} = h\nu_\alpha \int q_{1s2p}(T) n_e n_\text{\HI\,} \text{d}V \, ,
\end{equation}
where the temperature-dependent rate coefficient $q_{1s2p}(T)$ is taken from \citet{Scholz_1991}. We note that this term can be highly uncertain due to the exponential dependence on temperature around $T \sim 10^4\,\text{K}$, and thus requires a proper treatment of the thermal effects of ionizing radiation, e.g. as demonstrated in \citet{Faucher-Giguere_2010}.

As expected, during these redshifts the gas within the virial radius $R_\text{vir}$ is mostly ionized by volume but neutral by density. Specifically, after the post-processing calculations the time-averaged ionized fraction is $\langle x_\text{\HII} \rangle_\text{V} \approx 97.9$\%, when weighted by volume, and $\langle x_\text{\HII} \rangle_\rho \approx 4.7$\%, when weighted by density (Table~\ref{tab:time_avg_int}), where $x_\text{\HII} \equiv n_\text{\HII} / n_\text{H}$ is the ratio of ionized and total hydrogen number densities. The escape fraction of ionizing photons from high-$z$ galaxies can fluctuate significantly in response to the specific history of mergers, cosmological cold gas flows, and starburst activity. We calculate the time-averaged LyC escape fraction as $\langle f_\text{esc}^\text{LyC} \rangle \approx 13.3$\%, such that the ionizing luminosity escaping beyond the CGM ($\sim 2\,R_\text{vir}$) is $\langle L_\text{ion,esc} \rangle \approx 4.7 \times 10^{41}\,\text{erg\,s}^{-1}$.

\begin{figure}
  \centering
  \includegraphics[width=\columnwidth]{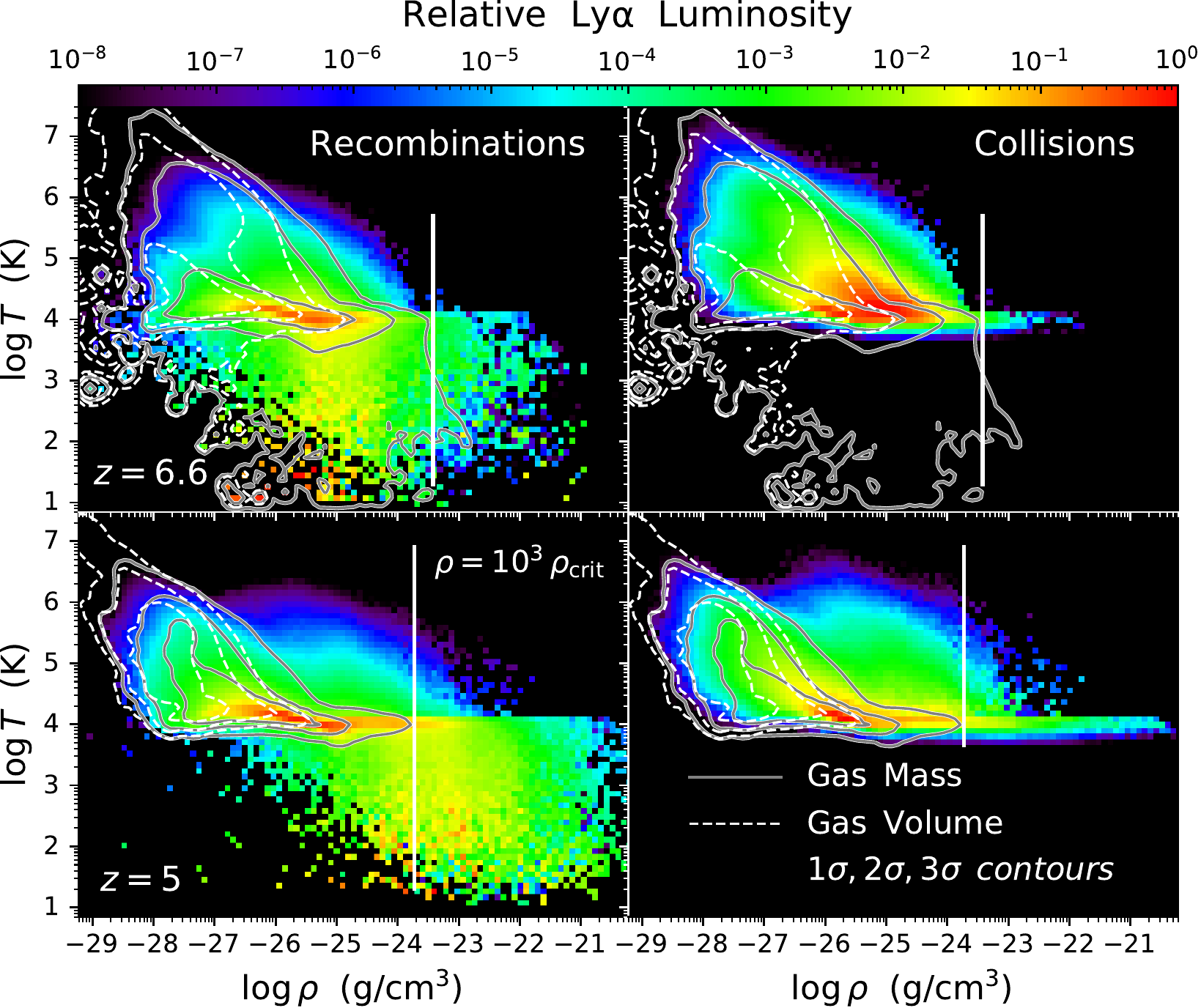}
  \caption{Relative Ly$\alpha$ luminosity in the density--temperature phase plane at $z =6.6$ and $z = 5$. Ly$\alpha$ photons are produced by recombination of ionized hydrogen and collisional excitation of hydrogen atoms by free electrons. The two mechanisms show significant overlap at warm temperatures ($\sim 10^4$\,K) and moderate densities ($\sim 10^{-27}$--$10^{-23}\,\text{g\,cm}^{-3}$), tracing photoionized \HII\ regions and shock heated gas in the circumgalactic medium (CGM). The recombination emission at $T < 10^4$\,K is likely associated with relic \HII\ region cooling and inefficient photoheating, whereas collisional excitation requires higher temperatures. However, see footnote~2 regarding the Ly$\alpha$ emissivity from under-heated \HII\ regions. The solid (dashed) contours show the phases containing most of the mass (volume).}
  \label{fig:phase_rho_T}
\end{figure}

The corresponding intrinsic Ly$\alpha$ luminosity within the $(4\,R_\text{vir})^3$ MCRT simulation domain based on Equations~(\ref{eq:L_alpha_rec})~and~(\ref{eq:L_alpha_col}) is $\langle L_\alpha \rangle \approx 1.3 \times 10^{42}\,\text{erg\,s}^{-1}$, but also fluctuates in proportion to the intrinsic ionizing luminosity with a peak of $L_\alpha \approx 10^{43}\,\text{erg\,s}^{-1}$ at $z \approx 5.63$. In Fig.~\ref{fig:phase_rho_T}, we illustrate the relative Ly$\alpha$ luminosity across the density--temperature phase space. Most of the emission comes from moderate density \HII\ regions with a temperature of $T \sim 10^4$\,K, characteristic of Ly$\alpha$ cooling. We have verified that our approach of modeling the recombination radiation directly is consistent with the luminosity calculated from star particles, as the LyC code is photon conserving. The spatial distribution of recombination emission is also more realistic within such resolved \HII\ region morphologies than emitting Ly$\alpha$ radiation from star particles directly.

\subsection{Ly\texorpdfstring{$\balpha$}{α} radiative transfer}
\label{sec:Lya_RT}
We perform Monte Carlo Ly$\alpha$ radiative transfer with the Cosmic Ly$\alpha$ Transfer code
% (\textsc{colt}) as described by \citet{Smith_2015}.
\citet[\textsc{colt};][]{Smith_2015}.
Photon packets are inserted according to the recombination and collisional luminosities from Equations~(\ref{eq:L_alpha_rec}) and (\ref{eq:L_alpha_col}). This is achieved by constructing the cumulative distribution function from individual cells, drawing a random number over the unit interval to find the index of the source, and assigning the emission position uniformly within the cell volume. This also allows us to link the final outcome of each photon trajectories to the original emission environment. The subsequent transport of photon packets follows the Monte Carlo procedure with a local Ly$\alpha$ absorption coefficient of
\begin{equation}
  k_\alpha = n_\text{\HI}\,\sigma_\alpha(\nu) \, ,
\end{equation}
where the Ly$\alpha$ cross-section $\sigma_\alpha(\nu)$ is proportional to the Voigt line profile. At line centre, $\nu_0 = 2.466 \times 10^{15}\,\text{Hz}$, the scattering cross-section obtains a maximum value of $\sigma_0 = 5.898 \times 10^{-14}\,T_4^{-1/2}\,\text{cm}^2$, where $T_4 \equiv T / (10^4\,\text{K})$. For a visual comparison, we also show the projected line centre absorption coefficient $k_{\alpha,0} = n_\text{\HI}\,\sigma_0$ in Fig.~\ref{fig:spatial}.

We follow the prescription of \citet{Laursen_2009} for calculating the dust content within each cell. Specifically, we compute the effective local dust absorption coefficient as
\begin{equation} \label{eq:k_dust}
  k_\text{d} = (x_\text{\HI} + f_\text{ion} x_\text{\HII}) \, n_\text{H}\,\sigma_\text{d,SMC}\,\frac{Z}{Z_\text{SMC}} \, ,
\end{equation}
where $n_\text{H}$ denotes the hydrogen number density such that $x_\text{\HI} \equiv n_\text{\HI} / n_\text{H}$ and $x_\text{\HII} \equiv n_\text{\HII} / n_\text{H}$ are the neutral and ionized fractions, respectively. We assume the dust survival fraction within ionized regions $f_\text{ion}$ is $1$\%, although in reality the dust abundance also depends on the temperature of the ionized medium. Furthermore, this model assumes Small Magellanic Cloud (SMC) type dust with an effective cross-section per hydrogen atom of $\sigma_\text{d,SMC} \approx 3.95 \times 10^{-22}\,\text{cm}^2$, where $\log(Z_\text{SMC}/\Zsun) \approx -0.6$ with $\Zsun \approx 0.0134$. Because this particular FIRE simulation does not include on-the-fly subgrid metal diffusion we use a smoothed version of the metallicity $Z$ based on a cubic spline kernel over the 32 nearest neighbor particles before mapping to the octree. Finally, the dust radiative transfer also follows the methodology outlined by \citet{Laursen_2009}, i.e. with a fiducial dust scattering albedo of $A = 0.32$ and Henyey-Greenstein phase function with an asymmetry parameter of $g = \langle \cos\theta \rangle = 0.73$. We also show the projected dust absorption coefficient in Fig.~\ref{fig:spatial}.

We now describe the binning method to obtain line-of-sight (LOS) statistics for a large number of directions. We perform separate \colt\ simulations for recombination and collisional excitation emission, each with $10^7$ photon packets. The final state of all escaped and absorbed photons is recorded, i.e. the frequency $\nu_i$, position $\bm{r}_i$, and direction $\hat{\bm{k}}_i$, where $i$ denotes the photon index. Although Poisson noise becomes significant at high angular and spatial resolution, we smooth the packet discretization by a weighting kernel with a specific estimated error criterion. For spatially integrated quantities in a given direction $\hat{\bm{n}}$, we calculate the directional cosine of each photon as $\mu_i \equiv \cos \theta_i = \hat{\bm{n}} \cdot \hat{\bm{k}}_i$ and $\Delta \mu_i \equiv 1 - \mu_i$. For reference, a tophat filter has constant weight if $\mu \geq \mu_\text{min}$ and zero elsewhere, for an effective photon number of $N_\text{ph,eff} \approx N_\text{ph} \Delta\mu_\text{min}/2$. Under an isotropic distribution the approximate error per pixel is $\approx 1/\sqrt{N_\text{ph,eff}}$, such that $1\%$ noise is obtained for $10^7$ photons with $\Delta\mu_\text{min} \approx 0.002$ or $\Delta\theta_\text{min} \approx 3\fdg6$ defining the effective resolution. However, in this paper we employ a Gaussian filter to increase the sensitivity and avoid edge effects, represented as
\begin{equation}
  W_\text{G}(\mu) \propto
  \begin{cases}
    \exp\left(-\frac{\Delta\mu^2}{2 \sigma_\mu^2}\right) & \mu \geq 0 \\
    0 & \mu < 0
  \end{cases} \, ,
\end{equation}
which when integrated gives an effective number of photons of $N_\text{ph,eff} \approx N_\text{ph} \sqrt{\frac{\upi}{2}} \frac{\sigma_\mu}{2} \text{erf}\left(\frac{1}{\sqrt{2}\sigma_\mu}\right)$. Numerically solving the analogous relation yields an approximate error per pixel of $1\%$ for $10^7$ photons with  a standard deviation of $\sigma_\mu \approx 0.0016$ or $\sigma_\theta \approx 3\fdg2$ defining the effective resolution. We find that this prescription generally provides reasonable control on the Poisson noise due to the Monte Carlo discretization. We also note that significant uncertainty from noise has the greatest impact on sightlines that are less likely to be observable.

\begin{table}
  \caption{Time-averaged quantities over the redshift interval $z = 5$--$7$, with additional subdivisions illustrating the redshift evolution. The type column denotes intrinsic properties of the galaxy (\textsc{int}) and quantities that rely on the ionizing radiative transfer calculations (\textsc{ion}). Specifically, $M_\text{gas}$, $M_\star$, $\dot{N}_\text{ion}$, $L_\text{ion}$, $\langle x_\text{\HII}\rangle_\text{V}$, and $\langle x_\text{\HII}\rangle_\rho$ are calculated for the gas within the virial radius, while all other quantities are for the $(4\,R_\text{vir})^3$ MCRT simulation domain to capture the extended Ly$\alpha$ halo. We also report the 1$\sigma$ standard deviation, defined as $\sigma_f^2 = \int (f(t)-\mu_f)^2\,\text{d}t / \int \text{d}t$, where the time-averaged value is $\mu_f = \int f(t)\,\text{d}t / \int \text{d}t$. In cases with LOS fluctuations we provide the median and asymmetric 1$\sigma$ (68.27\%) time-weighted confidence levels, based on binning the escaped photon packets into $3072$ healpix directions of equal solid angle.}
  \label{tab:time_avg_int}
  \addtolength{\tabcolsep}{-1.5pt}
  \renewcommand{\arraystretch}{1.1}
  \begin{tabular}{@{} l ccc @{}}
    \hline
    Quantity \hfill Type\hspace{-.1cm} & $z = 5$--$7$ & $z = 6$--$7$ & $z = 5$--$6$ \\
    \hline
    \vspace{.05cm}
    $\log M_\text{halo}$ \hfill [$\Msun$]\;\;\,\textsc{int} & $10.79\pm0.04$ & $10.77\pm0.01$ & $10.81\pm0.04$ \\
    $\log M_\text{gas}$ \hfill [$\Msun$]\;\;\,\textsc{int} & $9.58\pm0.1$ & $9.50\pm0.08$ & $9.64\pm0.06$ \\
    $\log M_\star$ \hfill [$\Msun$]\;\;\,\textsc{int} & $8.69\pm0.06$ & $8.65\pm0.01$ & $8.72\pm0.06$ \\
    $R_\text{vir}$ \hfill [$\text{kpc}$]\;\;\,\textsc{int} & $18.80\pm1.95$ & $16.89\pm0.70$ & $20.05\pm1.42$ \\
    $R_\text{max}$ \hfill [$\text{kpc}$]\;\;\,\textsc{int} & $11.24\pm1.92$ & $12.53\pm1.37$ & $10.39\pm1.75$ \\
    $V_\text{max}$ \hfill [$\text{km\,s}^{-1}$]\;\;\,\textsc{int} & $127.0\pm4.4$ & $126.9\pm5.6$ & $127.0\pm3.5$ \\
    $\log \dot{N}_\text{ion}$ \hfill [$\text{s}^{-1}$]\;\;\,\textsc{int} & $53.10\pm0.51$ & $52.83\pm0.39$ & $53.28\pm0.51$ \\
    $\log L_\text{ion}$ \hfill [$\text{erg\,s}^{-1}$]\;\;\,\textsc{int} & $42.65\pm0.52$ & $42.38\pm0.40$ & $42.83\pm0.51$ \\
    \hline
    $\langle x_\text{\HII}\rangle_\text{V}$ \hfill [\%]\;\;\,\textsc{ion} & $97.91\pm1.14$ & $97.44\pm1.03$ & $98.21\pm1.11$ \\
    $\langle x_\text{\HII}\rangle_\rho$ \hfill [\%]\;\;\,\textsc{ion} & $4.73\pm1.02$ & $5.22\pm0.96$ & $4.41\pm0.93$ \\
    $f_\text{esc}^\text{LyC}$ \hfill [\%]\;\;\,\textsc{ion} & $13.34\pm9.64$ & $16.65\pm9.35$ & $11.17\pm9.20$ \vspace{.02cm} \\
    $f_\text{esc}^\text{UV}$ \hfill [\%]\;\;\,\textsc{ion} & $92.80^{+3.94}_{-17.07}$ & $96.02^{+1.91}_{-3.23}$ & $87.05^{+7.57}_{-18.72}$ \vspace{.05cm} \\
    $M_\text{UV}$ \hfill \textsc{ion} & $-18.25^{+0.55}_{-0.95}$ & $-18.17^{+0.29}_{-0.76}$ & $-18.37^{+0.71}_{-0.84}$ \vspace{.05cm} \\
    $R_{1/2,\text{UV}}$ \hfill [$\text{kpc}$]\;\;\,\textsc{ion} & $2.86^{+0.77}_{-1.28}$ & $3.17^{+0.62}_{-0.69}$ & $2.43^{+1.04}_{-1.14}$ \vspace{.02cm} \\
    $\log L_\text{ion}$ \hfill [$\text{erg\,s}^{-1}$]\;\;\,\textsc{ion} & $41.67\pm0.68$ & $41.56\pm0.57$ & $41.74\pm0.73$ \\
    \hline
  \end{tabular}
  \addtolength{\tabcolsep}{1.5pt}
  \renewcommand{\arraystretch}{0.9090909090909090909}
\end{table}

\begin{figure}
  \centering
  \includegraphics[width=\columnwidth]{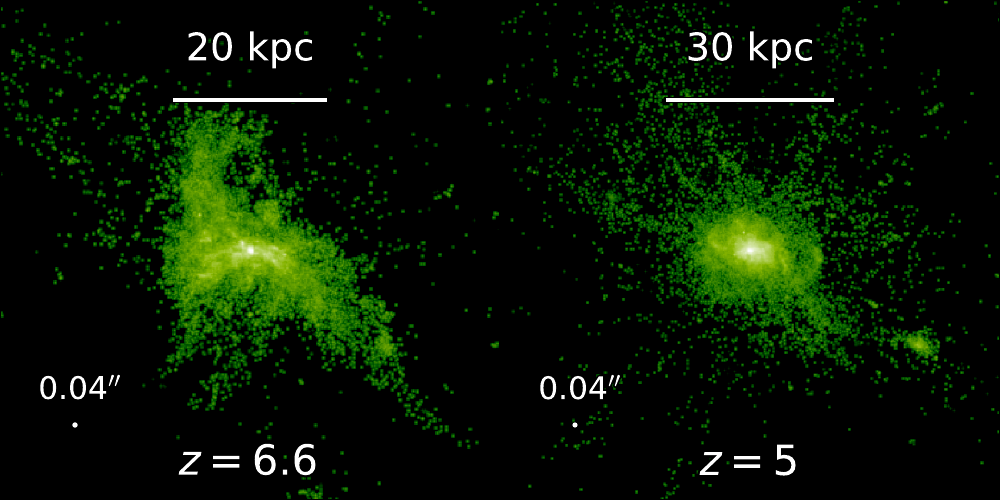}
  \includegraphics[width=\columnwidth]{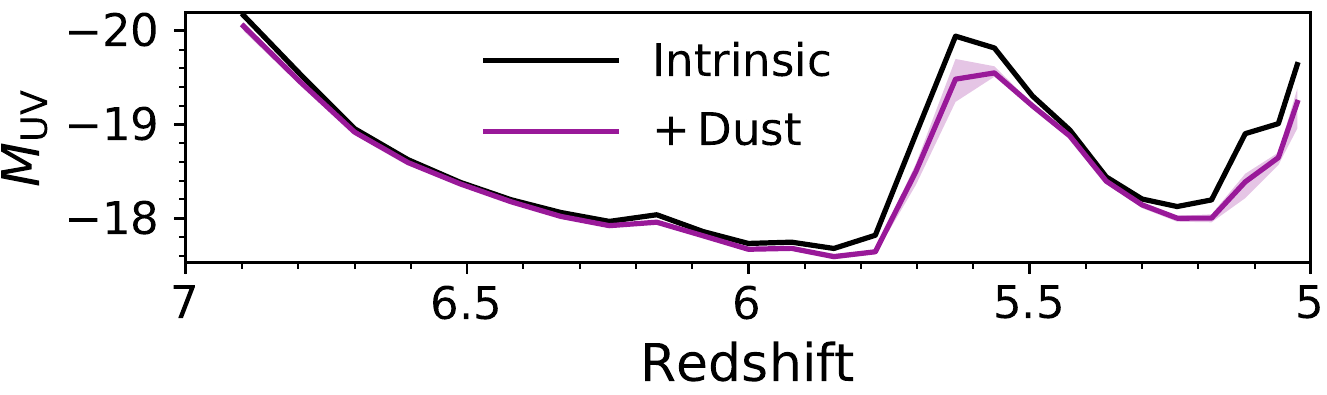}
  \caption{Spatial images of the UV continuum for the target galaxy at $z = 6.6$ and $z = 5$, showing the evolving distribution of stellar populations. The images are rendered after absorption by dust and Gaussian smoothing with $\text{FWHM} \approx 0\farcs04$ to simulate the \textit{JWST} NIRCam aperture. In the lower panel we show the total absolute magnitude, with 1$\sigma$ LOS variations.}
  \label{fig:M_UV}
\end{figure}

\subsection{IGM transmission}
\label{sec:IGM_transmission}
The \textsc{colt} output represents the emergent Ly$\alpha$ observables without accounting for subsequent scattering in the IGM. To include this important effect we use the frequency- and redshift-dependent transmission curves of \citet{Laursen_2011}, which have been kindly provided to us by the author. Specifically, we use their `benchmark' Model~1 results as described in relation to their figures~2~and~3. The curves provide the median and 1$\sigma$ statistics based on a large number of sightlines ($\gtrsim 10^3$) cast from several hundreds of galaxies through a simulated cosmological volume. Ly$\alpha$ photons are generally free streaming beyond a few virial radii, so in this paper `ISM' denotes escape from the $(4\,R_\text{vir})^3$ MCRT simulation domain, while `IGM' means that each photon is reweighted by $\exp[-\tau_\textsc{igm}(\nu,z)]$ from the transmission model. Any IGM model implemented in this manner should be considered in the context of nontrivial variations depending on the particular galaxy and sightline. We are not currently aware of any Ly$\alpha$ radiative transfer simulations that self-consistently follow the resonant scattering through both the highly resolved ISM of individual galaxies and the more diffuse IGM throughout cosmological volumes. Such a treatment is beyond the scope of the exploratory analysis of this paper but will be pursued in future work.

\subsection{UV continuum}
\label{sec:UV_continuum}
We also calculate the escape of non-ionizing UV continuum photons, which is necessary for observational comparison to the Ly$\alpha$ rest frame equivalent width $\text{EW}_{\text{Ly}\alpha,0}$. Given the uncertainties in this study it is sufficient to employ single-scattering monochromatic dust attenuation with an absorption coefficient given by Equation~(\ref{eq:k_dust}), and we expect a more accurate treatment with multiple scattering and frequency dependence to only slightly modify our results. Specifically, the intrinsic continuum luminosity, $L_\lambda$ in units of erg\,s$^{-1}$\,\AA$^{-1}$, is found by integrating the SED of each stellar population over a filter with an effective wavelength of $\lambda_\text{UV} = 1500\,\text{\AA}$. Images and other LOS observables in the $\hat{\bmath{n}}$ direction are calculated by direct ray-tracing from each star particle, such that $f_{\lambda,\text{obs}}(\hat{\bmath{n}}) \equiv \sum_i L_{\lambda,i} \exp(-\tau_{\text{d},i})/(4 \upi (1+z) d_\text{L}^2)$, where the dust optical depth from star $i$ is $\tau_{\text{d},i} \equiv \int_{i,\hat{\bmath{n}}} k_\text{d}\,\text{d}\ell$. In Fig.~\ref{fig:M_UV} we show the escaped UV continuum for the target galaxy at $z = 6.6$ and $z = 5$. We also apply Gaussian smoothing with a full width at half maximum of $\text{FWHM} \approx 0\farcs04$ to simulate the \textit{JWST} NIRCam aperture. We find the time-weighted median LOS escape fraction for non-ionizing UV continuum radiation to be $\langle f_\text{esc}^\text{UV} \rangle \approx 93\%$, which decreases at lower redshift as the mass and total dust content of the galaxy increases. The dust attenuated rest frame UV absolute magnitude is $\langle M_\text{UV} \rangle \approx -18.25$ with a half-light radius of $\langle R_{1/2,\text{UV}} \rangle \approx 3\,\text{kpc} \approx 0\farcs5$, which becomes brighter and more concentrated with time.

\begin{figure}
  \centering
  \includegraphics[width=\columnwidth]{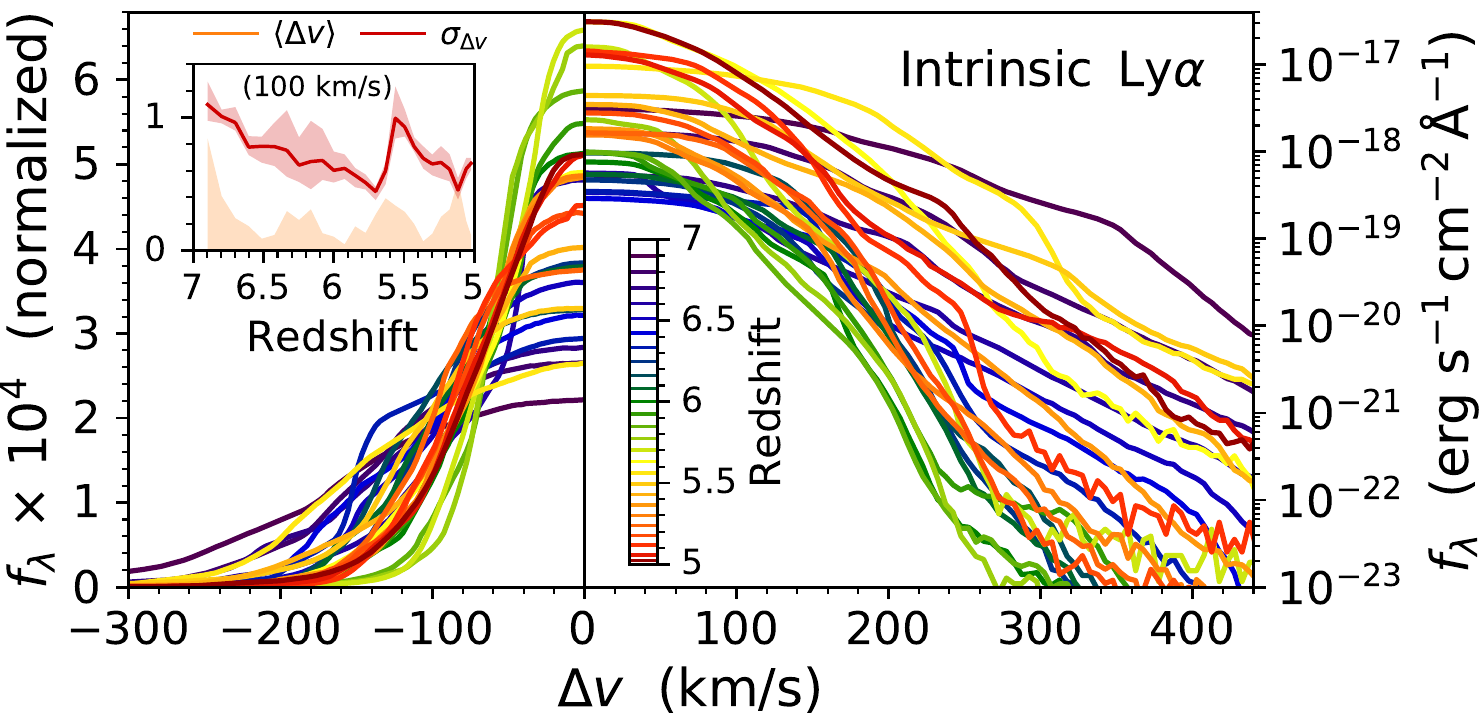}
  \includegraphics[clip, trim=1.11cm 0cm .12cm 0cm, width=.49\columnwidth]{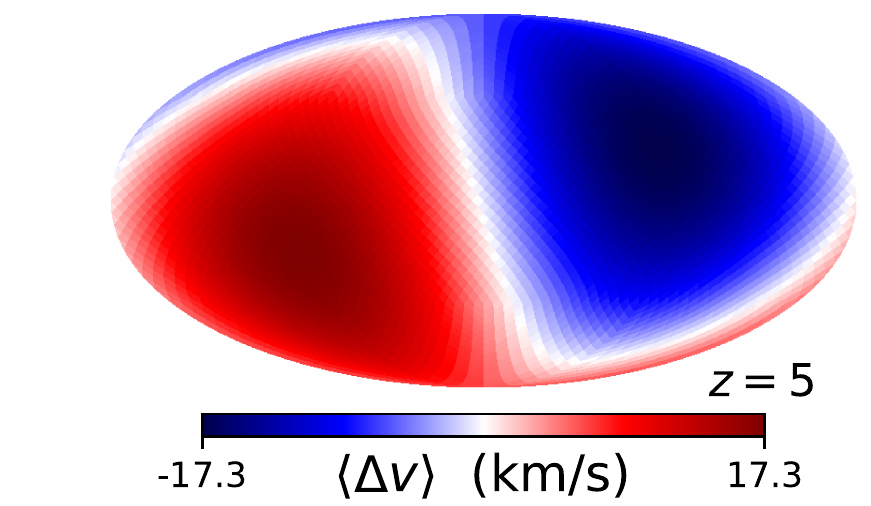}
  \includegraphics[clip, trim=1.11cm 0cm .12cm 0cm, width=.49\columnwidth]{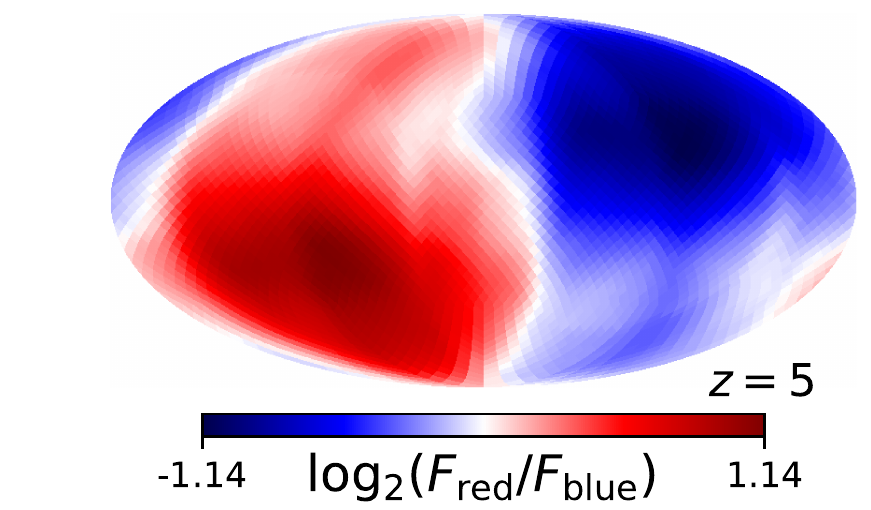}
  \includegraphics[clip, trim=1.11cm 0cm .12cm 0cm, width=.49\columnwidth]{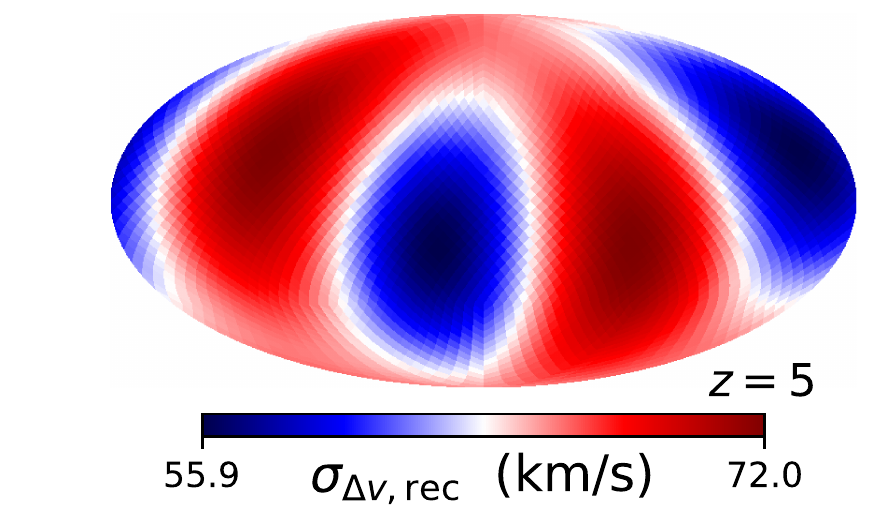}
  \includegraphics[clip, trim=1.11cm 0cm .12cm 0cm, width=.49\columnwidth]{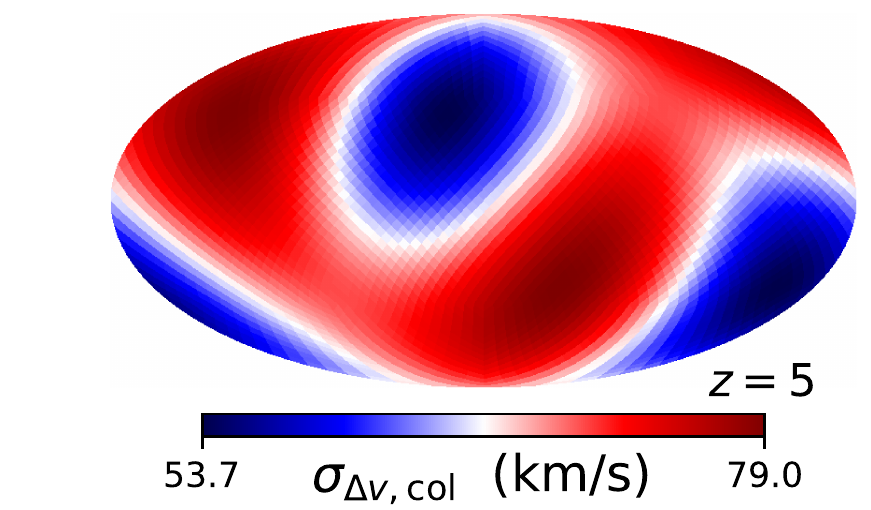}
  \caption{Angular-averaged intrinsic Ly$\alpha$ flux as a function of Doppler velocity $\Delta v = c \Delta \lambda / \lambda$ for each simulation snapshot. The mean profiles are symmetric about the line centre, however, observed spectra are more complex due to the aggregate LOS motion of the emitting gas. In the left panel we show the evolving shape of the normalized lines while the right panel shows the fluctuating luminosity. In the inset we include the redshift evolution of the flux-weighted frequency centroid $\langle \Delta v \rangle \equiv \int \Delta v f_\lambda\,\text{d}\lambda / \int f_\lambda\,\text{d}\lambda$ and standard deviation $\sigma_{\Delta v} \equiv (\langle \Delta v^2 \rangle - \langle \Delta v \rangle^2)^{1/2}$ as a measure of the velocity offset and width of the line, respectively. In the lower panels we illustrate the angular distribution of these quantities, which exhibit a dipole moment in $\langle \Delta v \rangle$ and $F_\text{red}/F_\text{blue}$ from the galaxy's proper motion and a quadrupole modulation of $\sigma_{\Delta v}$ from axially-asymmetric velocity dispersion. There is a slight kinematic misalignment between recombination and collisional emission.}
  \label{fig:H_alpha}
\end{figure}

\begin{figure}
  \centering
  \includegraphics[width=\columnwidth]{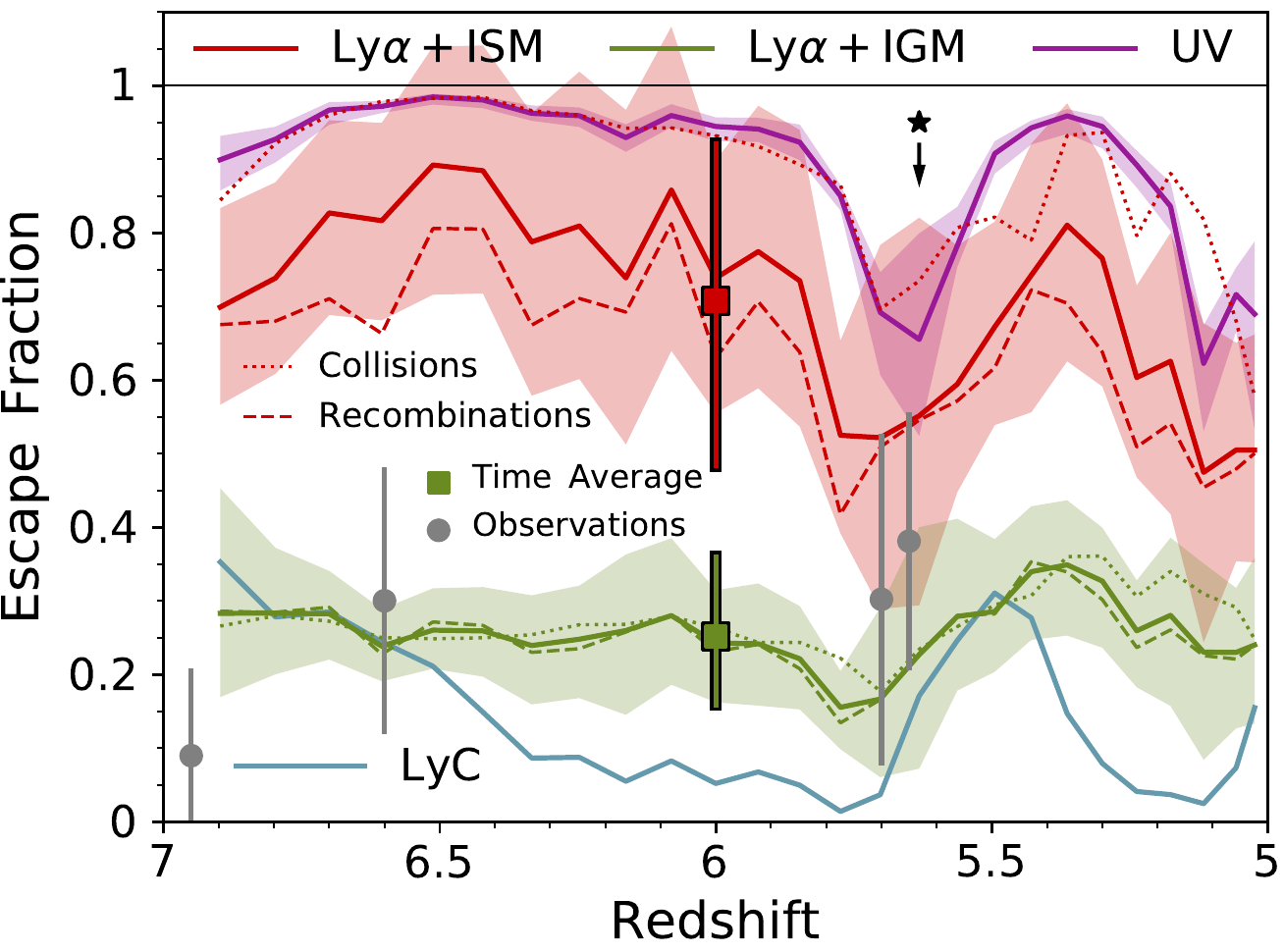}
  \caption{The redshift evolution of the Ly$\alpha$, non-ionizing UV (1500\,\AA), and LyC escape fractions from the target galaxy. The red curve is after dust absorption in the ISM while the green curves are after IGM transmission based on the models of \citet{Laursen_2011} (see text for details). The shaded regions show the 1$\sigma$ confidence levels considering different viewing angles. The gray data points are the observed Ly$\alpha$ escape fraction estimates reported by \citet{Hayes_2011}. The square points at $z = 6$ are the time-weighted averages from Tables~\ref{tab:time_avg_int}--\ref{tab:time_avg_igm}. Although the escape fraction of Ly$\alpha$ photons originating from collisions is higher than from recombinations, the transmitted escape fractions are roughly the same. The star symbol marks the main starburst.}
  \label{fig:f_esc}
\end{figure}

\begin{figure}
  \centering
  \includegraphics[width=\columnwidth]{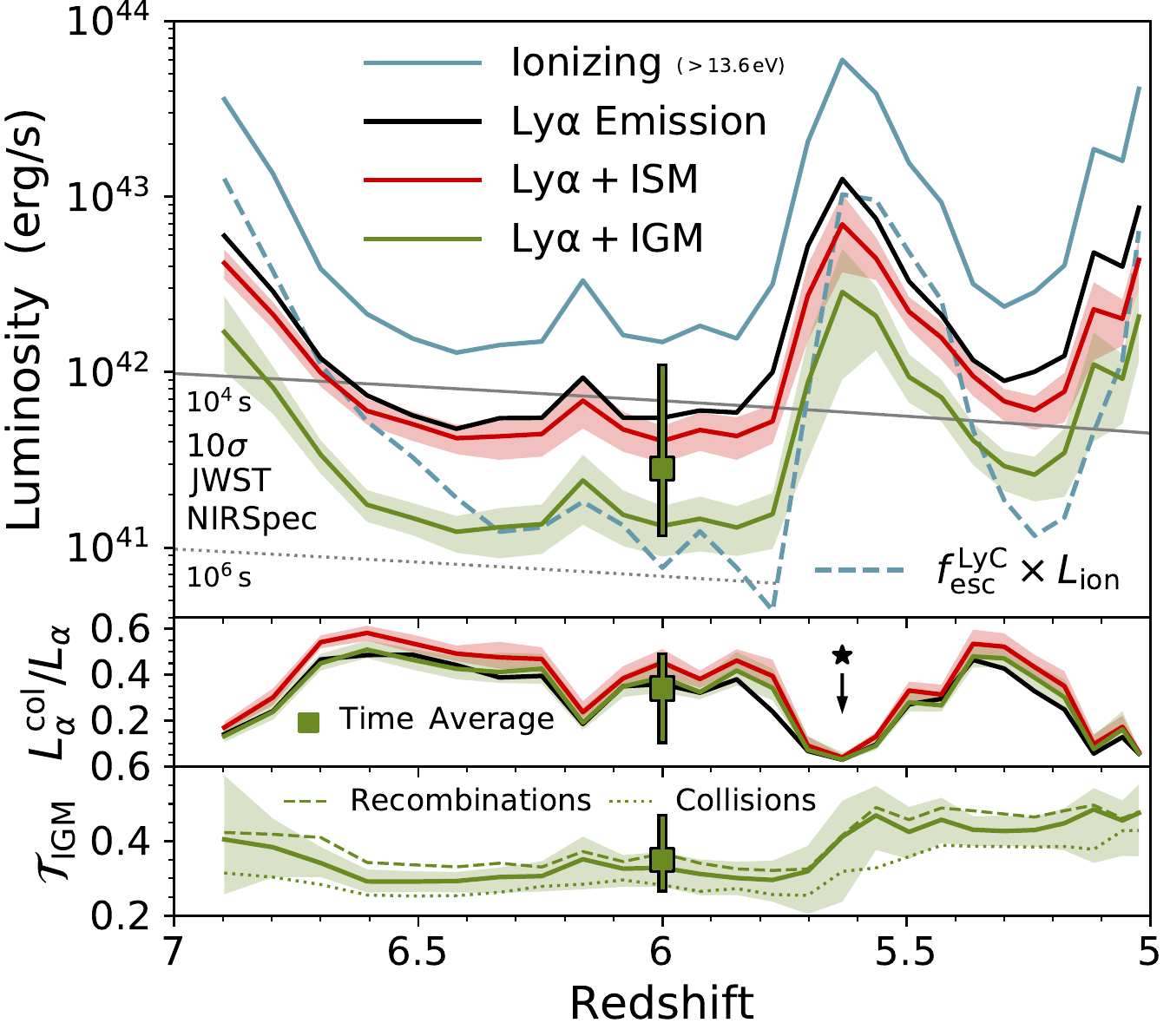}
  \caption{Redshift evolution of the luminosity for ionizing and Ly$\alpha$ radiation. The escaping luminosity corresponds to the respective escape fractions shown in Fig.~\ref{fig:f_esc} with similar styling. The lower panels illustrate the fraction of luminosity originating from collisional excitation and the overall transmission of Ly$\alpha$ photons through the IGM, defined as $\mathcal{T}_\textsc{igm} = f_{\text{esc,}\textsc{igm}}^{\text{Ly}\alpha} / f_{\text{esc,}\textsc{ism}}^{\text{Ly}\alpha}$. Ly$\alpha$ photons from collisions experience lower IGM transmission than from recombinations due to the overall bluer spectral shape, i.e. infall signatures with less resonant broadening. At $z \approx 5.63$ the galaxy experiences significant starburst activity, which is mirrored by a significant escaping LyC luminosity, $f_\text{esc}^\text{LyC} \times L_\text{ion}$. We also show approximate visibility curves for the \textit{JWST} NIRSpec instrument assuming a $10\sigma$ detection after $10^4$ and $10^6$\,s of exposure time.}
  \label{fig:L_esc}
\end{figure}

\section{Ly\texorpdfstring{$\balpha$}{α} radiative transfer results}
\label{sec:results}

\subsection{Intrinsic spectra}
We first discuss the spectral properties of the intrinsic Ly$\alpha$ emission and related optically-thin lines. In theory, under Case~B recombination the H$\alpha$\;(H$\beta$) Balmer transition of atomic hydrogen at $\lambda_{\text{H}\alpha} = 6562.8\;(4861.3)\;\text{\AA}$ closely traces the intrinsic Ly$\alpha$ recombination emission, albeit with a factor of $F_{\text{Ly}\alpha} / F_{\text{H}\alpha} \approx 0.68 \nu_{\text{Ly}\alpha} / 0.45 \nu_{\text{H}\alpha} \approx 8\;(23)$ reduction in flux \citep{Dijkstra_PASA_2014}. For methodological consistency we run $\colt$ without any absorption or scattering at high ($1$\,km\,s$^{-1}$) spectral resolution with $10^8$ photon packets. Interestingly, the proper motion.
% \footnote{The results in this paper are given in the center of mass velocity frame based on the dark matter and gas within a $3R_\text{vir}$ sphere.}
of the galaxy induces a substantial Doppler shift and complex spectral morphology for the LOS line, e.g. asymmetric peaks and plateaus. In fact, the peak frequency and full width at half maximum (FWHM) are not always robust characterizations of the line profile. Therefore, we instead calculate the flux-weighted frequency centroid defined as $\langle \Delta v \rangle \equiv \int \Delta v f_\lambda\,\text{d}\lambda / \int f_\lambda\,\text{d}\lambda$ and the standard deviation of the line as $\sigma_{\Delta v} \equiv (\langle \Delta v^2 \rangle - \langle \Delta v \rangle^2)^{1/2}$.

In Fig.~\ref{fig:H_alpha}, we show the redshift evolution of the angular-averaged intrinsic Ly$\alpha$ line profile. We note that although the angular-averaging generates symmetric profiles about the line centre, the observed spectra are much more complex due to the aggregate LOS motion of the emitting gas. We also illustrate the angular distribution of the velocity offset and line width with maps showing these quantities for 3072 healpix directions. There is a dipole moment\footnote{The results in this paper are given in the center of mass velocity frame based on the dark matter and gas within a $3R_\text{vir}$ sphere. We note that the dipole signal in the intrinsic flux depends on the reference frame so may not be a robust prediction. One could instead measure Ly$\alpha$ properties based on the systemic redshifts of optically-thin lines. However, the Ly$\alpha$ line couples to the CGM so without a more self-consistent treatment of IGM transmission the cosmological reference frame is the least biased over all sightlines.} ($\gtrsim 10$\,km\,s$^{-1}$) due to the proper motion of the galaxy, with higher-order residual angular variations ($\lesssim 1$\,m\,s$^{-1}$) from satellite galaxies and other substructures representing a small fraction of the total luminosity ($\lesssim 10^{-4}$). This is also mirrored in the red-to-blue-flux ratio $F_\text{red}/F_\text{blue}$, which varies by a factor of $\approx 4$. The line width also has a significant quadrupole modulation due to axially-asymmetric velocity dispersion, which induces broader lines along sightlines with higher directed turbulent motions. Specifically, for the simulated galaxy we calculate a time- and angular-averaged value of $\langle \sigma_{\Delta v} \rangle \approx 68 \pm 21\,\text{km\,s}^{-1}$ ($\pm 30\%$ of the mean), corresponding to $\langle \text{FWHM} \rangle \approx 2.355\,\langle \sigma_{\Delta v} \rangle \approx 161 \pm 50\,\text{km\,s}^{-1}$, assuming the line is approximately Gaussian. We also note that Ly$\alpha$ photons produced by recombination and collisional excitation have a kinematic misalignment of their $\langle \Delta v \rangle$ dipole and $\sigma_{\Delta v}$ quadrupole moments due to occupying slightly different regions of $\rho$--$T$ phase space (see Fig.~\ref{fig:phase_rho_T}).

\begin{table*}
  \contcaption{Time-averaged properties of the intrinsic Ly$\alpha$ emission over the redshift interval $z = 5$--$7$, with subdivisions summarizing the evolution. We also show separate columns for photons originating from recombinations and collisional excitation.}
  %%%%%%%%%%%%%%%%%%%%%%%%%%%%%
  % Table for All Lya Photons %
  %%%%%%%%%%%%%%%%%%%%%%%%%%%%%
  \addtolength{\tabcolsep}{-1.5pt}
  \renewcommand{\arraystretch}{1.25}
  \begin{tabular}{@{} l ccc @{}}
    & \multicolumn{3}{c}{All Ly$\alpha$ Photons} \\
    \hline
    Quantity \hfill Type\hspace{-.1cm} & $z = 5$--$7$ & $z = 6$--$7$ & $z = 5$--$6$ \\
    \hline
    $\log L_\alpha$ \hfill [$\text{erg\,s}^{-1}$]\;\;\,\textsc{ion} & $42.2\pm0.4$ & $41.9\pm0.3$ & $42.3\pm0.4$ \\
    $L_\alpha^\text{rec,col} / L_\alpha$ \hfill [\%]\;\;\,\textsc{ion} & -- & -- & -- \\
    $F_\text{red} / F_\text{blue}$ \hfill \textsc{ion} & $1^{+0.813}_{-0.449}$ & $1^{+0.715}_{-0.417}$ & $1^{+0.936}_{-0.484}$ \\
    $\langle \Delta v \rangle$ \hfill [$\text{km\,s}^{-1}$]\;\;\,\textsc{ion} & $0\pm21.09$ & $0\pm20.68$ & $0\pm21.37$ \\
    $\sigma_{\Delta v}$ \hfill [$\text{km\,s}^{-1}$]\;\;\,\textsc{ion} & $68.36^{+24.84}_{-17.29}$ & $80.22^{+20.81}_{-22.47}$ & $62.45^{+19.38}_{-14.04}$ \\
    $\text{EW}_{\text{Ly}\alpha,0}$ \hfill [$\text{\AA}$]\;\;\,\textsc{ion} & $106.7\pm55.8$ & $63.0\pm15.9$ & $135.4\pm54.0$ \\
    \hline
  \end{tabular}
  \hspace{-.45cm}
  %%%%%%%%%%%%%%%%%%%%%%%%%%%%%%%%
  % Table for Lya Recombinations %
  %%%%%%%%%%%%%%%%%%%%%%%%%%%%%%%%
  \begin{tabular}{@{} ccc @{}}
    \multicolumn{3}{c}{Ly$\alpha$ from Recombinations} \\
    \hline
    $z = 5$--$7$ & $z = 6$--$7$ & $z = 5$--$6$ \\
    \hline
    $42.0\pm0.5$ & $41.7\pm0.3$ & $42.2\pm0.5$ \\
    $70.8\pm13.4$ & $63.1\pm9.8$ & $75.8\pm13.1$ \\
    $1^{+1.301}_{-0.565}$ & $1^{+1.258}_{-0.557}$ & $1^{+1.337}_{-0.572}$ \\
    $0\pm27.12$ & $0\pm30.12$ & $0\pm25.48$ \\
    $66.03^{+27.63}_{-19.67}$ & $81.73^{+21.75}_{-26.67}$ & $58.66^{+22.04}_{-14.93}$ \\
    $83.0\pm59.4$ & $41.5\pm16.9$ & $110.2\pm61.5$ \\
    \hline
  \end{tabular}
  \hspace{-.45cm}
  %%%%%%%%%%%%%%%%%%%%%%%%%%%%%%%%%%%%%%%%
  % Table for Lya Collisional Excitation %
  %%%%%%%%%%%%%%%%%%%%%%%%%%%%%%%%%%%%%%%%
  \begin{tabular}{@{} ccc @{}}
    \multicolumn{3}{c}{Ly$\alpha$ from Collisional Excitation} \\
    \hline
    $z = 5$--$7$ & $z = 6$--$7$ & $z = 5$--$6$ \\
    \hline
    $41.5\pm0.2$ & $41.5\pm0.2$ & $41.6\pm0.2$ \\
    $29.2\pm13.4$ & $36.9\pm9.8$ & $24.2\pm13.1$ \\
    $1^{+0.419}_{-0.295}$ & $1^{+0.408}_{-0.290}$ & $1^{+0.427}_{-0.299}$ \\
    $0\pm13.62$ & $0\pm14.33$ & $0\pm13.07$ \\
    $71.74^{+17.19}_{-12.22}$ & $74.26^{+21.18}_{-15.41}$ & $70.78^{+11.68}_{-10.83}$ \\
    $23.8\pm7.0$ & $21.5\pm2.8$ & $25.3\pm8.3$ \\
    \hline
  \end{tabular}
  \addtolength{\tabcolsep}{1.5pt}
  \renewcommand{\arraystretch}{.8}
\end{table*}

\begin{table*}
  \caption{Time-averaged Ly$\alpha$ properties after accounting for ISM scattering (\textsc{ism}) with a similar layout as Table~\ref{tab:time_avg_int}.}
  \label{tab:time_avg_ism}
  %%%%%%%%%%%%%%%%%%%%%%%%%%%%%
  % Table for All Lya Photons %
  %%%%%%%%%%%%%%%%%%%%%%%%%%%%%
  \addtolength{\tabcolsep}{-1.5pt}
  \renewcommand{\arraystretch}{1.25}
  \begin{tabular}{@{} l ccc @{}}
    & \multicolumn{3}{c}{All Ly$\alpha$ Photons} \\
    \hline
    Quantity \hfill Type\hspace{-.1cm} & $z = 5$--$7$ & $z = 6$--$7$ & $z = 5$--$6$ \\
    \hline
    $\log L_\alpha$ \hfill [$\text{erg\,s}^{-1}$]\;\;\,\textsc{ism} & $41.86^{+0.56}_{-0.23}$ & $41.73^{+0.33}_{-0.15}$ & $42.03^{+0.51}_{-0.35}$ \\
    $L_\alpha^\text{rec,col} / L_\alpha$ \hfill [\%]\;\;\,\textsc{ism} & -- & -- & -- \\
    $f_\text{esc}^{\text{Ly}\alpha}$ \hfill [\%]\;\;\,\textsc{ism} & $70.77^{+21.32}_{-22.46}$ & $80.58^{+17.82}_{-17.83}$ & $63.06^{+23.02}_{-20.43}$ \\
    $F_\text{LOS} / F_\Omega$ \hfill \textsc{ism} & $1^{+0.2383}_{-0.2589}$ & $1^{+0.2087}_{-0.2194}$ & $1^{+0.2613}_{-0.2896}$ \\
    $F_\text{red} / F_\text{blue}$ \hfill \textsc{ism} & $1.00^{+0.38}_{-0.31}$ & $1.10^{+0.35}_{-0.24}$ & $0.93^{+0.39}_{-0.32}$ \\
    $R_{1/2}$ \hfill [$\text{kpc}$]\;\;\,\textsc{ism} & $7.168^{+5.696}_{-3.373}$ & $10.582^{+4.429}_{-4.666}$ & $5.904^{+4.054}_{-2.969}$ \\
    $R_{\alpha,\text{h}}$ \hfill [$\text{kpc}$]\;\;\,\textsc{ism} & $21.43^{+8.48}_{-4.46}$ & $22.19^{+9.12}_{-7.92}$ & $21.18^{+7.63}_{-3.45}$ \\
    $\langle \Delta v \rangle$ \hfill [$\text{km\,s}^{-1}$]\;\;\,\textsc{ism} & $-4.6^{+38.2}_{-46.6}$ & $6.1^{+32.4}_{-32.4}$ & $-13.7^{+42.3}_{-52.3}$ \\
    $\sigma_{\Delta v}$ \hfill [$\text{km\,s}^{-1}$]\;\;\,\textsc{ism} & $214.5^{+23.7}_{-24.7}$ & $206.1^{+16.5}_{-28.2}$ & $221.7^{+25.9}_{-25.0}$ \\
    $\text{EW}_{\text{Ly}\alpha,0}$ \hfill [$\text{\AA}$]\;\;\,\textsc{ism} & $68.84^{+47.43}_{-25.29}$ & $49.05^{+20.26}_{-12.85}$ & $83.17^{+69.64}_{-23.28}$ \\
    \hline
  \end{tabular}
  \hspace{-.45cm}
  %%%%%%%%%%%%%%%%%%%%%%%%%%%%%%%%
  % Table for Lya Recombinations %
  %%%%%%%%%%%%%%%%%%%%%%%%%%%%%%%%
  \begin{tabular}{@{} ccc @{}}
    \multicolumn{3}{c}{Ly$\alpha$ from Recombinations} \\
    \hline
    $z = 5$--$7$ & $z = 6$--$7$ & $z = 5$--$6$ \\
    \hline
    $41.61^{+0.72}_{-0.26}$ & $41.45^{+0.41}_{-0.17}$ & $41.81^{+0.68}_{-0.39}$ \\
    $60.66^{+25.35}_{-13.56}$ & $52.10^{+19.82}_{-7.81}$ & $65.67^{+24.32}_{-14.60}$ \\
    $62.55^{+21.98}_{-20.75}$ & $70.86^{+20.49}_{-18.50}$ & $56.22^{+22.60}_{-19.59}$ \\
    $1^{+0.2758}_{-0.3015}$ & $1^{+0.2472}_{-0.2642}$ & $1^{+0.3036}_{-0.3284}$ \\
    $1.18^{+0.57}_{-0.41}$ & $1.36^{+0.65}_{-0.38}$ & $1.07^{+0.49}_{-0.40}$ \\
    $6.615^{+4.128}_{-3.189}$ & $8.853^{+3.294}_{-3.803}$ & $5.553^{+3.606}_{-2.762}$ \\
    $20.85^{+8.43}_{-4.18}$ & $18.79^{+8.27}_{-3.75}$ & $21.85^{+8.74}_{-3.46}$ \\
    $15.0^{+44.3}_{-51.7}$ & $27.5^{+46.6}_{-37.7}$ & $4.2^{+42.0}_{-56.2}$ \\
    $208.3^{+25.1}_{-22.9}$ & $200.5^{+19.1}_{-20.1}$ & $215.2^{+26.9}_{-25.1}$ \\
    $40.67^{+56.19}_{-19.00}$ & $25.08^{+22.82}_{-8.35}$ & $52.63^{+84.41}_{-20.90}$ \\
    \hline
  \end{tabular}
  \hspace{-.45cm}
  %%%%%%%%%%%%%%%%%%%%%%%%%%%%%%%%%%%%%%%%
  % Table for Lya Collisional Excitation %
  %%%%%%%%%%%%%%%%%%%%%%%%%%%%%%%%%%%%%%%%
  \begin{tabular}{@{} ccc @{}}
    \multicolumn{3}{c}{Ly$\alpha$ from Collisional Excitation} \\
    \hline
    $z = 5$--$7$ & $z = 6$--$7$ & $z = 5$--$6$ \\
    \hline
    $41.43^{+0.31}_{-0.18}$ & $41.38^{+0.38}_{-0.16}$ & $41.46^{+0.27}_{-0.19}$ \\
    $39.34^{+13.56}_{-25.35}$ & $47.90^{+7.81}_{-19.82}$ & $34.33^{+14.60}_{-24.33}$ \\
    $87.85^{+20.21}_{-21.26}$ & $94.76^{+18.58}_{-17.07}$ & $82.25^{+21.22}_{-20.59}$ \\
    $1^{+0.2089}_{-0.2118}$ & $1^{+0.1899}_{-0.1865}$ & $1^{+0.2230}_{-0.2310}$ \\
    $0.77^{+0.22}_{-0.23}$ & $0.87^{+0.18}_{-0.17}$ & $0.68^{+0.24}_{-0.19}$ \\
    $8.374^{+6.406}_{-3.498}$ & $12.147^{+5.753}_{-4.764}$ & $6.852^{+4.485}_{-2.937}$ \\
    $21.48^{+10.97}_{-5.94}$ & $24.74^{+13.48}_{-12.11}$ & $20.47^{+8.18}_{-4.15}$ \\
    $-38.9^{+37.3}_{-47.6}$ & $-19.0^{+25.3}_{-34.2}$ & $-56.0^{+43.2}_{-43.6}$ \\
    $224.8^{+27.1}_{-30.4}$ & $208.3^{+19.8}_{-37.7}$ & $237.2^{+22.5}_{-25.6}$ \\
    $22.62^{+9.12}_{-6.95}$ & $20.98^{+5.87}_{-4.41}$ & $24.76^{+10.09}_{-10.17}$ \\
    \hline
  \end{tabular}
  \addtolength{\tabcolsep}{1.5pt}
  \renewcommand{\arraystretch}{.8}
\end{table*}

\begin{table*}
  \caption{Time-averaged Ly$\alpha$ properties after accounting for IGM transmission (\textsc{igm}) with a similar layout as Tables~\ref{tab:time_avg_int}~and~\ref{tab:time_avg_ism}.}
  \label{tab:time_avg_igm}
  %%%%%%%%%%%%%%%%%%%%%%%%%%%%%
  % Table for All Lya Photons %
  %%%%%%%%%%%%%%%%%%%%%%%%%%%%%
  \addtolength{\tabcolsep}{-1.5pt}
  \renewcommand{\arraystretch}{1.25}
  \begin{tabular}{@{} l ccc @{}}
    & \multicolumn{3}{c}{All Ly$\alpha$ Photons} \\
    \hline
    Quantity \hfill Type\hspace{-.1cm} & $z = 5$--$7$ & $z = 6$--$7$ & $z = 5$--$6$ \\
    \hline
    $\log L_\alpha$ \hfill [$\text{erg\,s}^{-1}$]\;\;\,\textsc{igm} & $41.45^{+0.57}_{-0.36}$ & $41.22^{+0.40}_{-0.18}$ & $41.62^{+0.52}_{-0.44}$ \\
    $L_\alpha^\text{rec,col} / L_\alpha$ \hfill [\%]\;\;\,\textsc{igm} & -- & -- & -- \\
    $f_\text{esc}^{\text{Ly}\alpha}$ \hfill [\%]\;\;\,\textsc{igm} & $25.16^{+10.81}_{-9.31}$ & $25.59^{+8.25}_{-7.29}$ & $24.72^{+12.90}_{-11.01}$ \\
    $\mathcal{T}_\text{IGM}$ \hfill [\%]\;\;\,\textsc{igm} & $34.99^{+10.93}_{-7.41}$ & $30.86^{+6.37}_{-3.96}$ & $39.91^{+8.50}_{-11.01}$ \\
    $F_\text{LOS} / F_\Omega$ \hfill \textsc{igm} & $1^{+0.3117}_{-0.3518}$ & $1^{+0.2832}_{-0.2938}$ & $1^{+0.3372}_{-0.4007}$ \\
    $\log(F_\text{red} / F_\text{blue})$ \hfill \textsc{igm} & $3.50^{+1.85}_{-2.08}$ & $4.75^{+8.92}_{-0.74}$ & $2.12^{+1.42}_{-0.94}$ \\
    $R_{1/2}$ \hfill [$\text{kpc}$]\;\;\,\textsc{igm} & $6.022^{+5.137}_{-3.103}$ & $8.753^{+4.272}_{-4.191}$ & $5.155^{+3.648}_{-2.773}$ \\
    $R_{\alpha,\text{h}}$ \hfill [$\text{kpc}$]\;\;\,\textsc{igm} & $17.52^{+10.79}_{-4.58}$ & $17.46^{+11.64}_{-8.55}$ & $17.54^{+10.10}_{-3.32}$ \\
    $\langle \Delta v \rangle$ \hfill [$\text{km\,s}^{-1}$]\;\;\,\textsc{igm} & $201.9^{+30.2}_{-34.8}$ & $209.6^{+40.0}_{-15.3}$ & $193.8^{+27.9}_{-41.1}$ \\
    $\sigma_{\Delta v}$ \hfill [$\text{km\,s}^{-1}$]\;\;\,\textsc{igm} & $126.5^{+27.4}_{-19.4}$ & $114.6^{+25.9}_{-13.6}$ & $133.4^{+33.5}_{-18.0}$ \\
    $\Delta v_\text{peak}$ \hfill [$\text{km\,s}^{-1}$]\;\;\,\textsc{igm} & $132.3^{+32.7}_{-32.2}$ & $134.7^{+21.4}_{-15.8}$ & $125.8^{+46.8}_{-36.3}$ \\
    $\text{FWHM}$ \hfill [$\text{km\,s}^{-1}$]\;\;\,\textsc{igm} & $157.7^{+34.3}_{-28.8}$ & $145.9^{+30.5}_{-19.8}$ & $166.1^{+33.4}_{-31.9}$ \\
    $\text{EW}_{\text{Ly}\alpha,0}$ \hfill [$\text{\AA}$]\;\;\,\textsc{igm} & $24.74^{+22.15}_{-11.95}$ & $15.11^{+10.20}_{-4.78}$ & $32.79^{+30.55}_{-12.83}$ \\
    \hline
  \end{tabular}
  \hspace{-.45cm}
  %%%%%%%%%%%%%%%%%%%%%%%%%%%%%%%%
  % Table for Lya Recombinations %
  %%%%%%%%%%%%%%%%%%%%%%%%%%%%%%%%
  \begin{tabular}{@{} ccc @{}}
    \multicolumn{3}{c}{Ly$\alpha$ from Recombinations} \\
    \hline
    $z = 5$--$7$ & $z = 6$--$7$ & $z = 5$--$6$ \\
    \hline
    $41.24^{+0.71}_{-0.38}$ & $40.98^{+0.48}_{-0.20}$ & $41.44^{+0.66}_{-0.47}$ \\
    $65.93^{+21.97}_{-13.50}$ & $58.35^{+19.09}_{-7.62}$ & $70.37^{+21.85}_{-15.05}$ \\
    $24.17^{+12.14}_{-9.85}$ & $25.00^{+9.94}_{-8.13}$ & $23.51^{+13.76}_{-11.31}$ \\
    $38.37^{+11.03}_{-8.76}$ & $34.45^{+8.10}_{-5.40}$ & $42.49^{+8.96}_{-12.08}$ \\
    $1^{+0.3548}_{-0.3924}$ & $1^{+0.3206}_{-0.3469}$ & $1^{+0.3845}_{-0.4340}$ \\
    $3.58^{+1.96}_{-2.08}$ & $4.80^{+9.18}_{-0.69}$ & $2.25^{+1.38}_{-1.02}$ \\
    $5.550^{+3.816}_{-2.899}$ & $7.451^{+3.083}_{-3.392}$ & $4.753^{+3.203}_{-2.477}$ \\
    $15.66^{+10.77}_{-5.23}$ & $14.14^{+10.88}_{-6.68}$ & $16.26^{+11.25}_{-3.83}$ \\
    $205.4^{+32.7}_{-32.7}$ & $216.1^{+40.8}_{-16.8}$ & $194.9^{+29.2}_{-36.3}$ \\
    $128.2^{+27.2}_{-18.5}$ & $117.8^{+28.0}_{-12.6}$ & $132.8^{+31.3}_{-15.8}$ \\
    $132.0^{+35.7}_{-35.6}$ & $135.7^{+28.9}_{-18.5}$ & $124.3^{+48.7}_{-37.6}$ \\
    $150.4^{+35.0}_{-28.7}$ & $145.3^{+30.3}_{-22.6}$ & $154.7^{+36.2}_{-34.2}$ \\
    $16.01^{+22.58}_{-8.81}$ & $8.62^{+9.89}_{-3.15}$ & $21.92^{+34.61}_{-10.10}$ \\
    \hline
  \end{tabular}
  \hspace{-.45cm}
  %%%%%%%%%%%%%%%%%%%%%%%%%%%%%%%%%%%%%%%%
  % Table for Lya Collisional Excitation %
  %%%%%%%%%%%%%%%%%%%%%%%%%%%%%%%%%%%%%%%%
  \begin{tabular}{@{} ccc @{}}
    \multicolumn{3}{c}{Ly$\alpha$ from Collisional Excitation} \\
    \hline
    $z = 5$--$7$ & $z = 6$--$7$ & $z = 5$--$6$ \\
    \hline
    $40.92^{+0.35}_{-0.24}$ & $40.82^{+0.40}_{-0.18}$ & $41.01^{+0.29}_{-0.30}$ \\
    $34.07^{+13.50}_{-21.98}$ & $41.65^{+7.62}_{-19.09}$ & $29.63^{+15.05}_{-21.86}$ \\
    $26.79^{+9.00}_{-7.46}$ & $25.65^{+7.38}_{-5.61}$ & $27.80^{+10.03}_{-9.19}$ \\
    $29.75^{+9.79}_{-5.26}$ & $27.17^{+3.83}_{-3.27}$ & $34.73^{+6.84}_{-9.30}$ \\
    $1^{+0.2627}_{-0.2698}$ & $1^{+0.2454}_{-0.2295}$ & $1^{+0.2741}_{-0.3024}$ \\
    $3.40^{+1.81}_{-2.11}$ & $4.70^{+8.73}_{-0.81}$ & $1.87^{+1.55}_{-0.89}$ \\
    $7.545^{+6.070}_{-3.056}$ & $10.664^{+5.584}_{-4.670}$ & $6.379^{+4.349}_{-2.891}$ \\
    $17.56^{+12.51}_{-5.60}$ & $19.65^{+16.11}_{-11.82}$ & $16.88^{+9.84}_{-4.03}$ \\
    $196.6^{+29.8}_{-42.0}$ & $200.6^{+36.6}_{-16.1}$ & $193.6^{+26.6}_{-57.0}$ \\
    $123.0^{+38.3}_{-22.4}$ & $106.6^{+22.2}_{-14.8}$ & $135.4^{+49.2}_{-24.8}$ \\
    $135.1^{+33.2}_{-27.6}$ & $134.6^{+20.3}_{-15.3}$ & $135.9^{+44.4}_{-38.8}$ \\
    $160.3^{+36.8}_{-33.6}$ & $137.3^{+33.4}_{-21.0}$ & $173.0^{+35.0}_{-27.0}$ \\
    $6.79^{+4.10}_{-2.50}$ & $5.64^{+2.07}_{-1.38}$ & $8.22^{+4.44}_{-3.89}$ \\
    \hline
  \end{tabular}
  \addtolength{\tabcolsep}{1.5pt}
  \renewcommand{\arraystretch}{.8}
\end{table*}

\subsection{Escape fraction}
\label{sec:Lya_escape_fraction}
One of the primary motivations for this study is to determine how the Ly$\alpha$ escape fraction $f_\text{esc}^{\text{Ly}\alpha}$ evolves for a typical galaxy during the epoch of reionization. In Fig.~\ref{fig:f_esc}, we show the fraction of photons emerging from the $(4\,R_\text{vir})^3$ region centered on the galaxy. The red curve represents the situation after absorption by dust during scattering through the ISM, while the green curve takes into account the subsequent transmission through the IGM, based on the median from all of the \citet{Laursen_2011} galaxies (see Section~\ref{sec:IGM_transmission} for details). To understand the uncertainty in their IGM model, we compared the results obtained from three subsamples based on the circular velocity of the host galaxy environment as being ``small'' ($V_c < 55\,\text{km\,s}^{-1}$), ``intermediate'', and ``large'' ($V_c > 80\,\text{km\,s}^{-1}$), finding that they yield similar results. Thus, for simplicity we employ the model based on their entire sample of galaxies. For reference, in Fig.~\ref{fig:f_esc} we also show the escape fraction of 1500\,\AA\ UV photons $f_\text{esc}^\text{UV}$ (purple curve) and the escape fraction of ionizing photons $f_\text{esc}^\text{LyC}$ (blue curve).

\begin{figure}
  \centering
  \includegraphics[width=\columnwidth]{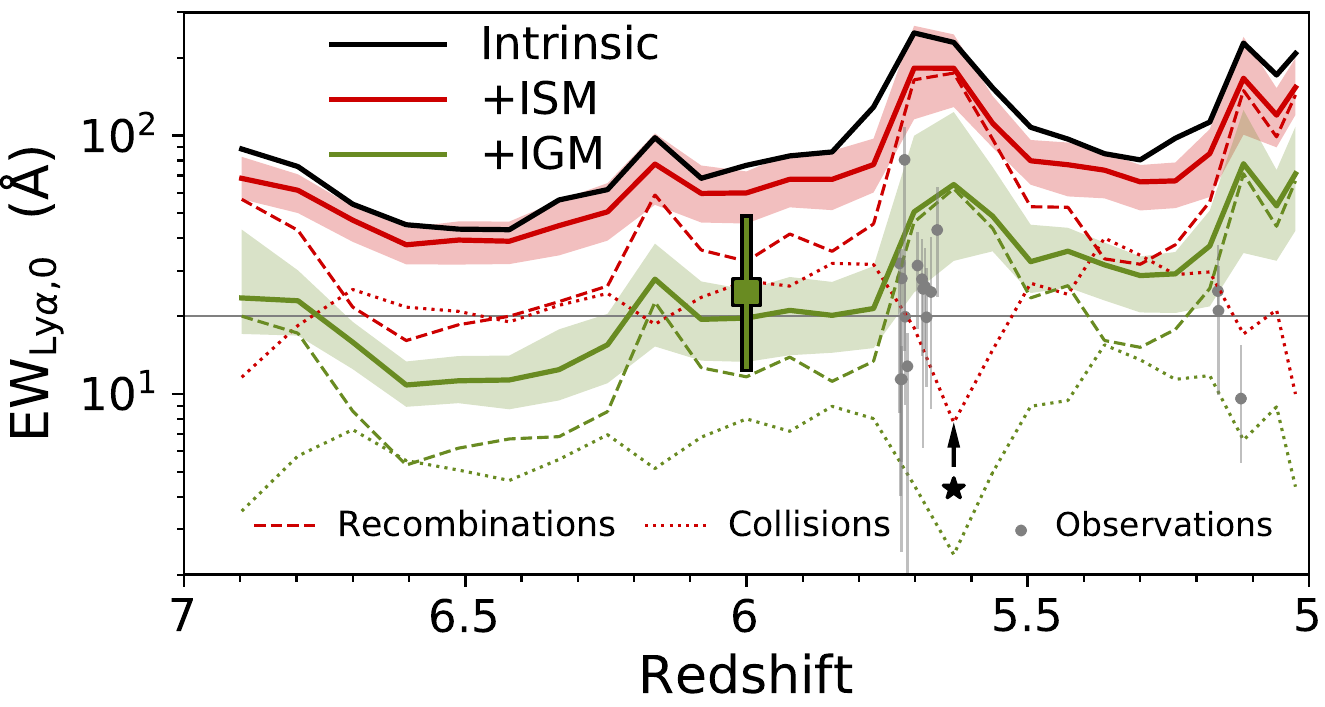}
  \includegraphics[width=\columnwidth]{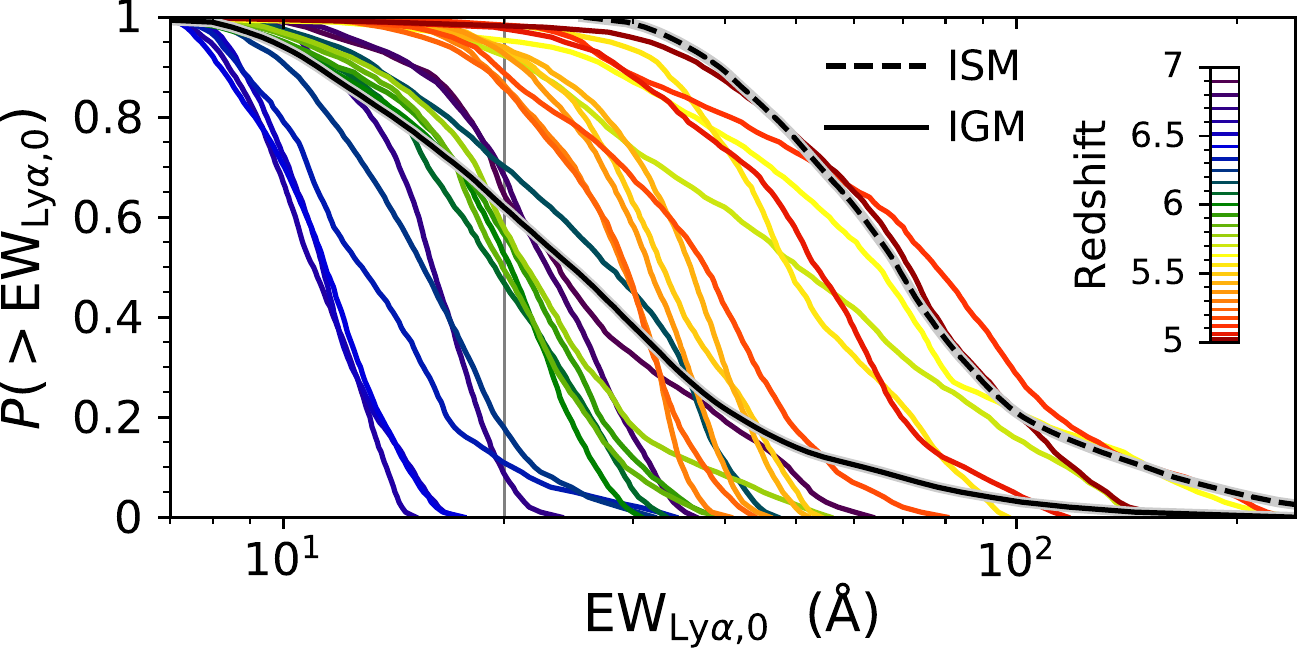}
  \includegraphics[width=\columnwidth]{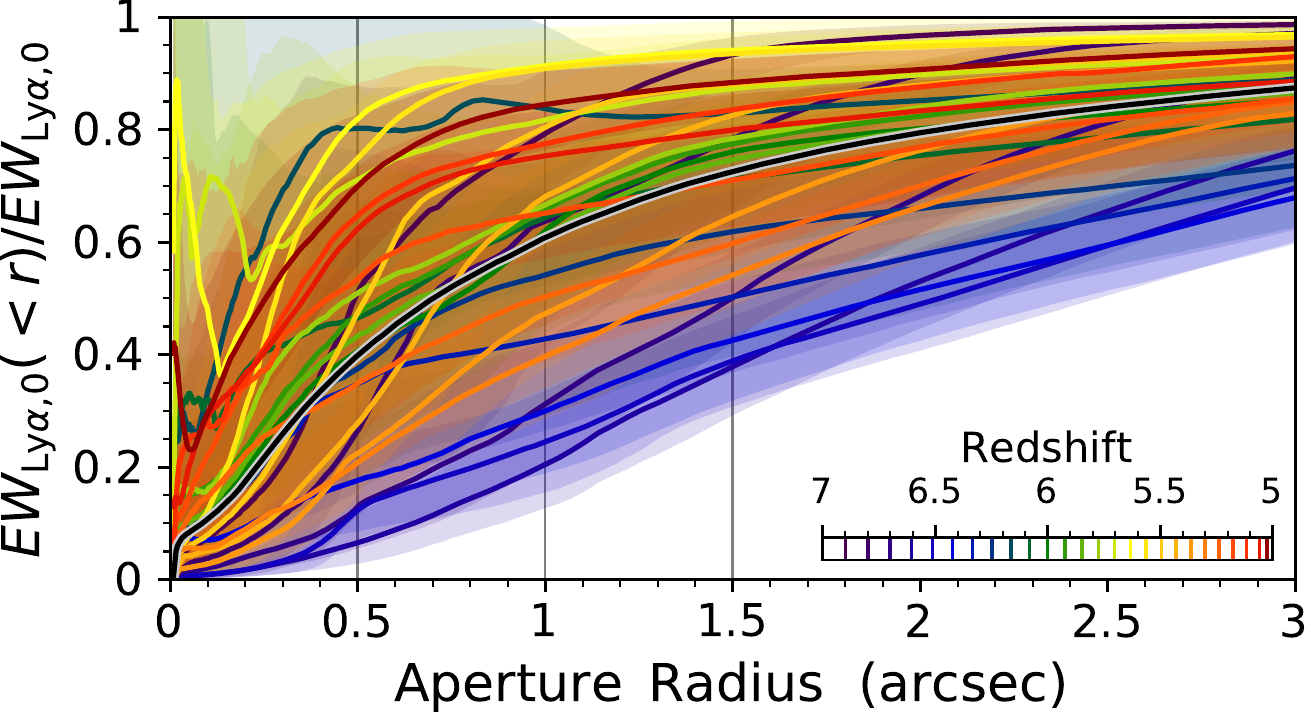}
  \caption{The redshift evolution of the Ly$\alpha$ rest frame equivalent width for the target galaxy. The black curve is the intrinsic value, while the red and green curves are for median sightlines after ISM scattering and IGM transmission, respectively. The shaded regions show the 1$\sigma$ confidence levels considering different viewing angles. For illustration purposes we also show several observed values from \citet{Mallery_2012} selected with $z > 5$ and $\log(M_\star/\Msun) < 9.7$ as the gray data points. In the middle panel we show the cumulative probability for observing an equivalent width greater than a given value based on viewing angle. The solid (dashed) black curve is the time-weighted distribution after (before) IGM transmission. Finally, in the lower panel we demonstrate the impact of aperture size on the observed equivalent widths, which are greatly reduced in the central $\sim 1''$ regions.}
  \label{fig:EW}
\end{figure}

The shaded regions in Fig.~\ref{fig:f_esc} illustrate the 1$\sigma$ variation ($16^\text{th}$ and $84^\text{th}$ percentiles) based on binning the escaped photon packets into 3072 sightlines, corresponding to equal area healpix directions of the unit sphere. Including multiple viewing angles allows us to mimic observations of different galaxies with similar properties. Although the scatter due to orientation is not equivalent to cosmic variance, e.g. environmental and large-scale structure effects, which induce further variation across different parts of the sky, it is certainly useful to consider both effects and their corresponding biases. Finally, we include several observed escape fraction estimates over our redshift range from \citet{Hayes_2011} as gray data points along with the reported errors, which is based on samples with comparable galaxy halo masses. Given the large uncertainties in these measurements and the variation between sightlines in our simulation, both sets of values are consistent over the redshift range considered.

\begin{figure*}
  \centering
  \includegraphics[width=\textwidth]{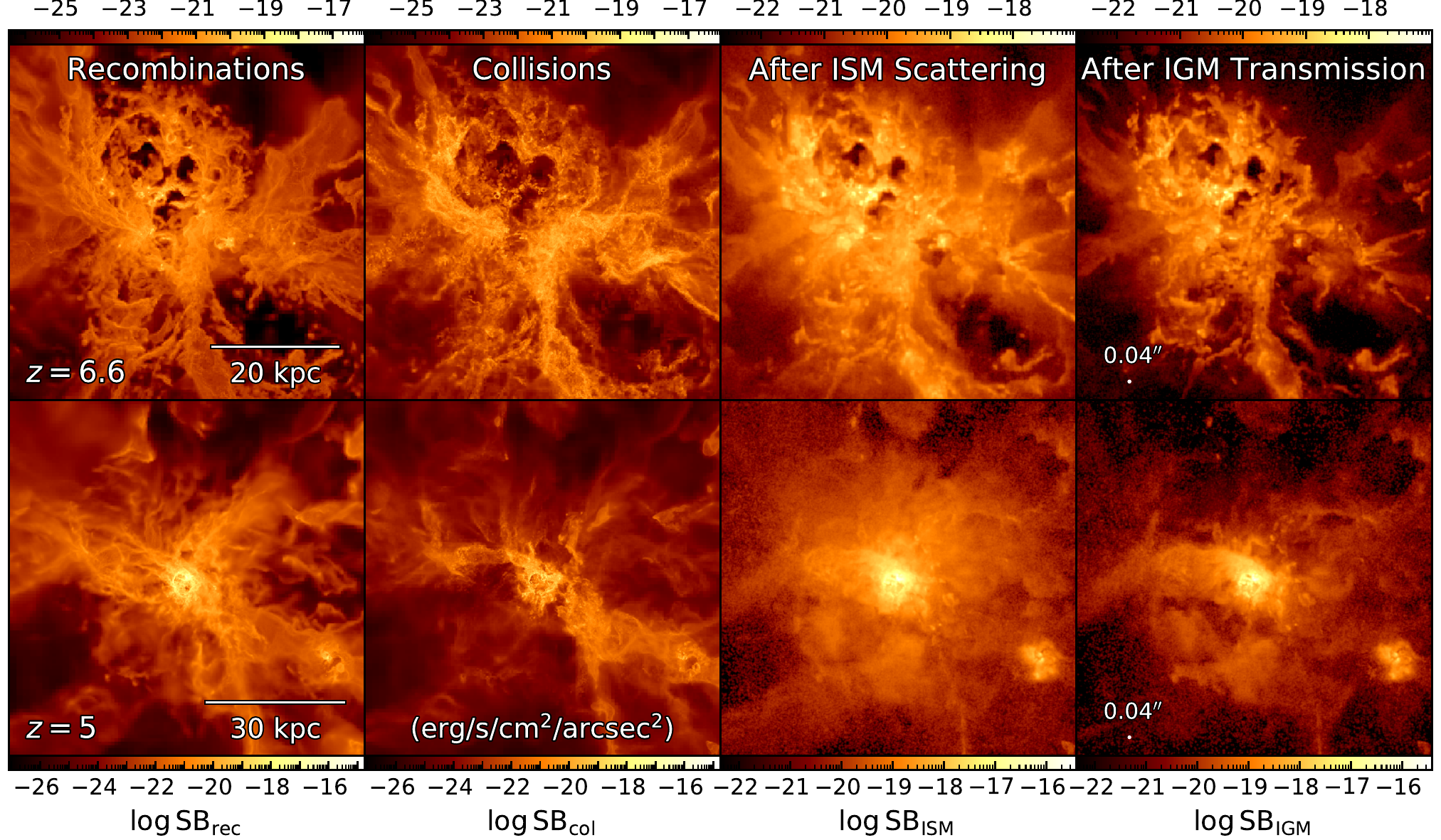}
  \caption{The Ly$\alpha$ surface brightness at three stages of the escape process: intrinsic recombination and collisional emission, total escaped radiation after ISM scattering, and observed surface brightness after IGM transmission. This illustrates the impact of resonant scattering as the final images are noticeably fainter and more diffuse than the intrinsic emission. The images are shown at $z = 6.6$ and $z = 5$, to compare the evolving morphologies. The panels for ISM scattering and IGM transmission use the same colour scale.}
  \label{fig:L_alpha}
\end{figure*}

Furthermore, the escape fraction exhibits significant fluctuation across redshift, lagging slightly behind the star formation activity. In Fig.~\ref{fig:L_esc} we illustrate the evolution of the observed Ly$\alpha$ luminosity along with the fraction of luminosity originating from collisional excitation and the overall transmission of Ly$\alpha$ photons through the IGM, defined as $\mathcal{T}_\textsc{igm} = f_{\text{esc,}\textsc{igm}}^{\text{Ly}\alpha} / f_{\text{esc,}\textsc{ism}}^{\text{Ly}\alpha}$. To summarize the \textsc{colt} results, we also provide the corresponding time- and angular-averaged quantities in Tables~\ref{tab:time_avg_ism}~and~\ref{tab:time_avg_igm}, again over full and half intervals to indicate any evolution. Although the perceived trends are primarily due to the dynamical changes of our target galaxy, the evolution is representative of the expected behavior of similar high-$z$ galaxies. It is significant that the intrinsic (ISM) and transmitted (IGM) Ly$\alpha$ escape fractions are relatively high compared to ionizing photons, with median values of $\langle f_\text{esc}^{\text{Ly}\alpha} \rangle_\textsc{ism} \approx 71$\% and $\langle f_\text{esc}^{\text{Ly}\alpha} \rangle_\textsc{igm} \approx 25$\%, such that the IGM transmission is $\langle \mathcal{T}_\textsc{igm} \rangle \approx 35$\%. The average observed luminosity is $\langle L_{\alpha,\text{obs}} \rangle \approx 3 \times 10^{41}\,\text{erg\,s}^{-1}$ with a peak of $L_{\alpha,\text{obs}} \approx 3 \times 10^{42}\,\text{erg\,s}^{-1}$ corresponding to the $z \approx 5.6$ starburst. This implies that a given galaxy can go in and out of visibility depending on the observational sensitivity. In Fig.~\ref{fig:L_esc}, we include curves for the \textit{JWST} NIRSpec instrument capabilities, assuming a $10\sigma$ point source detection after $10^4$ and $10^6$\,s of exposure time at a medium spectral resolution of $R \equiv \lambda/\Delta\lambda \approx 1000$ \citep[see][]{Gardner_2006}.

\subsection{Equivalent width}
Combining the results of Sections~\ref{sec:UV_continuum}~and~\ref{sec:Lya_escape_fraction} yields the Ly$\alpha$ equivalent width, characterizing the strength of the observed Ly$\alpha$ line relative to the continuum flux, defined as:
\begin{equation}
  \text{EW}_{\text{Ly}\alpha} \equiv \int \frac{f_{\lambda\,,\text{Ly}\alpha} - f_{\lambda\,,\text{UV}}}{f_{\lambda\,,\text{UV}}}\,\text{d}\lambda \approx \frac{F_{\text{Ly}\alpha}}{f_{\lambda\,,\text{UV}}} \, ,
\end{equation}
where the bolometric Ly$\alpha$ flux is $F_{\text{Ly}\alpha} \equiv \int f_{\lambda\,,\text{Ly}\alpha}\,\text{d}\lambda$, and in the rest frame $\text{EW}_{\text{Ly}\alpha,0} = \text{EW}_{\text{Ly}\alpha} / (1+z)$. We calculate an intrinsic time-averaged Ly$\alpha$ rest frame equivalent width of $\langle \text{EW}_{\text{Ly}\alpha,0} \rangle_\textsc{int} \approx 110\,\text{\AA}$ (with no dust extinction). The median LOS value decreases to $\langle \text{EW}_{\text{Ly}\alpha,0} \rangle_\textsc{ism} \approx 70\,\text{\AA}$ after ISM scattering (with dust) and $\langle \text{EW}_{\text{Ly}\alpha,0} \rangle_\textsc{igm} \approx 25\,\text{\AA}$ after IGM transmission (assuming no effect for continuum photons). In Fig.~\ref{fig:EW}, we show the redshift evolution of the rest frame equivalent width, which is primarily driven by the star formation activity in our simulations. We also show several representative observations from the study of \citet{Mallery_2012} as gray data points, using selection criteria of $z > 5$ and $\log(M_\star/\Msun) < 9.7$ for a similar mass range as our simulations. To illustrate the equivalent width further, we also provide the cumulative probability distributions for observing an equivalent width greater than a given value. During and after the main starbust, we identify a non-negligible number of high equivalent-width sightlines ($\text{EW}_{\text{Ly}\alpha,0} \gtrsim 100\,\text{\AA}$), even after accounting for transmission through the IGM. Finally, we also demonstrate the impact of aperture size on the observed equivalent width, which can be greatly reduced within the central $\sim 1''$ regions \citep{Laursen_2018}. However, the fraction $\text{EW}_{\text{Ly}\alpha,0}(<r) / \text{EW}_{\text{Ly}\alpha,0}$ converges fairly quickly when the source is brightest due to the corresponding reduced Ly$\alpha$ half-light radius.

Our target galaxy has a moderate virial mass $M_\text{halo} < 10^{11}\,\Msun$, such that the star formation efficiency is regulated by strong feedback, resulting in $M_\star / M_\text{gas} < 5\%$. The dominant starburst corresponds to $z \approx 5.6$, although the luminosity is also quite strong at $z \approx 7$ and $z \approx 5$ (see Fig.~\ref{fig:L_esc}). Specifically, the galaxy has duty cycles for the IGM transmitted (intrinsic) equivalent width with $\text{EW}_{\text{Ly}\alpha,0} > 20\,\text{\AA}$ of $62\%$ ($100\%$) and the luminosity with $L_\alpha > 10^{42}\,\text{erg\,s}^{-1}$ of $17\%$ ($52\%$). The equivalent width distribution for more massive galaxies may be driven by different mechanisms. For example, star forming regions enshrouded by dusty gas could disproportionately attenuate UV continuum photons leading to an equivalent width enhancement \citep{Finkelstein_2008,Finkelstein_2009}, especially if the Ly$\alpha$ resonant scattering process avoids such pathways \citep{Neufeld_1991,Hansen_Oh_2006}. Furthermore, Ly$\alpha$ photons originate from more diffuse emission and subsequent scattering in the CGM leads to relatively isotropic low surface brightness Ly$\alpha$ haloes.

Finally, the equivalent widths due to recomnination and collisional excitation are anti-correlated. During a starburst the cooling radiation only contributes a small fraction of the total budget, while there is a delayed boost to around $\sim 50\%$ afterwards (compare $L_\alpha^\text{col} / L_\alpha$ over $z \approx 5.6$--$5.3$ in Fig.~\ref{fig:L_esc}). The time variability is mainly regulated by young, massive stars that dominate the production of ionizing photons. However, the deposition of energy by post-starburst galactic winds also enhances the collisional contribution, as predicted for UV metal lines in the CGM \citep{Sravan_2016}. Therefore, it is possible that a subset of LAEs is powered by cooling radiation rather than nebular emission, although this may require a different environment, halo mass, assembly history, or redshift than the galaxy in these simulations.

\begin{figure}
  \centering
  \includegraphics[width=\columnwidth]{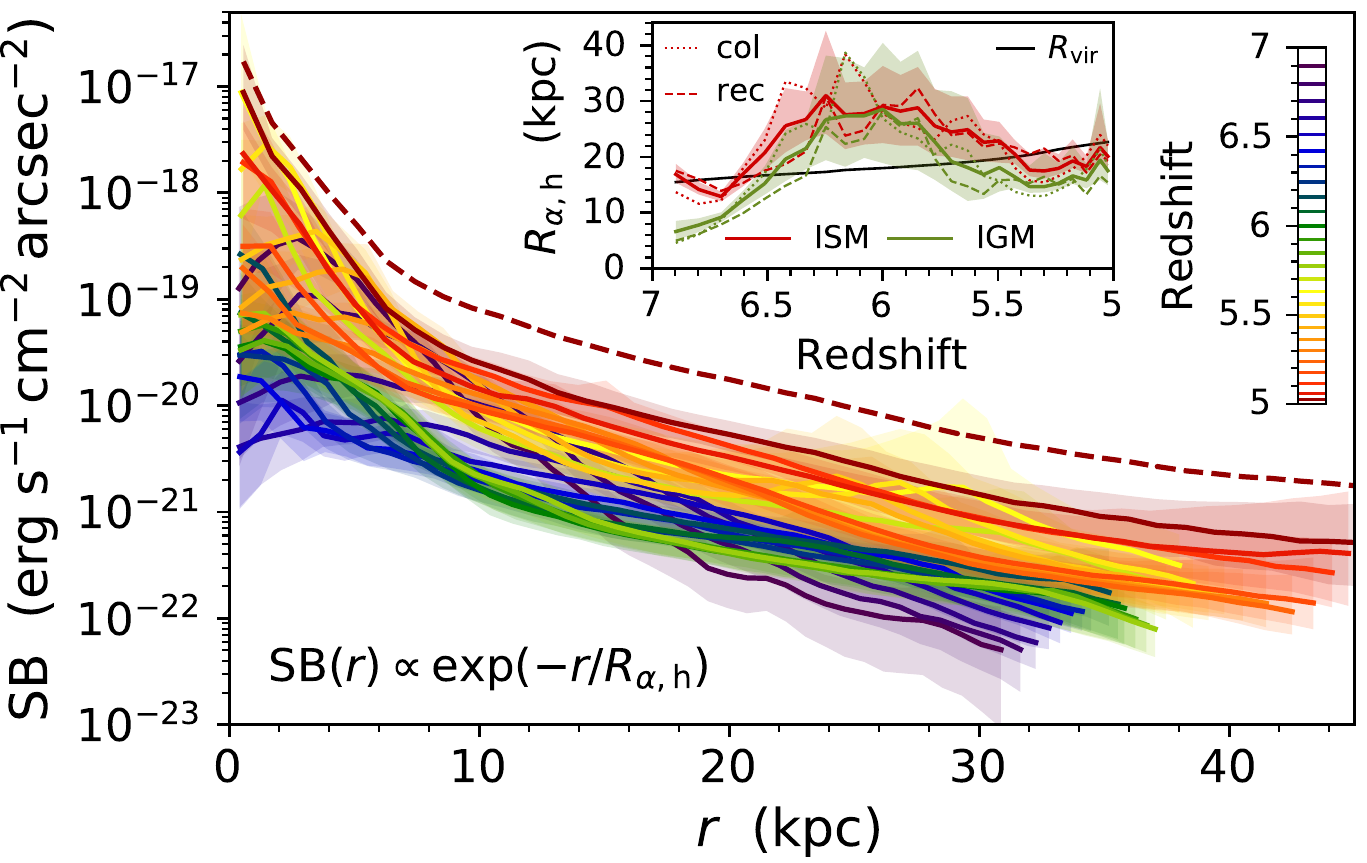}
  \includegraphics[width=\columnwidth]{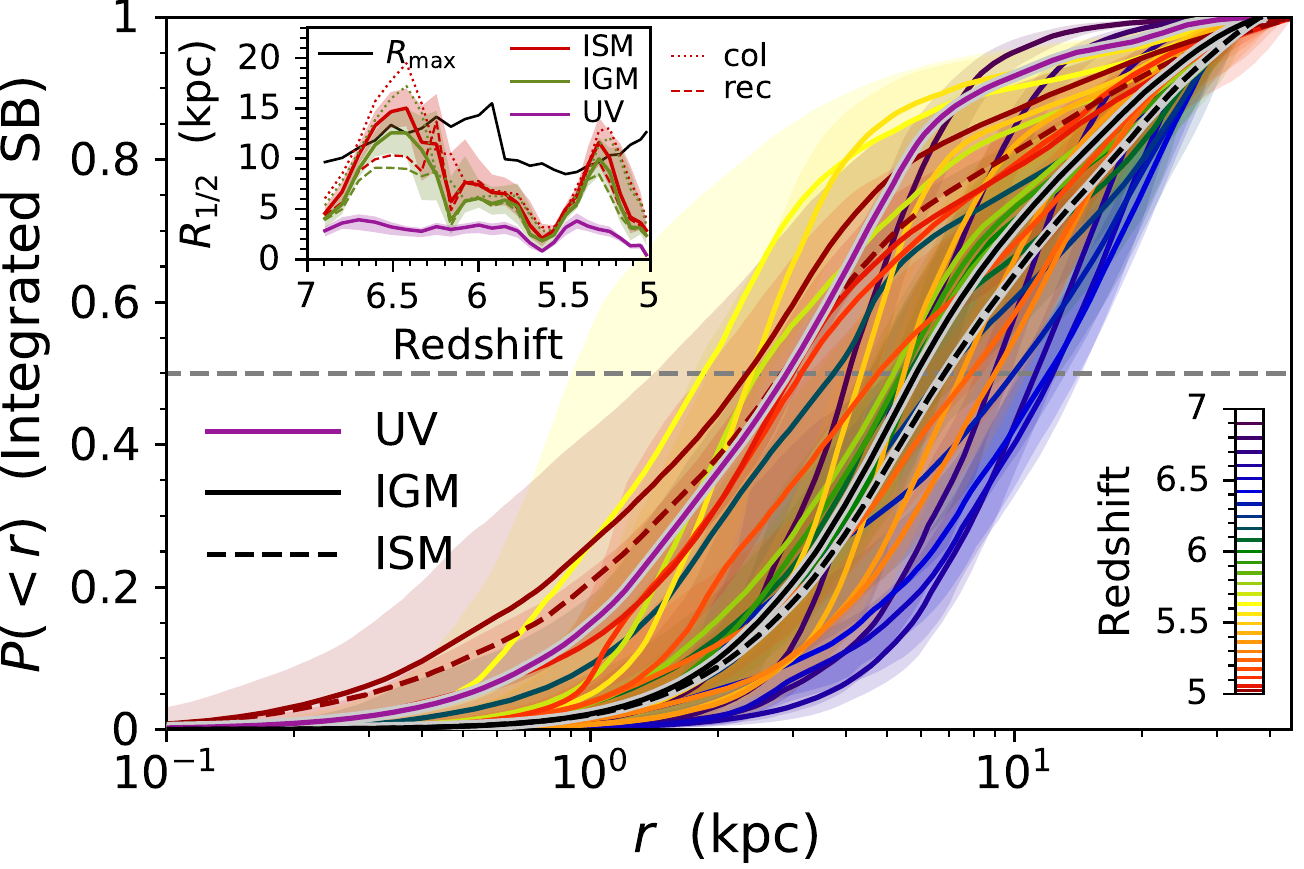}
  \caption{Radial Ly$\alpha$ surface brightness profiles and integrated light within a given radius, $P(<r) \propto \int_0^r \text{SB}(r') r' \text{d}r'$, shown after transmission through the IGM. The curves represent median LOS values with 1$\sigma$ confidence regions, all colored according to redshift. The upper inset shows the redshift evolution of the Ly$\alpha$ halo scale length $R_{\alpha,\text{h}}$, which is calculated by fitting to an exponential function and is comparable to the virial radius $R_\text{vir}$. The lower inset shows the half-light radius, defined as $P(<R_{1/2}) = 1/2$, and is within the radius of maximal circular velocity $R_\text{max}$.}
  \label{fig:SB_r}
\end{figure}

\begin{figure}
  \centering
  \includegraphics[width=\columnwidth]{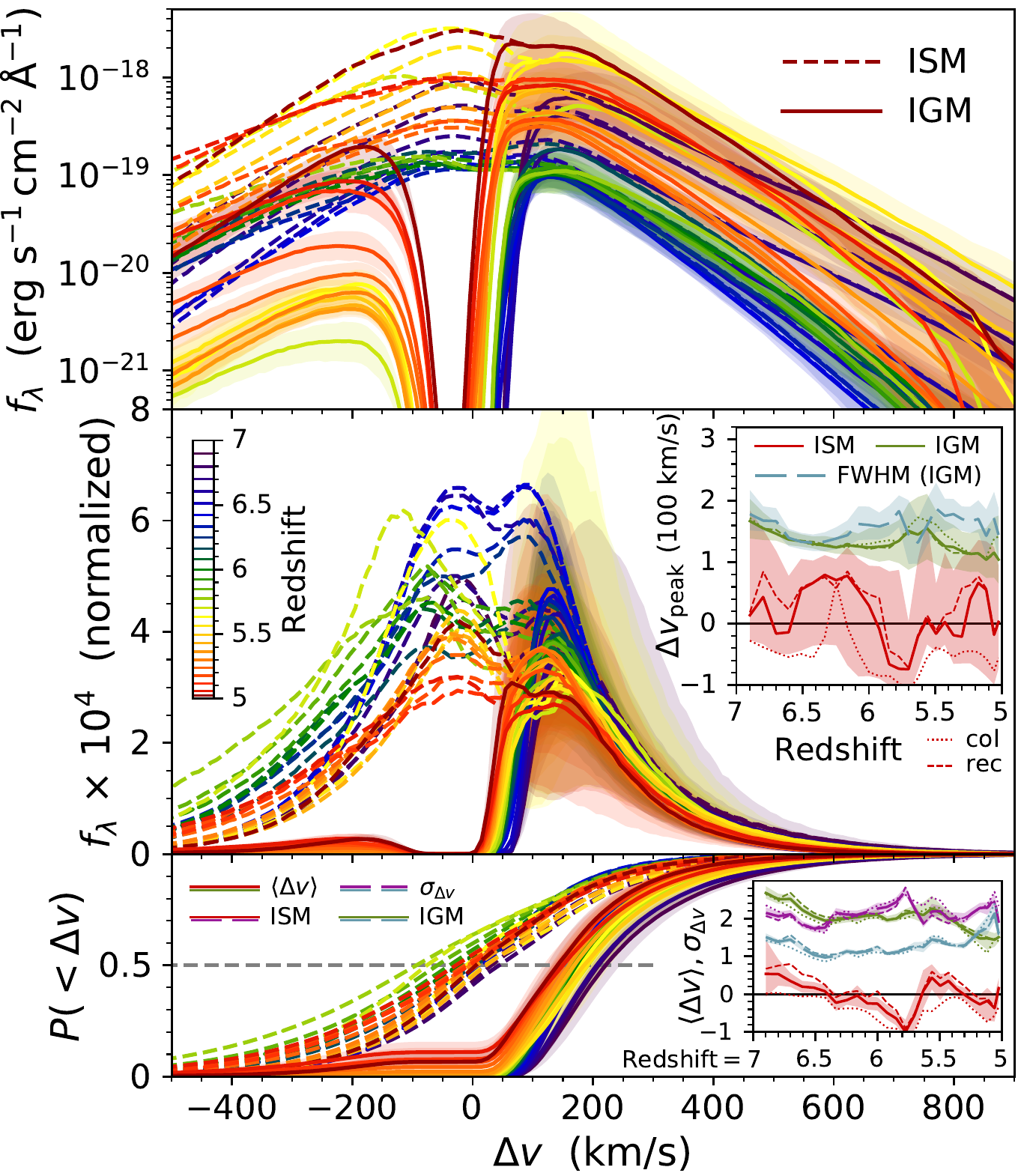}
  \caption{Median LOS Ly$\alpha$ flux density as a function of Doppler velocity $\Delta v = c \Delta \lambda / \lambda$ for each simulation snapshot after ISM scattering and IGM transmission. The top panel shows the fluctuating flux while the middle panel illustrates the evolving shape of the (IGM) normalized lines. The bottom panel shows the corresponding cumulative distribution functions. We also include insets showing the redshift dependence of the peak velocity offset $\Delta v_\text{peak}$, full width at half maximum (FWHM), flux-weighted frequency centroid $\langle \Delta v \rangle$, and standard deviation $\sigma_{\Delta v}$, all given as median LOS quantities.}
  \label{fig:flux_esc}
\end{figure}

\subsection{Surface brightness}
\label{sec:SB}
We employ the next-event estimator method in \textsc{colt} to construct LOS surface brightness images for the target galaxy. The simulated images are composed of square pixels at the resolution of the \textit{JWST} NIRCam instrument corresponding to $\Omega_\text{pix,NIRCam} \approx 10^{-3}\,\text{arcsec}^2$ for photometry. When also considering the full surface brightness spectral density we employ a Doppler resolution of $\Delta v \approx 10\,\text{km\,s}^{-1}$, corresponding to a spectral resolution of $R \equiv \lambda / \Delta \lambda \approx 30\,000$. This is achievable with large-aperture ground-based telescopes equipped with adaptive optics, but significantly exceeds the capability of the NIRSpec instrument aboard the \textit{JWST}. Thus, our resolution is sufficient for comparison with both photometric and spectroscopic observations.

In Fig.~\ref{fig:L_alpha} we illustrate the different stages of Ly$\alpha$ escape from high-$z$ galaxies with spatially resolved images of the (i) intrinsic Ly$\alpha$ recombination and collisional luminosity, (ii) escaped Ly$\alpha$ radiation after scattering through the ISM, and (iii) observed Ly$\alpha$ surface brightness after transmission through the IGM. The observed morphology changes in a nontrivial manner at each stage, with the final surface brightness being much more diffuse and fainter than the intrinsic emission due to the resonant scattering process. For some viewing angles there are dark fluffy streaks illustrating obscuration from high neutral hydrogen column density clouds blocking particular sightlines. Such inhomogeneities also provide preferred channels of escape. To quantify the net effect of Ly$\alpha$ directional boosting, we calculate the time-averaged deviation from isotropy as $\langle F_\text{LOS}/F_\Omega \rangle_\textsc{ism} \approx 1 \pm 0.25$ after ISM scattering and $\langle F_\text{LOS}/F_\Omega \rangle_\textsc{igm} \approx 1 \pm 0.33$ after IGM transmission, where $F_\text{LOS}$ and $F_\Omega$ denote the spatially integrated LOS and angular-averaged flux, respectively. We also note that there is a slight increase in anisotropy with time as the gas structure becomes more defined and the CGM is more ionized (Tables~\ref{tab:time_avg_ism}~and~~\ref{tab:time_avg_igm}).

\begin{figure}
  \centering
  \includegraphics[width=\columnwidth]{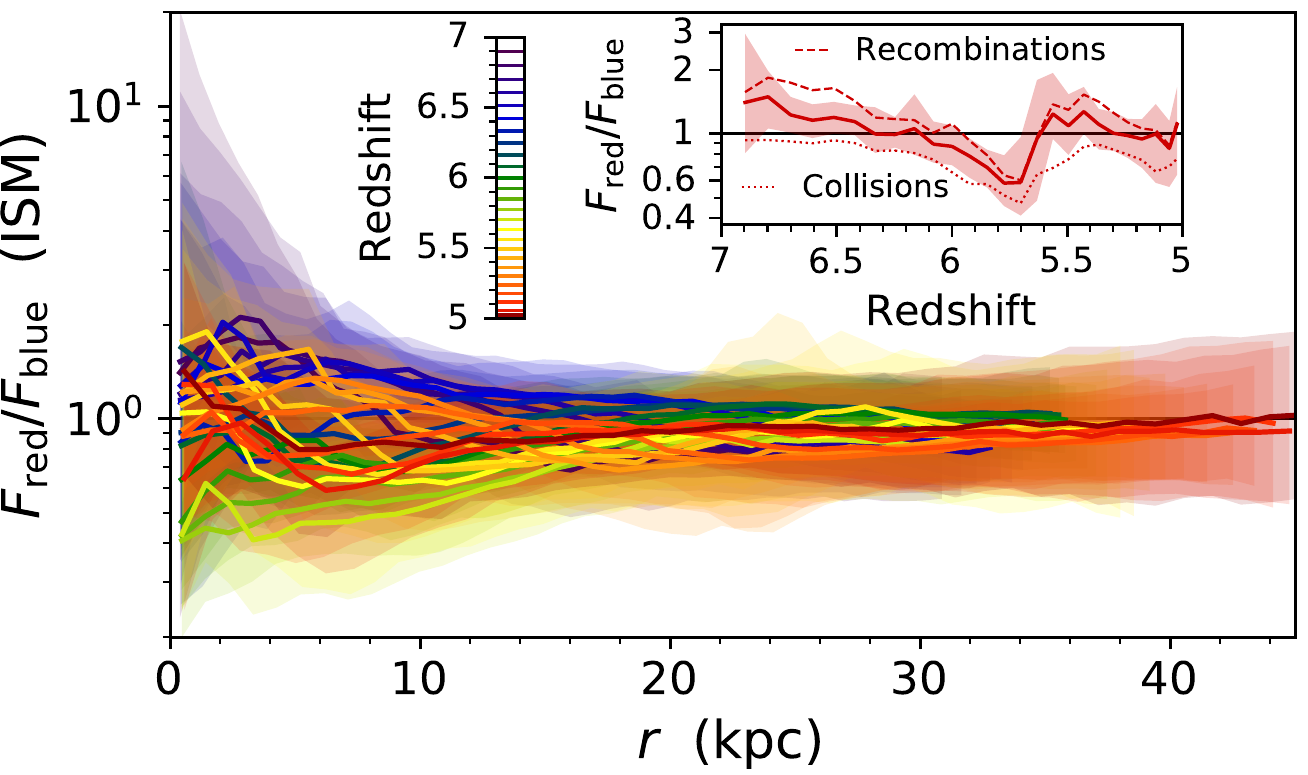}
  \includegraphics[width=\columnwidth]{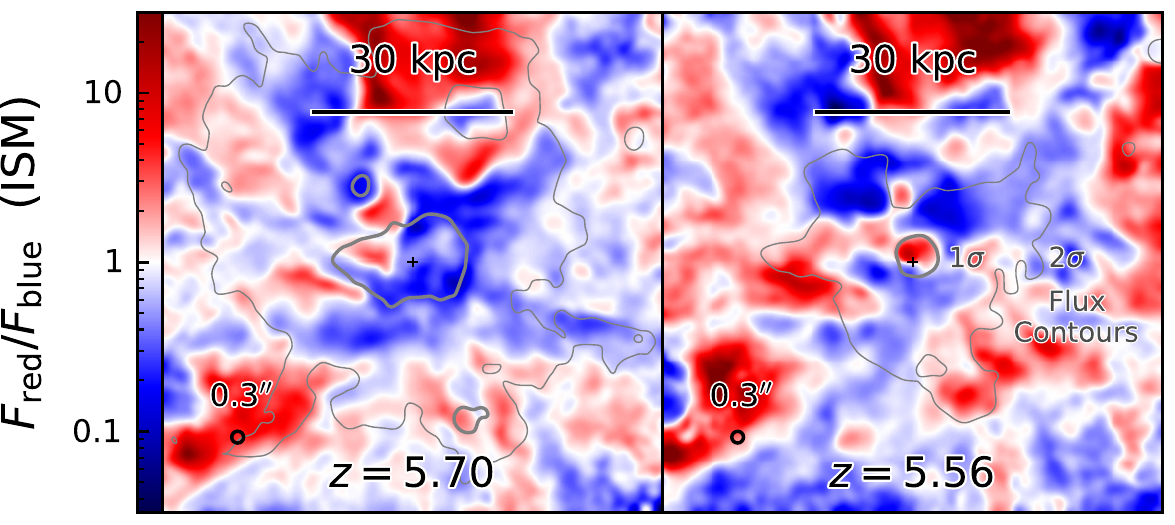}
  \caption{Radial red-to-blue-flux ratio $F_\text{red}/F_\text{blue}$ before transmission through the IGM. The curves are median LOS values with 1$\sigma$ confidence regions, all coloured according to redshift. The inset shows the redshift evolution of the spatially-integrated flux ratio. The lower panel illustrates the extreme change from a blue (infall) to red (outflow) dominated signature within the 1$\sigma$ Ly$\alpha$ surface brightness contours before and after the main starburst. For presentation purposes, the red and blue channel maps have been smoothed to simulate a $\text{FWHM} \approx 0\farcs3$ aperture.}
  \label{fig:F_rb_r}
\end{figure}

\begin{figure}
  \centering
  \includegraphics[width=\columnwidth]{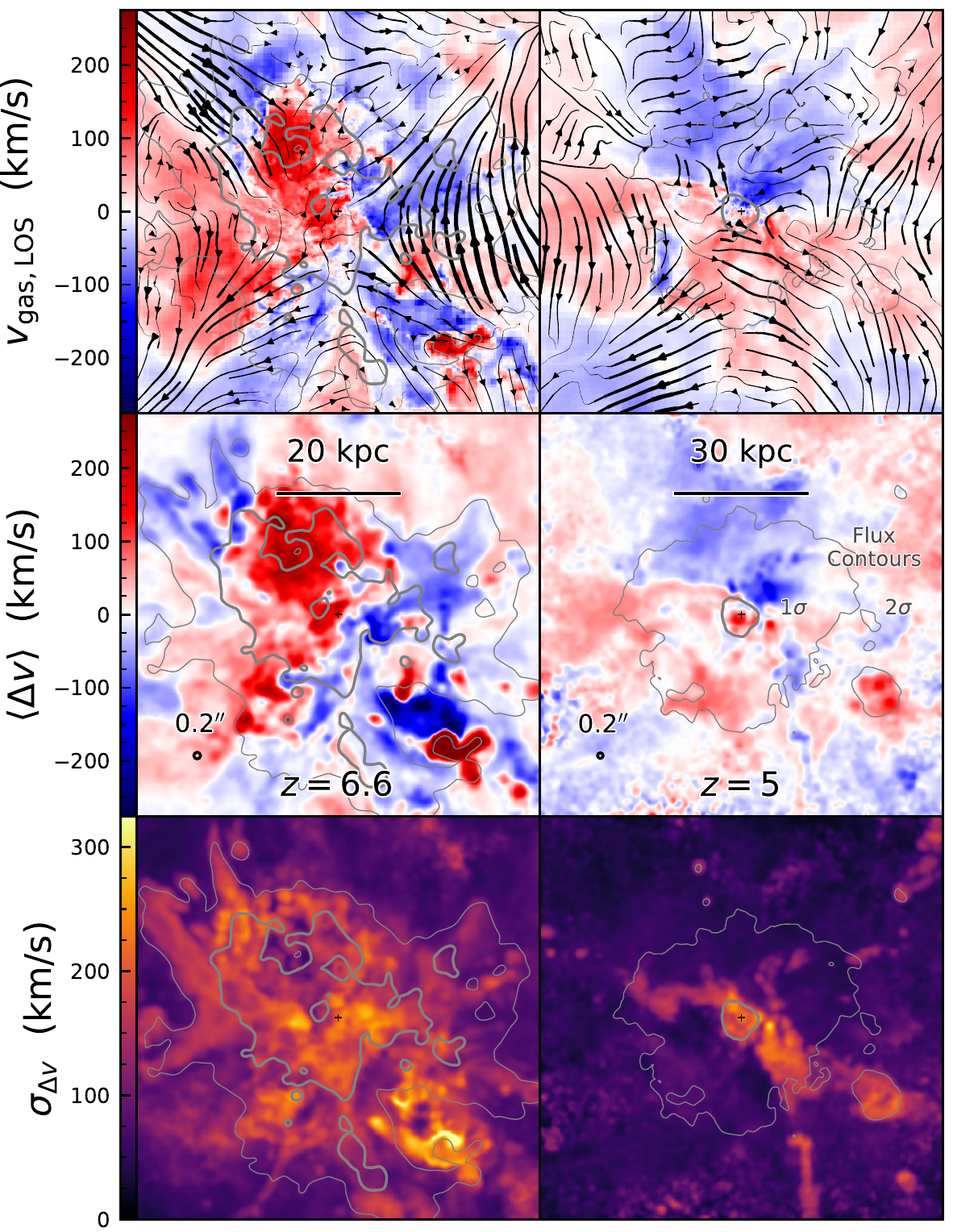}
  \caption{Spatial images of the LOS gas velocity, flux-weighted frequency centroid $\langle \Delta v \rangle$, and standard deviation $\sigma_{\Delta v}$ of the escaped Ly$\alpha$ radiation at $z = 6.6$ and $z = 5$, before transmission through the IGM. There are hints of effects due to CGM scale bulk rotation, radial inflows and outflows, and significant line broadening from resonant scattering in the optically thick regions. For reference, we include $1\sigma$ and $2\sigma$ Ly$\alpha$ surface brightness contours. Likewise, the streamlines illustrate the perpendicular flow, with the thickness denoting the relative magnitude.}
  \label{fig:moments}
\end{figure}

In Fig.~\ref{fig:SB_r} we show the median radial surface brightness after IGM transmission for each snapshot colored according to redshift. The shaded regions represent the 1$\sigma$ confidence levels based on the variation between different sightlines for the 3072 healpix directions. The radial profiles illustrate the cuspy nature of the central Ly$\alpha$ source and the transition to an exponentially damped halo. We fit the logarithmic slopes of the halos as $\text{SB}(r) \propto \exp(-r / R_{\alpha,\text{h}})$ beyond 10\,kpc, plotting the characteristic scale lengths of the Ly$\alpha$ haloes $R_{\alpha,\text{h}}$ both after ISM scattering and IGM transmission as an inset within the figure. In Fig.~\ref{fig:SB_r} we also present the median normalized integrated light within a given radius, which may be thought of as $P(<r) \propto \int_0^r \text{SB}(r') r' \text{d}r'$. We employ linear radial bins with $\sim 1\,\text{kpc}$ resolution for $\text{SB}(r)$ and higher resolution logarithmic radial bins to calculate the cumulative distribution function $P(<r)$. As with the radial surface brightness profiles we also provide the 1$\sigma$ confidence levels based on viewing angle uncertainties. We also show the redshift evolution of the half-light radius, defined as $P(<R_{1/2}) = 1/2$ normalized such that $P(<2R_\text{vir}) = 1$, both after ISM scattering and IGM transmission as an inset within the figure.
For reference, the Ly$\alpha$ halo scale length is comparable to the virial radius, i.e. $R_{\alpha,\text{h}} \gtrsim R_\text{vir}$, and the half-light radius is within the radius of maximal circular velocity, i.e. $R_{1/2} \lesssim R_\text{max}$. Specifically, the halo slopes are steeper after IGM transmission with time-averaged values of $\langle R_{\alpha,\text{h}} \rangle_\textsc{ism} \approx 21.4\,\text{kpc}$ and $\langle R_{\alpha,\text{h}} \rangle_\textsc{igm} \approx 17.5\,\text{kpc}$, also displaying slight flattening with redshift. Likewise, the half-light radius decreases after IGM transmission with time-averaged values of $\langle R_{1/2} \rangle_\textsc{ism} \approx 7.2\,\text{kpc}$ and $\langle R_{1/2} \rangle_\textsc{igm} \approx 6.0\,\text{kpc}$, becoming increasingly concentrated with redshift (see Tables~\ref{tab:time_avg_ism}~and~\ref{tab:time_avg_igm}). The Ly$\alpha$ haloes are significantly more extended than the UV continuum emission with $\langle R_{1/2} / R_{1/2,\text{UV}} \rangle_\textsc{ism} \approx 2.7$ and $\langle R_{1/2} / R_{1/2,\text{UV}} \rangle_\textsc{igm} \approx 2.2$. After accounting for the $(1+z)^{-4}$ dependence of the Ly$\alpha$ surface brightness along with potential variations due to environmental factors and IGM reprocessing, our predicted profiles are consistent with the Ly$\alpha$ haloes generically observed around star-forming galaxies at lower redshifts, e.g. $R_{\alpha,\text{h}} \approx 25\,\text{kpc}$ in \citet{Steidel_2011}.

\subsection{Line flux}
\label{sec:line_flux}
The spectral line profile is perhaps the most distinguishing observable feature from Ly$\alpha$ emitting galaxies. To summarize the redshift evolution of its characteristic properties, in Fig.~\ref{fig:flux_esc} we display the median LOS emergent Ly$\alpha$ flux density with 1$\sigma$ confidence regions, as a function of Doppler velocity $\Delta v = c \Delta \lambda / \lambda$ for each simulation snapshot considering both ISM scattering (dashed) and IGM transmission (solid). The top panel employs a logarithmic scale to show the time-dependent luminosity while the middle panel utilizes a linear scale to illustrate the evolving shape of the normalized lines. The bottom panel shows the cumulative distribution function of each profile. To further condense the information, we also include the redshift dependence of the peak velocity offset $\Delta v_\text{peak}$ shown with the viewing angle uncertainties (shaded regions), as well as the full width of the observed line at half of the maximum (FWHM), both given as median LOS quantities. Specifically, after reprocessing by the IGM these quantities are $\langle \Delta v_\text{peak} \rangle_\textsc{igm} \approx 132\,\text{km\,s}^{-1}$ and $\langle \text{FWHM} \rangle_\textsc{igm} \approx 158\,\text{km\,s}^{-1}$. We also consider the redshift dependence of the flux-weighted frequency centroid $\langle \Delta v \rangle \equiv \int \Delta v f_\lambda\,\text{d}\lambda / \int f_\lambda\,\text{d}\lambda$ and standard deviation of the line $\sigma_{\Delta v} \equiv (\langle \Delta v^2 \rangle - \langle \Delta v \rangle^2)^{1/2}$ (see the inset). The time-averaged values after escaping the target galaxy are $\langle \langle \Delta v \rangle \rangle_\textsc{ism} \approx -5\,\text{km\,s}^{-1}$ and $\langle \sigma_{\Delta v} \rangle_\textsc{ism} \approx 215\,\text{km\,s}^{-1}$, or $\langle \langle \Delta v \rangle \rangle_\textsc{igm} \approx 202\,\text{km\,s}^{-1}$ and $\langle \sigma_{\Delta v} \rangle_\textsc{igm} \approx 126\,\text{km\,s}^{-1}$ after IGM transmission. We note that there is some evolution towards broader and less offset lines with decreasing redshift, mostly due to the changing IGM transmission (see Tables~\ref{tab:time_avg_ism}~and~\ref{tab:time_avg_igm}).

\subsubsection{Red and blue flux ratio}
The ratio of red and blue fluxes $F_\text{red}/F_\text{blue}$ can potentially act as a diagnostic for the inflow/outflow kinematics of galaxies, particularly at lower redshifts where the blue peak is less affected by IGM transmission. Therefore, we focus our discussion on the red-to-blue flux ratio of Ly$\alpha$ photons immediately escaping from the target galaxy. In Fig.~\ref{fig:F_rb_r} we show the median LOS values as a function of radius and redshift. Interestingly, there is an apparent evolution towards lower red-to-blue flux ratios with time, due to the decaying influence of feedback between star forming episodes. In fact, prior to the starburst at $z \approx 5.6$, the infall feature shows significant blue dominance, changing to red dominance as radiative feedback and supernovae launch new galactic winds. To highlight this transition in Fig.~\ref{fig:F_rb_r} we illustrate the spatial distribution of the red-to-blue-flux ratio before and after the starburst, providing 1$\sigma$ and 2$\sigma$ (68\% and 95\%) Ly$\alpha$ surface brightness contours for reference. For presentation purposes, the red and blue channel maps have each been smoothed to simulate a $\text{FWHM} \approx 0\farcs3$ aperture prior to taking the ratio. Finally, we note that the more extended collisional cooling emission almost always exhibits a blue signature (inflow), while the more centralized recombination emission exhibits a red signature (outflow). Specifically, we calculate $\langle F_\text{red}/F_\text{blue} \rangle_\textsc{ism} \approx 1$ with $\langle F_\text{red}/F_\text{blue} \rangle_{\textsc{ism},\text{rec}} \approx 1.2$ and $\langle F_\text{red}/F_\text{blue} \rangle_{\textsc{ism},\text{col}} \approx 0.8$, explaining the lower transmission of collisional emission through the IGM (see Table~\ref{tab:time_avg_ism}). These results are consistent with an observation by \citet*{Erb_2018} of a low mass ($M_\star \approx 5 \times 10^8\,\Msun$), low metallicity ($Z \approx 0.25\,\Zsun$) star-forming galaxy at $z = 2.3$, which exhibits a dominant red peak in the centre with $F_\text{red} \approx F_\text{blue}$ in the outskirts of the Ly$\alpha$ halo.

\subsubsection{Moment maps}
Images of the frequency moments before IGM transmission provide additional insight beyond the spatially resolved ratio of red and blue fluxes. In Fig.~\ref{fig:moments} we show the flux-weighted frequency centroid $\langle \Delta v \rangle$ and standard deviation $\sigma_{\Delta v}$ of the escaped Ly$\alpha$ radiation at $z = 6.6$ and $z = 5$, for direct comparison with many of the other figures in this paper. Despite the substantial morphological differences between the two redshifts, both seem to exhibit evidence for effects due to CGM scale bulk rotation, radial inflows and outflows, and significant line broadening from resonant scattering in the optically thick regions. There is a very close correspondence between the LOS gas velocity and frequency moments.

\begin{figure}
  \centering
  \includegraphics[width=\columnwidth]{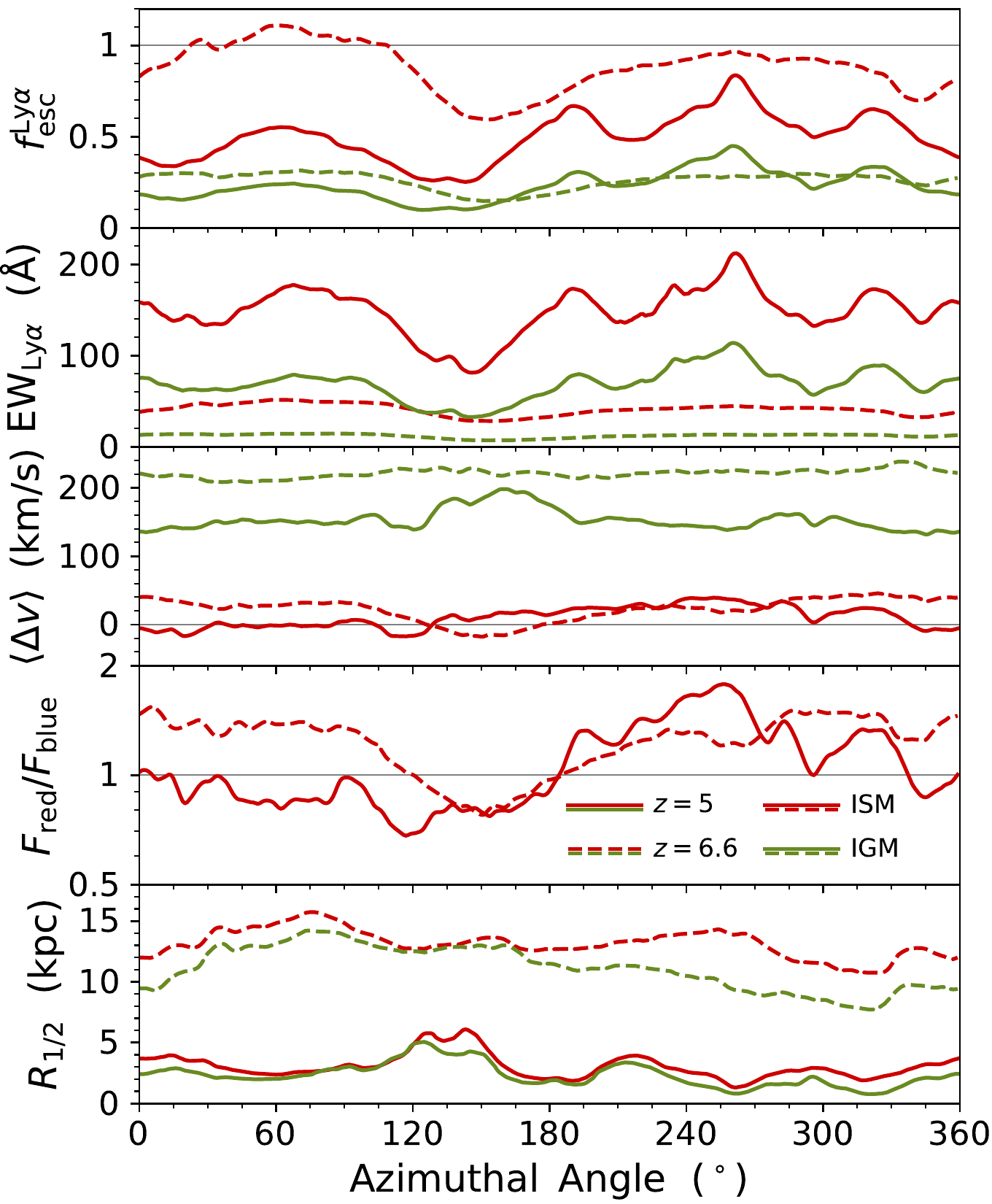}
  \vspace{-.25cm} \\
  \includegraphics[width=\columnwidth]{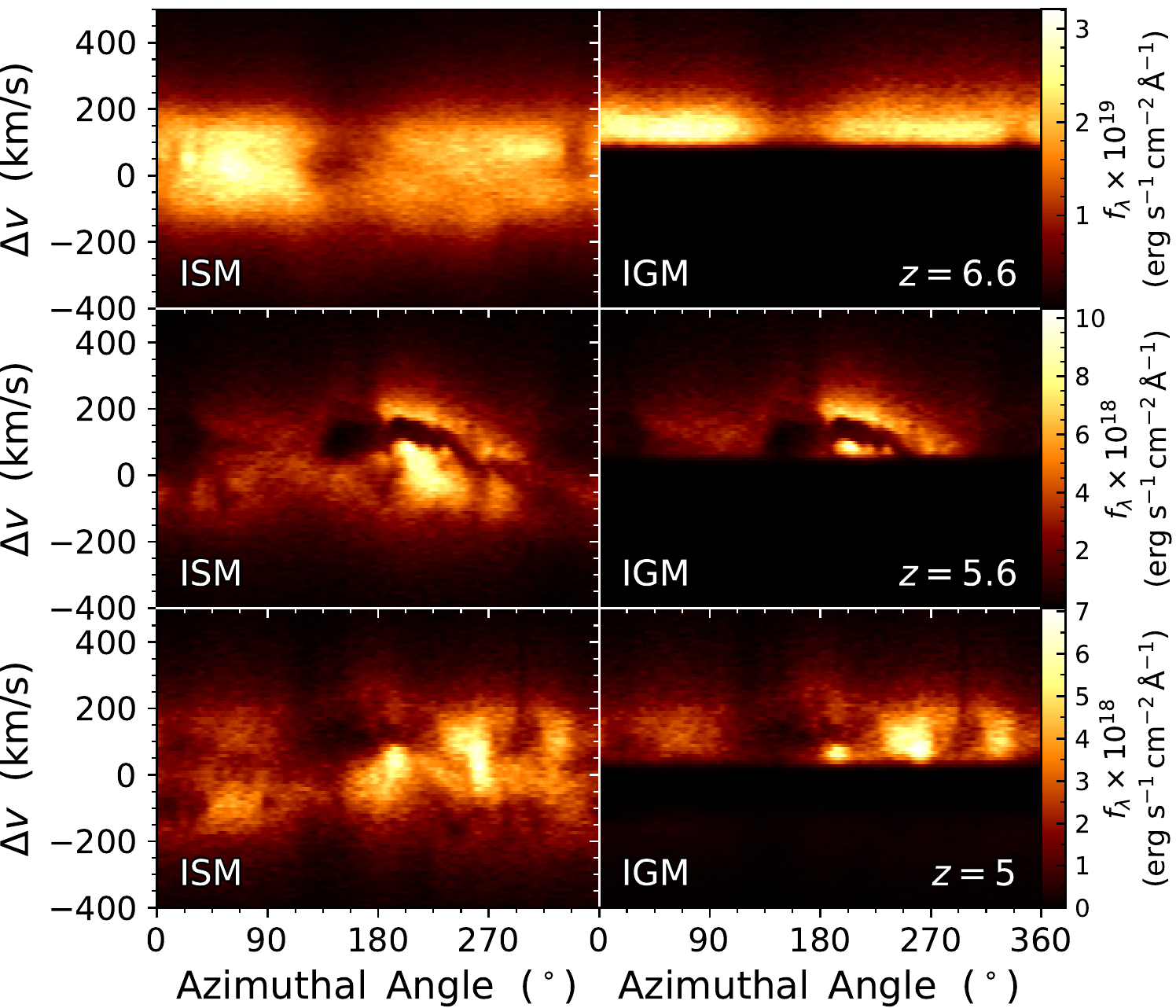}
  \caption{Continuous dependence of Ly$\alpha$ properties for equatorial sightlines as a function of azimuthal angle, including the escape fraction $f_\text{esc}^{\text{Ly}\alpha}$, rest frame equivalent width $\text{EW}_{\text{Ly},\alpha}$, flux-weighted velocity centroid $\langle \Delta v \rangle$, red-to-blue flux ratio $F_\text{red} / F_\text{blue}$, and half-light radius $R_{1/2}$. The lower figure shows the complex diversity of spectral line profiles along the rotation axis after ISM scattering and IGM transmission at $z = \{5, 5.6, 6.6\}$.}
  \label{fig:azimuthal}
\end{figure}

\subsubsection{Azimuthal variations}
\label{sec:azimuthal}
To explore the line profile in further detail, in Fig.~\ref{fig:azimuthal} we plot the Ly$\alpha$ escape fraction $f_\text{esc}^{\text{Ly}\alpha}$, rest frame equivalent width $\text{EW}_{\text{Ly}\alpha,0}$, flux-weighted velocity centroid $\langle \Delta v \rangle$, red-to-blue flux ratio $F_\text{red} / F_\text{blue}$, and half-light radius $R_{1/2}$, as a function of azimuthal angle for an arbitrary rotation axis. We also show how the full spectral line profile smoothly changes across these viewing angles. In particular, the linear colour scale illustrates the impact of directional dependence on the peak velocity offset due to neutral hydrogen cloud obscuration, dust absorption, and Doppler shifting from bulk velocity flows, which affects the escape fraction both after ISM scattering and IGM transmission.

\subsection{Angular power spectra}
\begin{figure}
  \centering
  \includegraphics[clip, trim=1.11cm 0cm .12cm 0cm, width=.49\columnwidth]{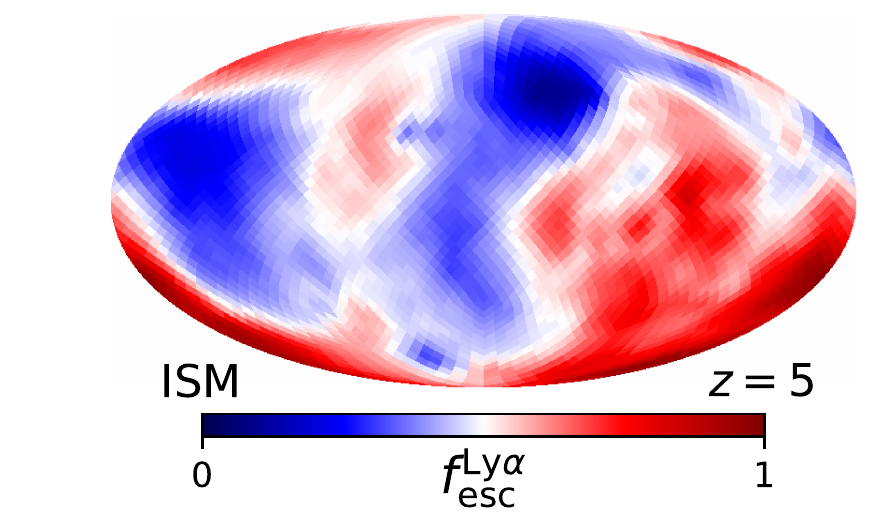}
  \includegraphics[clip, trim=1.11cm 0cm .12cm 0cm, width=.49\columnwidth]{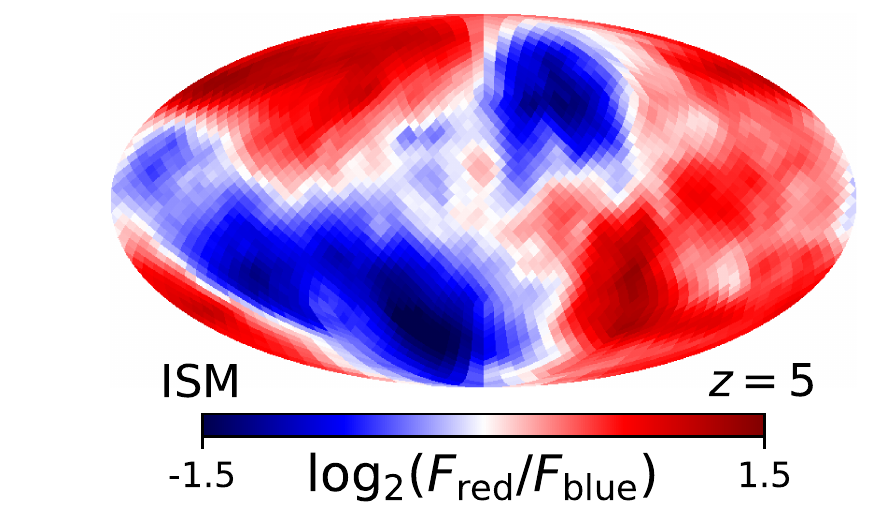}
  \includegraphics[clip, trim=1.11cm 0cm .12cm 0cm, width=.49\columnwidth]{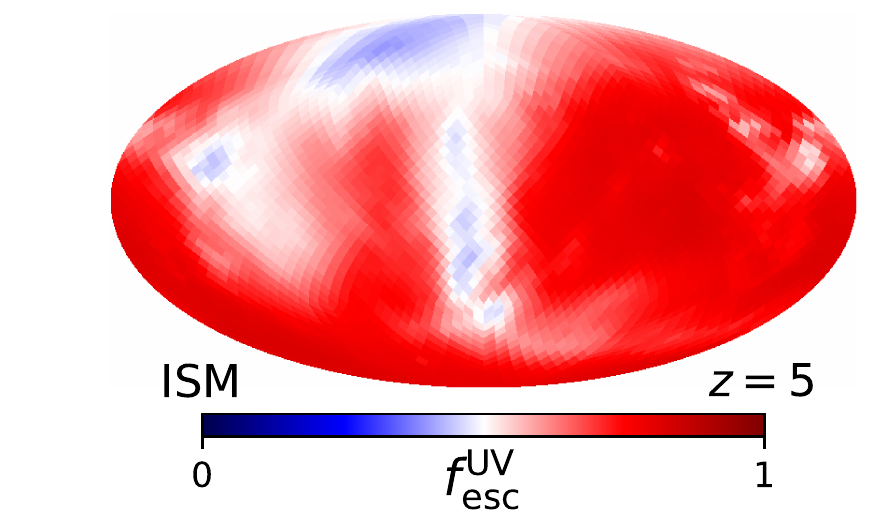}
  \includegraphics[clip, trim=1.11cm 0cm .12cm 0cm, width=.49\columnwidth]{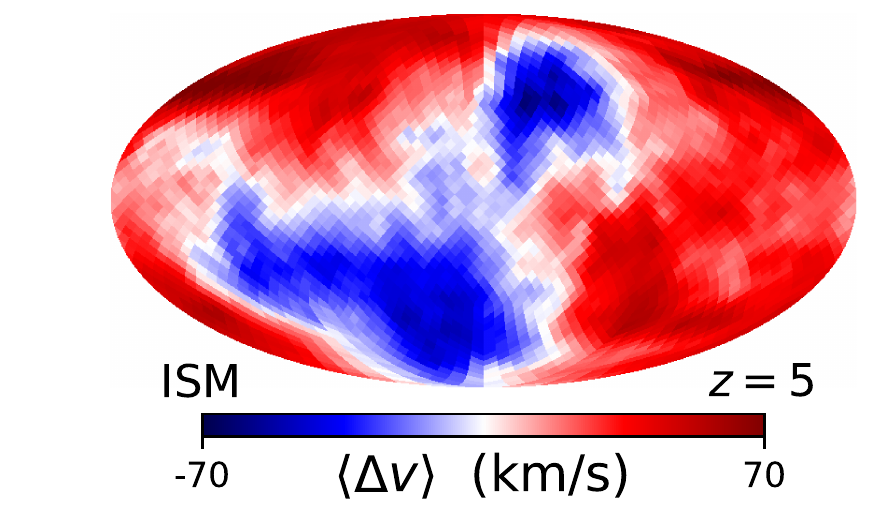}
  \includegraphics[clip, trim=1.11cm 0cm .12cm 0cm, width=.49\columnwidth]{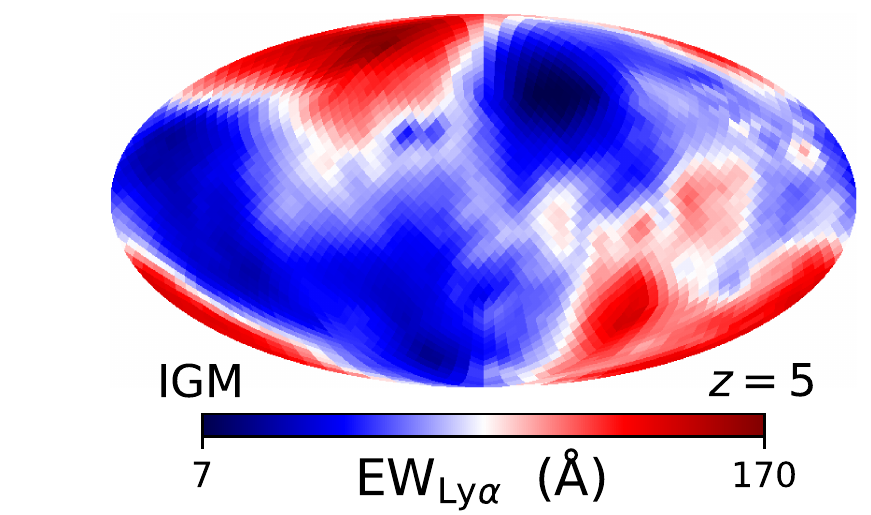}
  \includegraphics[clip, trim=1.11cm 0cm .12cm 0cm, width=.49\columnwidth]{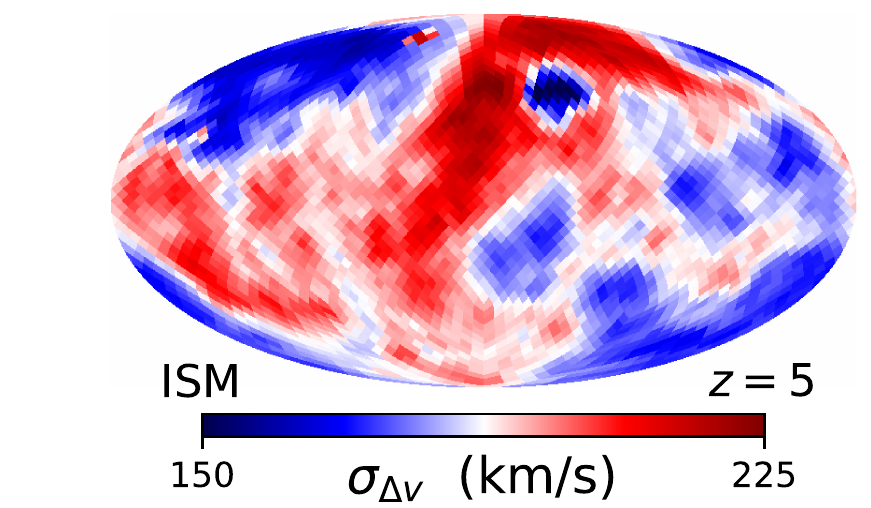}
  \includegraphics[width=\columnwidth]{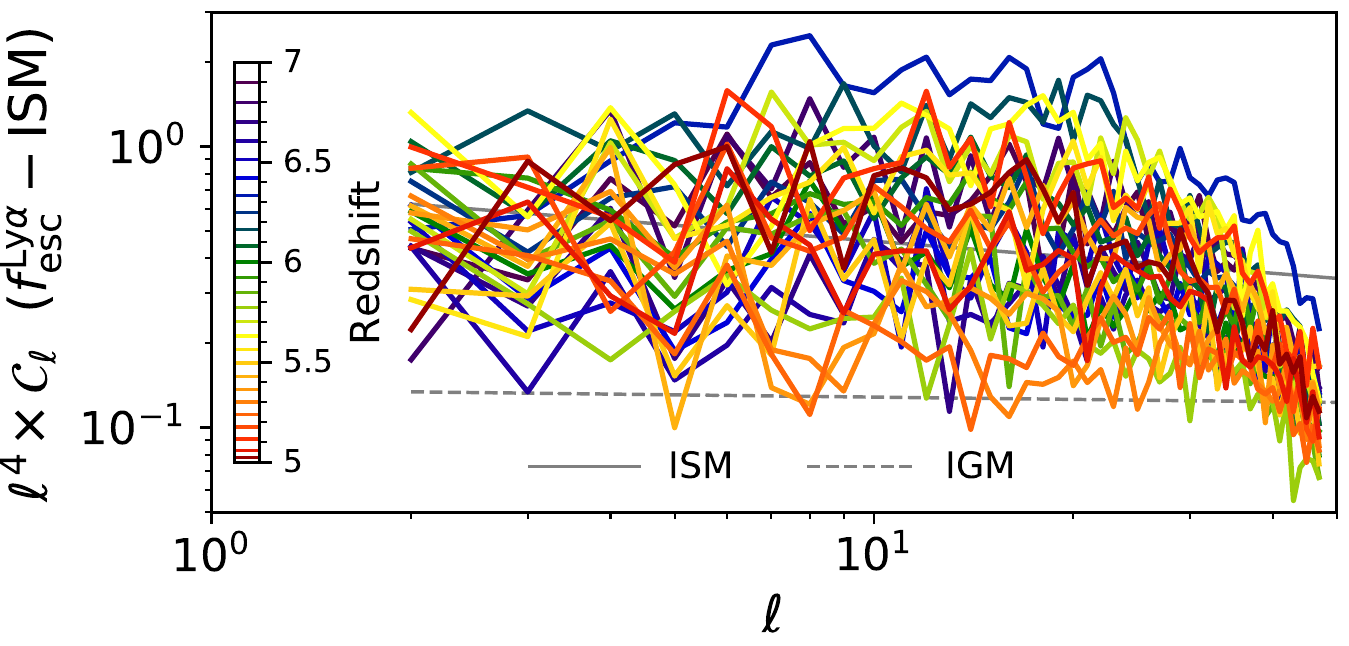}
  \caption{The angular distributions of various Ly$\alpha$ properties after scattering through the ISM, including $f_\text{esc}^{\text{Ly}\alpha}$, $f_\text{esc}^\text{UV}$, $\text{EW}_{\text{Ly}\alpha,0}$ (after IGM transmission), $F_\text{red} / F_\text{blue}$, $\langle \Delta v \rangle$, and $\sigma_{\Delta v}$. There are hints of connections between galaxy kinematics, dust absorption, and line broadening due to resonant scattering. The highest rest frame equivalent widths also correspond to outflowing regions with narrower (or single peaked) lines and greater UV continuum attenuation. In the lower panel we also show the angular power spectrum of the Ly$\alpha$ escape fraction before transmission through the IGM. The downturn at the highest $\ell$ modes is caused by numerical damping from the Gaussian filter with an effective resolution of $\sigma_\theta \approx 3\fdg2$, corresponding to $\sigma_\ell \approx 56$, used in the directional binning of photon packets.}
  \label{fig:angular}
\end{figure}

We now consider the angular variations of the spatially integrated flux for all lines of sight. In Fig.~\ref{fig:angular} we show the escape fraction of Ly$\alpha$ photons and other quantities in each of the 3072 healpix directional bins. There are hints of correlations between the galaxy kinematics, dust absorption, and line broadening due to resonant scattering. Although the detailed radiative transfer is far from trivial, it is not surprising that the highest rest frame equivalent widths also correspond to outflowing regions with a narrow dominant red peak and greater UV continuum attenuation.

For the most part, significant differences between pixels are mainly apparent on large scales. We quantify this structure by considering the spherical harmonic decomposition of a given quantity,
\begin{equation}
  f(\theta,\varphi) = \sum_{\ell=0}^\infty \sum_{m=-\ell}^\ell a_{\ell m} Y_{\ell m}(\theta,\varphi) ,
\end{equation}
where the spectral coefficients $a_{\ell m}$ represent the contribution from each of the real harmonics $Y_{\ell m}$ and are found by integration as $a_{\ell m} = \int f(\theta,\varphi) Y_{\ell m}(\theta,\varphi)\,\text{d}\Omega$. We calculate the angular power spectrum as
\begin{equation}
  C_\ell = \frac{1}{2 \ell + 1} \sum_{m=-\ell}^\ell |a_{\ell m}|^2 \, ,
\end{equation}
which is well approximated by a power law with time-averaged slopes of $\text{d}\log C_\ell / \text{d}\log f_\text{esc}^{\text{Ly}\alpha} \approx -4.2$ after ISM scattering and $\approx -4.0$ after IGM transmission (shown in Fig.~\ref{fig:angular}). Similarly, $\text{d}\log C_\ell / \text{d}\log(F_\text{red}/F_\text{blue}) \approx -3.7$ (ISM), $\text{d}\log C_\ell / \text{d}\log f_\text{esc}^\text{UV} \approx -3.6$, $\text{d}\log C_\ell / \text{d}\log R_{1/2} \approx -3.9$ (ISM) and $\approx -3.6$ (IGM), and $\text{d}\log C_\ell / \text{d}\log \text{EW}_{\text{Ly}\alpha,0} \approx -4.1$ (ISM) and $\approx -4.0$ (IGM). We restrict the fit range to $2 \leq \ell \leq 35$, which minimizes the bias due to numerical damping from the Gaussian filter used in the directional binning of photon packets. Heuristically, we can associate each $\ell$ mode with a typical angular size of $\theta \sim 180\degr / \ell$. Thus, the damping of the angular power spectrum means that at increasingly small scales the deviations in $f_\text{esc}^{\text{Ly}\alpha}$ are decreasing even faster. Essentially, the global escape is smoothed out on small angular scales by the multiple scatterings during the transition from optically thick gas to the free streaming regime beyond the circumgalactic medium. The low $\ell$ mode fluctuations may also be viewed as the projection of galactic and cosmic web structures. For example, regions with the lowest escape fractions likely correspond to cosmological filaments and to a lesser degree high opacity clouds that produce a shadowing effect. In this sense, we find that the covering factor is quite small, such that the global Ly$\alpha$ escape is fairly isotropic. However, the rest frame equivalent width can vary on much smaller angular scales as UV continuum photons do not undergo the same multiple scattering.

\begin{figure}
  \centering
  \includegraphics[width=\columnwidth]{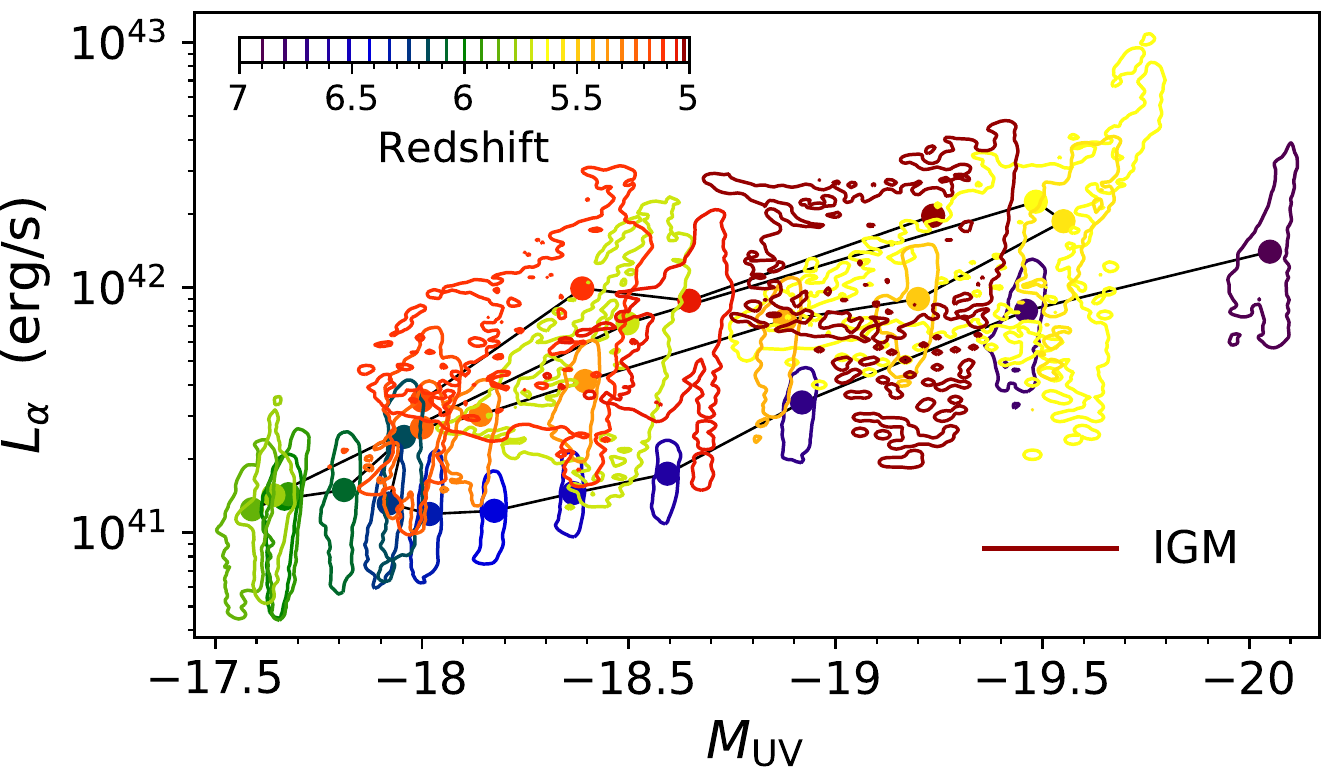}
  \includegraphics[width=\columnwidth]{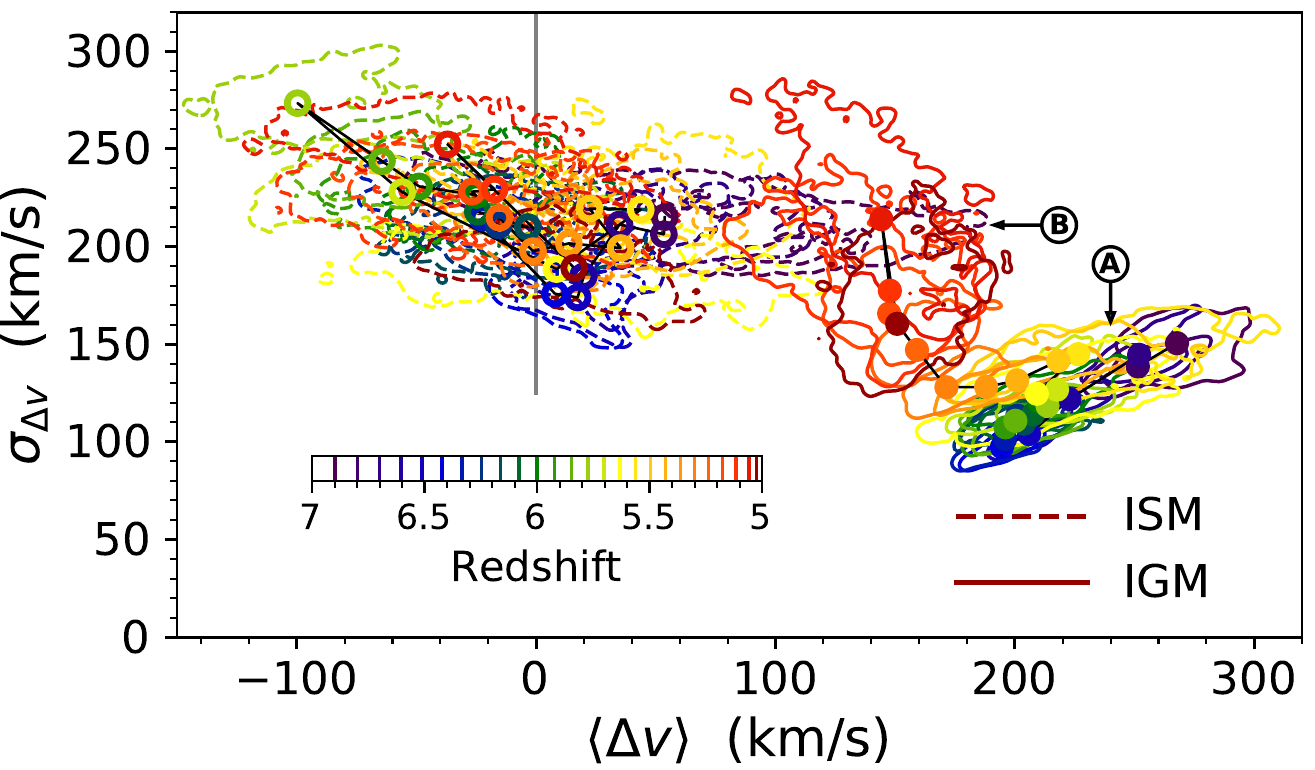}
  \caption{The redshift evolution for potentially correlated observables shown as the median LOS values along with $1\sigma$ contours illustrating the variation across 3072 sightlines. The upper panel shows the Ly$\alpha$ luminosity $L_\alpha$ and UV continuum absolute magnitude $M_\text{UV}$, highlighting the role of the starburst in the visibility. The lower panel shows the line width standard deviation $\sigma_{\Delta v}$ and frequency centroid $\langle \Delta v \rangle$, which could help constrain models of galaxy kinematics and IGM transmission. The dashed (open) and solid (filled) contours (medians) denote values after ISM scattering and IGM transmission, respectively. The regions labeled `A' and `B' are to guide the discussion in the text.}
  \label{fig:contours}
\end{figure}

\section{Observability}
\label{sec:observability}

\subsection{Correlations between observables}
The relationships between the various observables considered in this paper are highly nontrivial, and would likely be most meaningful for large samples of simulated and observed LAEs covering a wide range of possible masses, environments, and star formation histories. Therefore, we simply attempt to validate our intuition regarding the redshift evolution and uncertainties from our target galaxy. In particular, in Fig.~\ref{fig:contours} we show the median values with $1\sigma$ LOS contours for the observed Ly$\alpha$ luminosity $L_\alpha$ and UV continuum absolute magnitude $M_\text{UV}$. This illustrates the role of the starburst in LAE visibility. There are also hints of a positive correlation between different sightlines at each redshift, however, at low metallicity the spread in $f_\text{esc}^{\text{Ly}\alpha}$ dominates over the spread in $f_\text{esc}^\text{UV}$. Still, the median observed $L_\alpha$ can vary by almost an order of magnitude between different redshifts with the same $M_\text{UV}$. Thus, due to the efficient production of Ly$\alpha$ photons from young massive stars, there is an overall increase in the equivalent width while ramping up to the peak brightness of a starburst and a corresponding decrease in the aftermath with the aging of the star populations (see also Fig.~\ref{fig:EW}).

\begin{figure}
  \centering
  \includegraphics[width=\columnwidth]{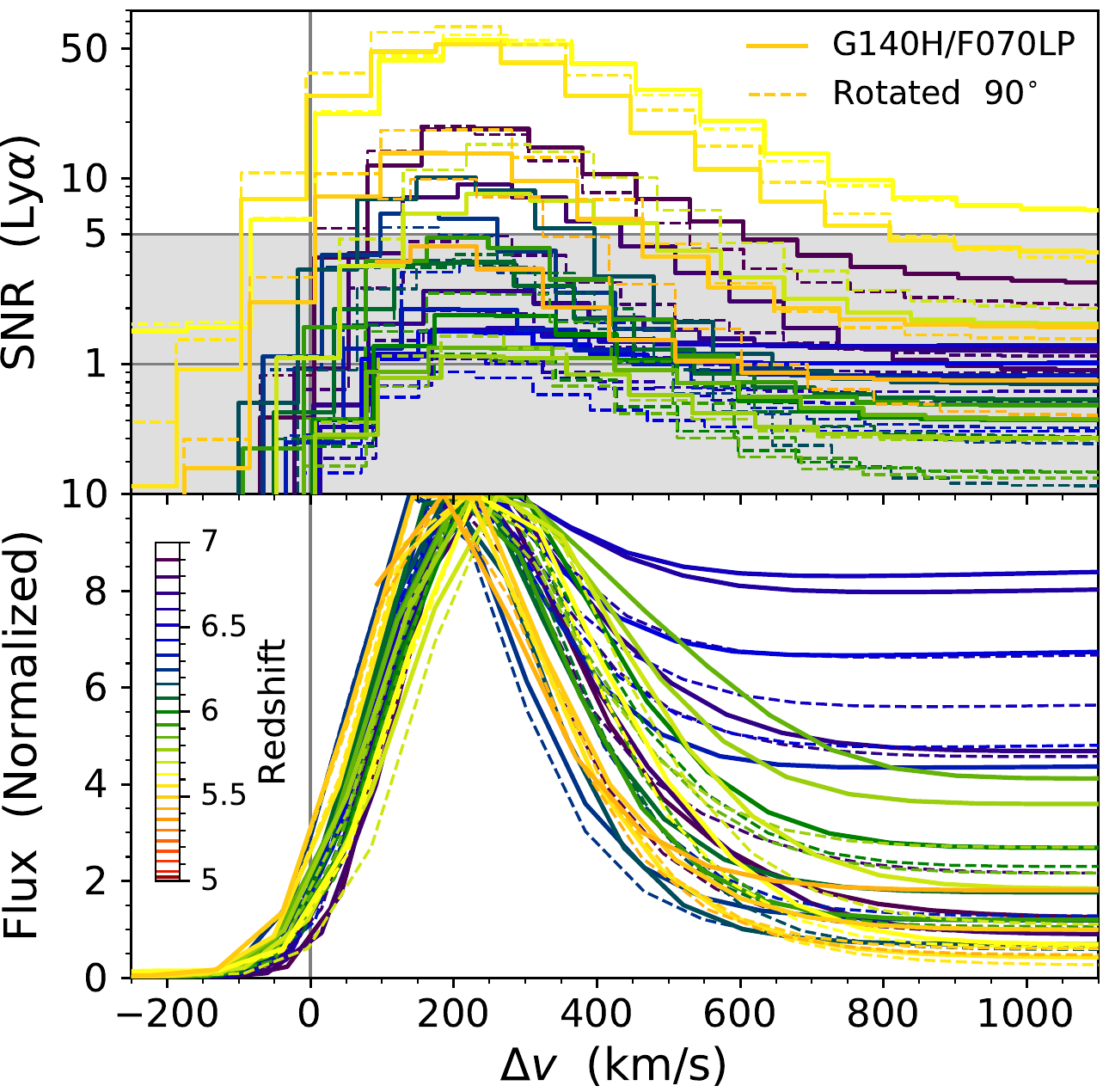}
  \caption{Spectral signal to noise ratio (SNR) after $10^4$ seconds of exposure time for the \textit{JWST} three shutter MSA high resolution G140H/F070LP spectral configuration. For comparison, we also show the flux normalized to the peak value as a function of Doppler velocity $\Delta v = c \Delta \lambda / \lambda$ for each simulation snapshot (see also Fig.~\ref{fig:flux_esc}). The dashed curves correspond to the same direction but with the slit rotated by $90\degr$ to demonstrate the approximate variation due to orientation.}
  \label{fig:SNR_flux}
\end{figure}

\begin{figure}
  \centering
  \includegraphics[width=\columnwidth]{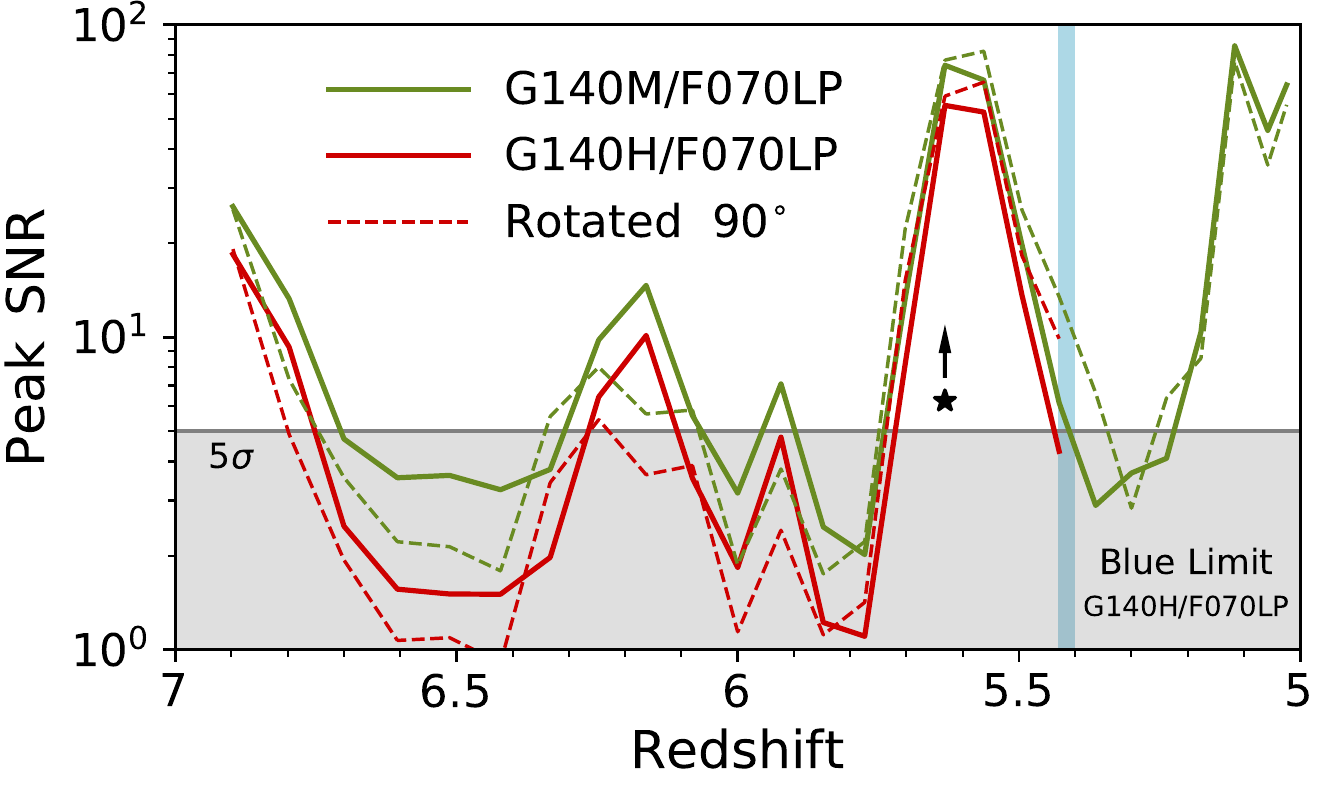}
  \includegraphics[width=\columnwidth]{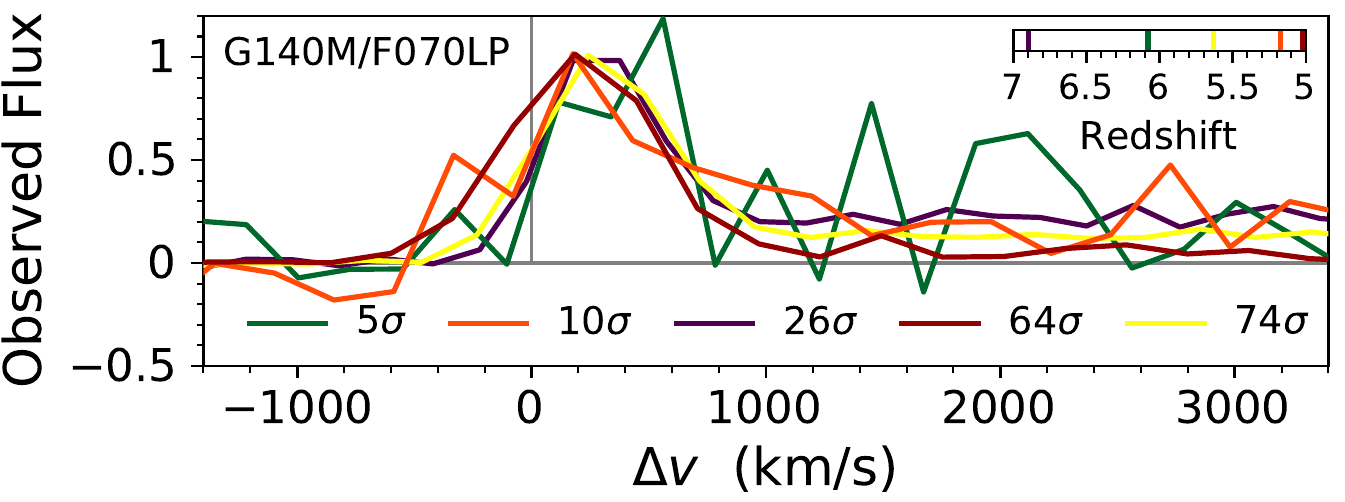}
  \caption{Redshift evolution of the peak SNR corresponding to Fig.~\ref{fig:SNR_flux}, which is closely correlated with the luminosities from Fig.~\ref{fig:L_esc}. The restricted wavelength coverage of the high resolution G140H/F070LP configuration implies a blue lower limit of $z \gtrsim 5.4$ for LAE detections, while the medium resolution G140M/F070LP mode has higher SNRs and allows coverage down to $z \gtrsim 4.8$. In the lower panel we show example observed spectra inclusive of noise, selected to illustrate both high and low peak SNRs.}
  \label{fig:max_SNR}
\end{figure}

In Fig.~\ref{fig:contours} we also show a potential relationship between the line width standard deviation $\sigma_{\Delta v}$ and frequency centroid $\langle \Delta v \rangle$, which could help constrain models of galaxy kinematics and IGM transmission. However, because our IGM model is statistical we warn against over interpretation as the most robust aspect is that IGM transmission reprocesses the Ly$\alpha$ line to be fainter, redder, and narrower. Still, there is an intriguing correlation between the IGM transmitted observables with higher velocity offsets also yielding broader lines (see the region marked ``A''), which follows our intuition from analytical considerations \citep{Neufeld_1990}, empirical results from simulations \citep{Schaerer_2011,Zheng_Wallace_2014}, and observations of LAEs across a wide redshift range \citep{Verhamme_2018}. It seems that for the most realistic radiative transfer models the velocity offset and line width correlation is primarily driven by the \HI\ opacity of the ISM and CGM, with secondary effects from galaxy kinematics, 3D geometry, and transmission through the IGM. The upturn in Fig.~\ref{fig:contours} towards broader lines at lower redshifts is due to a significant increase in the fraction of double peaked spectra (see the region marked ``B''), in which case one might use the separation of blue and red peaks to infer the systemic redshifts of the observed galaxies.

\subsection{\textit{JWST} NIRSpec observations}
\label{sec:NIRSpec}
At high redshifts there is a significant reduction in surface brightness as Ly$\alpha$ photons scatter to larger distances from the galaxy. This has important implications for deciding on optimal Ly$\alpha$ observation strategies. However, one of the preferred observing modes with the upcoming \textit{JWST} will be NIRSpec multi-object spectroscopy (MOS) using the micro-shutter assembly (MSA), as LAE spectroscopic surveys will undoubtedly benefit from the capability of obtaining simultaneous spectra of multiple targets within each $3.6' \times 3.4'$ field of view exposure. Therefore, in this section we focus on a three shutter MSA configuration with a size of $(3 \times 0.46)'' \times 0.2'' = 0.276\,\sq''$ centered on the UV continuum flux centroid. We note that other modes with larger collecting areas could provide higher signal to noise ratios for individual Ly$\alpha$ follow-up observations, e.g. the S400A1 and S1600A1 fixed slits have areas that are respectively $5.5$ and $9.3$ times larger than the three shutter MSA configuration.

In Fig.~\ref{fig:SNR_flux} we illustrate the spectral signal to noise ratio (SNR) and line flux normalized to the peak value for each simulation snapshot, while in Fig.~\ref{fig:max_SNR} we show the peak SNR as a function of redshift. The curves were calculated with the \textit{JWST} exposure time calculator using the total flux extracted from the MSA region after applying a Gaussian point spread function with $\text{FWHM} \approx 0\farcs04$ \citep{Pontoppidan_2016}. Each curve corresponds to the same direction and orientation, although we also show the results obtained when the slit is rotated by $90\degr$ to demonstrate the approximate variation due to the slit orientation. We find that the SNR and peak locations are fairly robust as this is dominated by the central emission. However, the orientation can induce subtle differences in the UV continuum flux due to the obliquity of stellar orbits, and as seen in Section~\ref{sec:line_flux} the details of the line profile are affected by the density and kinematics of the observed regions.

The SNR and flux profiles are for the high resolution G140H/F070LP spectral configuration, which provides a nominal resolving power of $R \approx 2,700$ over a wavelength range of $\lambda = 0.78$--$1.27\,\umu\text{m}$. In Fig.~\ref{fig:max_SNR} we also show the results under the medium resolution G140M/F070LP configuration with $R \approx 1,000$ down to $\lambda \approx 0.7\,\umu\text{m}$. We note that the restricted wavelength coverage implies a blue lower limit for the Ly$\alpha$ line corresponding to $z \gtrsim 5.4\ (4.8)$ for the high (medium) resolution mode, although at wavelengths shorter than $0.9\,\umu\text{m}$ the filter throughput is already below $20\%$. For our calculations we have chosen an exposure time of $10^4$\,seconds, which is a realistic depth that nicely demonstrates the source coming in and out of visibility. Unfortunately, it seems that spectroscopic continuum detections for relatively low mass LAEs, such as the one in this study, would require significantly deeper exposures. However, these measurements can still be obtained with \textit{JWST} NIRCam photometry or the low resolution PRISM/CLEAR configuration with $R \approx 100$. In any case, with a time-averaged peak SNR of $18\sigma$ ($11\sigma$) and a $> 5\sigma$ detection duty cycle of $37\%$ ($25\%$) for $10^4$\,seconds of exposure time using the medium (high) resolution configuration, it seems that Ly$\alpha$ detections and spectroscopy of high-$z$ galaxies is quite feasible. The medium (high) resolution configuration only achieves $\Delta v \approx 300$ ($100$)\,$\text{km\,s}^{-1}$, roughly the width of the IGM transmitted lines (see the spectra with noise in Fig.~\ref{fig:max_SNR}). Ly$\alpha$ surveys at higher spectral resolution are possible with large aperture ground-based telescopes, however, even with next-generation facilities follow-up observations are subject to the night sky background. The high likelihood of encountering a sky emission line typically restricts LAE searches to narrow redshift windows of reduced contamination. Still, diagnostics from other lines and cross-correlation studies will also help unravel the detailed properties of high-$z$ LAEs.

\begin{figure}
  \centering
  \includegraphics[width=\columnwidth]{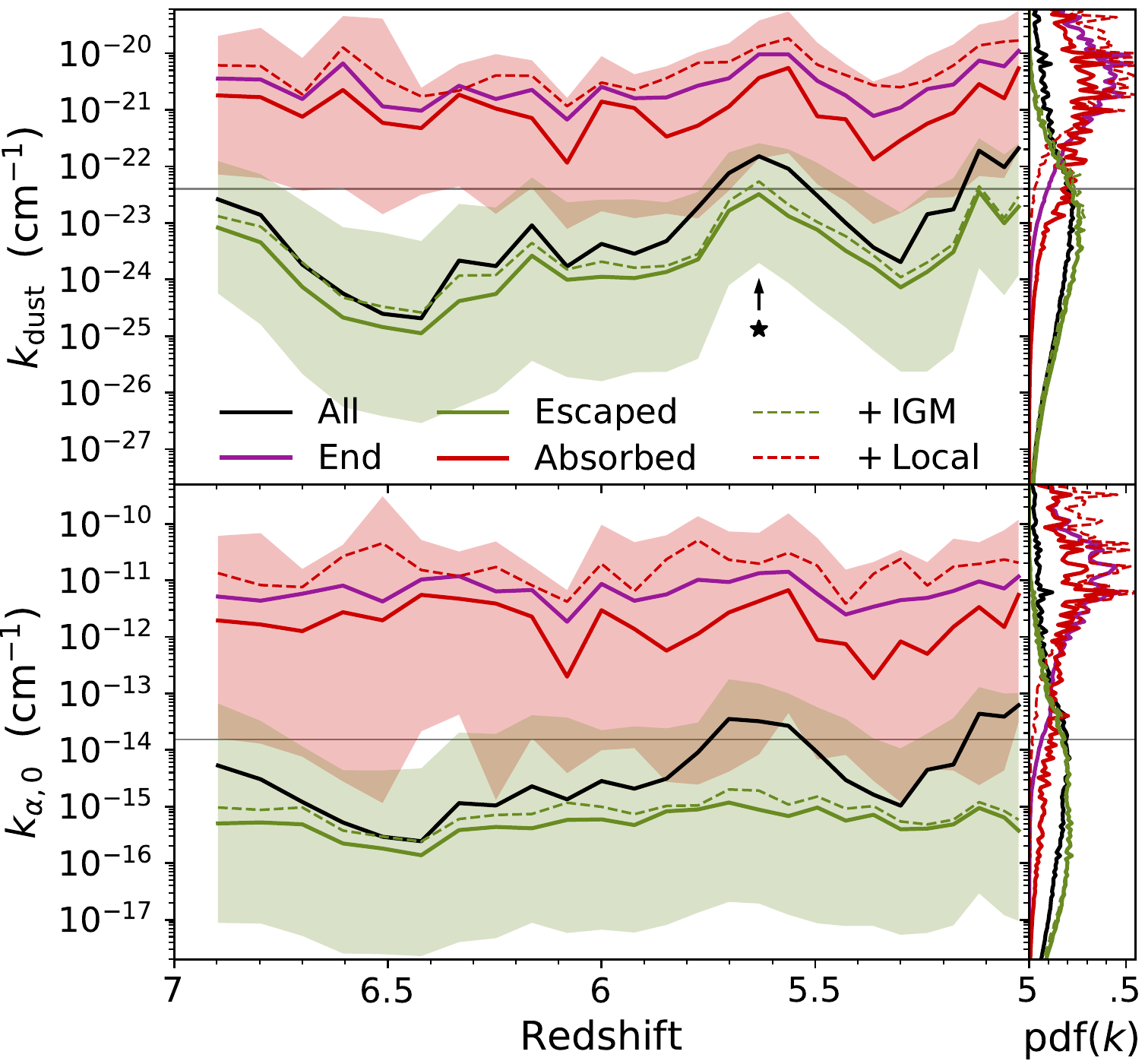}
  \caption{The local absorption coefficients for dust $k_\text{dust}$ and Ly$\alpha$ $k_{\alpha,0}$ at line centre are fairly accurate predictors for the Ly$\alpha$ escape fraction. The median curves show the evolution and time-averaged probability distribution functions for the categories described in Section~\ref{sec:insights} and summarized in Table~\ref{tab:photons}, which also provides statistics on the probability bisect (grey line), overlap, and separation distance for the absorbed and escaped distributions.}
  \label{fig:k_source}
\end{figure}

\section{Predicting the escape of individual photons}
\label{sec:insights}
The Monte Carlo procedure also naturally generates a mapping from the local emission of each photon packet to the corresponding outcome, motivating predictive models for the Ly$\alpha$ escape fractions of galaxies in cosmological simulations. The full radiative transfer calculations only need to be carried out for select training datasets and then the computationally inexpensive local model can be applied to a larger number of simulated galaxies to probe the statistical properties of LAEs, although we defer this particular application to a future study.

\begin{table*}
  \caption{Time-averaged median and 1$\sigma$ statistics over the redshift interval $z = 5$--$7$ for (i) `All' = local environment of all emitted Ly$\alpha$ photon packets, (ii) `Escaped' = subset for escaped photons, (iii) `+IGM' = escaped photons weighted by their IGM transmission, (iv) `Absorbed' = subset for absorbed photons, (v) `+Local' = photons absorbed in the same cell from which they were emitted, and (vi) `End' = environment at the absorption locations. We illustrate the evolution and separation of the absorbed and escaped distributions in Fig.~\ref{fig:k_source} for the absorption coefficients for dust $k_\text{dust}$ and Ly$\alpha$ $k_{\alpha,0}$ at line centre. We also provide the symmetric probability bisect $x_\text{B}$, the minimum overlap $\underline{\omega}$, $L^2$-norm overlap $\omega_2$, Hellinger distance $H$, first Wasserstein distance $W_1$, and number of standard deviations $N^\sigma_1$ separating the absorbed and escaped distributions (see Equations~\ref{eq:bisect}--\ref{eq:sigma_1}). The rows are in decreasing order of discriminating significance.}
  \label{tab:photons}
  \addtolength{\tabcolsep}{-1.5pt}
  \renewcommand{\arraystretch}{1.25}
  \begin{tabular}{@{} l cccccccccccc @{}}
    \hline
    Quantity & All & Escaped & +IGM & Absorbed & +Local & End & Bisect & $\underline{\omega}$ & $\omega_2$ & $H$ & $W_1$ & $N^\sigma_1$ \\
    \hline
    $\log k_\text{dust}$ \hfill [$\text{cm}^{-1}$] & $-23.24^{+1.85}_{-2.02}$ & $-23.62^{+1.40}_{-1.83}$ & $-23.40^{+1.33}_{-1.75}$ & $-21.07^{+1.15}_{-1.51}$ & $-20.32^{+0.77}_{-0.82}$ & $-20.61^{+0.86}_{-1.11}$ & $-22.40$ & $0.39$ & $0.30$ & $0.58$ & $2.56$ & $1.74$ \\
    $\log k_{\alpha,0}$ \hfill [$\text{cm}^{-1}$] & $-14.57^{+2.14}_{-2.23}$ & $-15.31^{+1.80}_{-1.83}$ & $-15.11^{+1.82}_{-1.81}$ & $-11.77^{+1.44}_{-2.50}$ & $-10.82^{+1.02}_{-1.04}$ & $-11.20^{+0.97}_{-1.55}$ & $-13.82$ & $0.40$ & $0.26$ & $0.56$ & $3.15$ & $1.67$ \\
    $\log \rho$ \hfill [$\text{g\,cm}^{-3}$] & $-23.78^{+1.68}_{-1.17}$ & $-23.90^{+1.33}_{-1.14}$ & $-23.71^{+1.31}_{-1.17}$ & $-21.92^{+0.67}_{-1.10}$ & $-21.65^{+0.53}_{-0.71}$ & $-21.96^{+0.70}_{-1.17}$ & $-22.78$ & $0.41$ & $0.36$ & $0.54$ & $1.73$ & $1.68$ \\
    $\log T$ \hfill [$\text{K}$] & $4.04^{+0.24}_{-0.79}$ & $4.02^{+0.23}_{-0.55}$ & $4.00^{+0.21}_{-0.63}$ & $3.13^{+0.74}_{-1.19}$ & $2.75^{+0.67}_{-1.21}$ & $3.06^{+0.77}_{-1.13}$ & $3.66$ & $0.44$ & $0.66$ & $0.48$ & $0.92$ & $1.39$ \\
    $\log x_\text{\HI}$ & $-0.91^{+0.77}_{-2.37}$ & $-1.90^{+1.63}_{-1.78}$ & $-1.86^{+1.60}_{-1.82}$ & $-0.35^{+0.26}_{-2.29}$ & $-0.18^{+0.11}_{-0.30}$ & $-0.11^{+0.10}_{-0.25}$ & $-0.87$ & $0.62$ & $0.58$ & $0.37$ & $0.95$ & $0.69$ \\
    $\log r$ \hfill [$\text{kpc}$] & $0.72^{+0.56}_{-0.49}$ & $0.73^{+0.59}_{-0.51}$ & $0.70^{+0.56}_{-0.50}$ & $0.45^{+0.38}_{-0.42}$ & $0.44^{+0.36}_{-0.43}$ & $0.46^{+0.38}_{-0.42}$ & $0.59$ & $0.72$ & $0.77$ & $0.28$ & $0.28$ & $0.56$ \\
    $\log Z$ \hfill [$\Zsun$] & $-1.38^{+0.41}_{-0.67}$ & $-1.37^{+0.41}_{-0.72}$ & $-1.32^{+0.38}_{-0.65}$ & $-1.21^{+0.27}_{-0.40}$ & $-1.17^{+0.25}_{-0.36}$ & $-1.19^{+0.27}_{-0.39}$ & $-1.27$ & $0.80$ & $0.83$ & $0.24$ & $0.27$ & $0.53$ \\
    \hline
  \end{tabular}
  \addtolength{\tabcolsep}{1.5pt}
  \renewcommand{\arraystretch}{0.8}
\end{table*}

\begin{figure}
  \centering
  \includegraphics[width=\columnwidth]{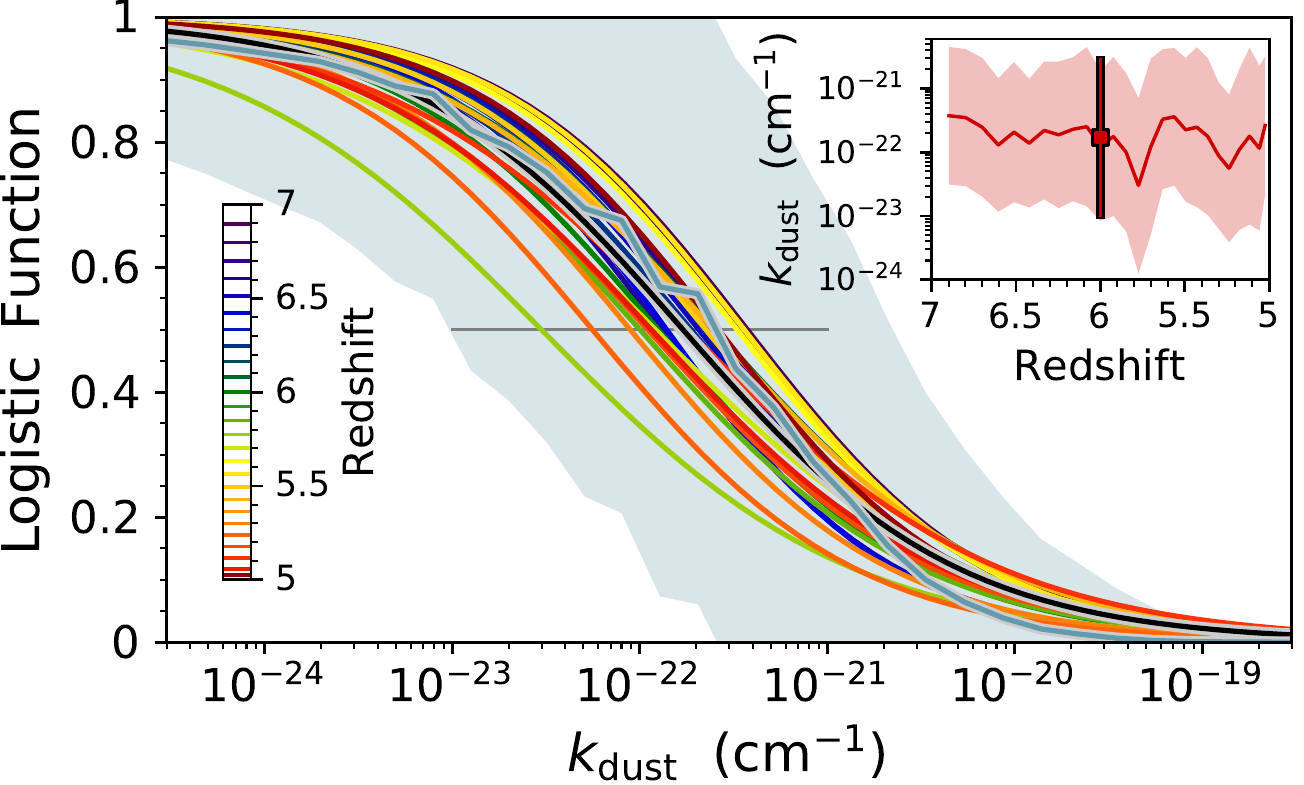}
  \caption{Logistic functions for the Ly$\alpha$ (ISM) escape fraction at different redshifts for a model based on the local dust absorption coefficient $k_\text{dust}$ as a single predictor. The black curve represents the time-weighted model. The blue curve illustrates the average and standard deviation of the simulated escape fractions after binning according to $k_\text{dust}$. The inset shows the redshift dependence of the 50\% transition region with 1$\sigma$ confidences.}
  \label{fig:logistic_kdust}
\end{figure}

For the analysis in this section we track the initial position and cell index of each photon packet, along with the final states of escaping and absorbed photons. At this point we differentiate between six categories of photon trajectories: (i) `All', to describe the local environments of all emitted photon packets, (ii) `Escaped', to select only escaped photons, (iii) `+IGM', to further weight escaped photons according to the IGM transmission based on the escaping frequency, (iv) `Absorbed', to select only absorbed photons, (v) `+Local', to further distinguish between photons that are absorbed in the same cell from which they were emitted, and (vi) `End', to describe the environments where photons are eventually absorbed. In Fig.~\ref{fig:k_source}, we illustrate the properties of each of these categories by considering the local absorption coefficients for dust $k_\text{dust}$ and Ly$\alpha$ $k_{\alpha,0}$ at line centre. We show the redshift evolution of the median values for each category, along with the 1$\sigma$ confidence levels for absorbed (red) and escaped (green) photons. We note that the final absorption environment typically exhibits higher absorption coefficients (purple), especially for photons absorbed ``locally'' (dashed red). Also, the total distribution represents an escape fraction weighted average between the absorbed and escaped distributions. Finally, we include the time-averaged normalized probability distribution functions for each category within inset right panels.

Based on the apparent separation between the absorbed and escaped distributions, it seems reasonable to employ the local dust content as a predictor of the global Ly$\alpha$ escape fraction. However, it is important to understand the validity and robustness of such a procedure. To this end, we consider several statistical quantities to compare the time-averaged absorbed and escaped probability distribution functions, which we denote by $p_\text{abs}(x)$ and $p_\text{esc}(x)$ with $x$ the random variable, e.g. $x = k_\text{dust}$ for the dust absorption coefficient. First, we define the symmetric probability bisect $x_\text{B}$ as the point dividing equal chances of absorption at lower values and escape at higher values,
\begin{equation} \label{eq:bisect}
  \text{Probability~Bisect:} \;\; \int_{-\infty}^{x_\text{B}} p_\text{abs}(x)\,\text{d}x = \int_{x_\text{B}}^\infty p_\text{esc}(x)\,\text{d}x \, .
\end{equation}

\begin{table}
  \caption{Time-averaged logistic functions for models with predicator variables in the set $\{k_\text{dust},k_{\alpha,0},\rho,T\}$ along with the relative error of the estimates. See Equation~(\ref{eq:logistic}) for the relation to the conditional escape probability and Fig.~\ref{fig:logistic_kdust} for an illustration.}
  \label{tab:logistic}
  \begin{tabular}{@{} l cc @{}}
    \hline
    Quantity & Logit: $f(\bm{x}) = \beta \log x + \beta_0$ & $\langle |1 - f_\text{esc,est}^{\text{Ly}\alpha}/f_\text{esc}^{\text{Ly}\alpha}| \rangle$ \\
    \hline
    $k_\text{dust}$ \hfill [$\text{cm}^{-1}$] & $-1.370 \log k_\text{dust} - 29.82$ & 4.91\% \\
    $k_{\alpha,0}$ \hfill [$\text{cm}^{-1}$] & $-0.899 \log k_{\alpha,0} - 11.49$ & 6.66\% \\
    $\rho$ \hfill [$\text{g\,cm}^{-3}$] & $-1.656 \log \rho - 37.00$ & 5.69\% \\
    $T$ \hfill [$\text{K}$] & $2.202 \log T - 6.97$ & 8.99\% \\
    \hline
  \end{tabular}
\end{table}

We next consider the overlap of the distributions, based on their smoothed histograms to ensure common support in $x$. This statistic provides the probability that samples could be represented by either outcome, and is therefore a value in the unit interval. The minimum overlap $\underline{\omega}$ is intuitively the intersection between probability densities,
\begin{equation}
  \underline{\omega} \equiv \int_{-\infty}^\infty \min\big(p_\text{abs}(x), p_\text{esc}(x)\big)\,\text{d}x \, .
\end{equation}
On the other hand, the $L^2$-norm overlap $\omega_2$ considers regions of mutually significant probability with higher weight, which provides information about the similarity of their shapes,
\begin{equation}
  \omega_2 \equiv \left(\int_{-\infty}^\infty p_\text{abs}(x)\,p_\text{esc}(x)\,\text{d}x\right)^{1/2} \, .
\end{equation}
Likewise, the (squared) Hellinger distance also quantifies the similarity between the probability distributions,
\begin{equation}
  H^2 \equiv \frac{1}{2} \int_{-\infty}^\infty \left(\sqrt{p_\text{abs}(x)} - \sqrt{p_\text{esc}(x)}\right)^2\text{d}x \, .
\end{equation}

\begin{figure*}
  \centering
  \includegraphics[width=\textwidth]{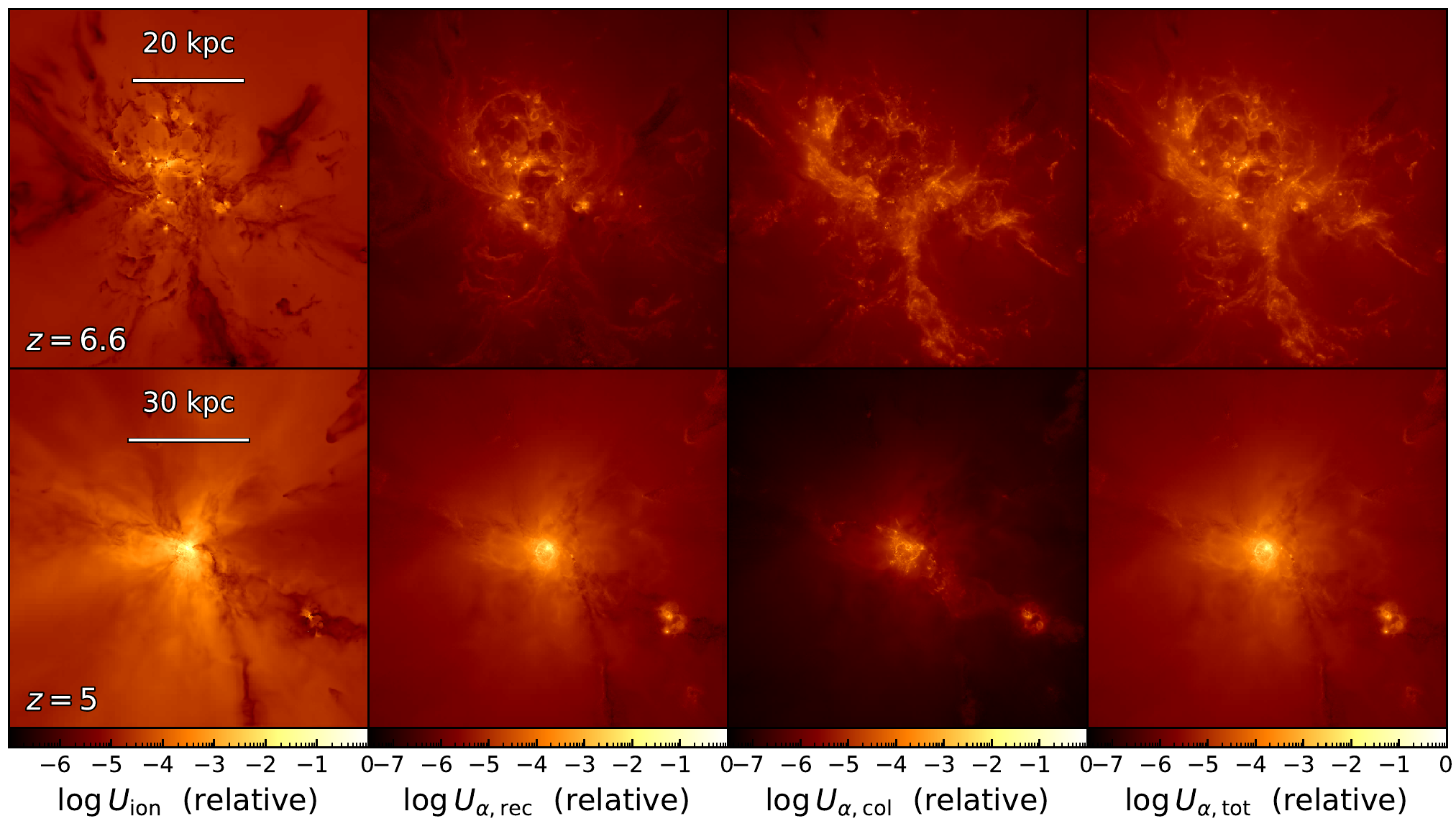}
  \caption{Projected mass-weighted energy density for ionizing and Ly$\alpha$ photons at $z = 6.6$ and $z = 5$, with the recombination and collisional emission colour scaled to the maximum projected $U_{\alpha,\text{tot}}$. Although these quantities are not directly observable, they provide intuition as ionizing and Ly$\alpha$ radiation have significantly different photon trajectories, resulting in visually distinctive escape morphologies. Furthermore, Ly$\alpha$ from recombinations is more concentrated and experiences more trapping than Ly$\alpha$ from collisional excitation. The ionization rate includes local stellar sources, collisional ionization, and a uniform redshift-dependent meta-galactic ionizing background.}
  \label{fig:energy_density}
\end{figure*}

However, neither quantity encapsulates a measure of how far apart the distributions are. Thus, we also compute the first Wasserstein distance~$W_1$, also known as the earth mover's distance as it can be viewed as the minimum amount of work in terms of distribution weight multiplied by distance to transform $p_\text{abs}(x)$ into $p_\text{esc}(x)$. A practical definition for computing this metric is
\begin{equation}
  W_1 = \int_0^1 \left| P_\text{abs}^{-1}(t) - P_\text{esc}^{-1}(t)\right| \text{d}t \, ,
\end{equation}
where $P_\text{abs}(t)^{-1}$ and $P_\text{esc}(t)^{-1}$ denote inverse cumulative distribution functions or quartile functions of $p_\text{abs}(x)$ and $p_\text{esc}(x)$, respectively \citep{Ramdas_2017}. Finally, we transform this distance into units of significance by normalizing according to the average standard deviations, i.e. we define
\begin{equation} \label{eq:sigma_1}
  N^\sigma_1 \equiv \frac{2\,W_1}{\sigma_\text{abs} + \sigma_\text{esc}} \, .
\end{equation}
We interpret this as the approximate number of standard deviations separating the absorbed and escaped distributions, and employ $N^\sigma_1$ to rank the power of a given local quantity to predict the fate of Ly$\alpha$ photons.

In Table~\ref{tab:photons}, we summarize each of these statistics for the following quantities in order of decreasing discriminating significance based on $N^\sigma_1$: dust absorption coefficient $k_\text{dust}$, Ly$\alpha$ absorption coefficient at line centre $k_{\alpha,0}$, density $\rho$, temperature $T$, neutral fraction $x_\text{\HI} \equiv n_\text{\HI} / n_\text{H}$, radial distance from the galactic centre $r$, and metallicity $Z$.

Finally, we employ a linear logistic regression analysis to find a function representing the conditional probability of escape given the local emission environment. Specifically, we consider a model with one or more predictor variables in the set $\bm{X} = \{k_\text{dust},k_{\alpha,0},\rho,T\}$, with one binary target variable $Y \in \{0,1\}$ which is described by a (sigmoid) logistic function,
\begin{equation} \label{eq:logistic}
  P(Y=1|\bm{X}=\bm{x}) = \sigma(f(\bm{x})) = \frac{1}{1 + \exp(-f(\bm{x}))} \,
\end{equation}
where the logit (log odds) is given by $f(\bm{x}) = \sum \beta_i \log x_i + \beta_0$.
An illustration of this is given in Fig.~\ref{fig:logistic_kdust} for a model based on the local dust absorption coefficient $k_\text{dust}$ as a single predictor. We report the time-weighted results obtained from the \texttt{scikit-learn} logistic regression python package in Table~\ref{tab:logistic}. The logistic function may be used as a classifier by choosing a cutoff value, e.g. the 50\% transition region is defined via $f(\bm{x}) > 0$. Additionally, the global escape fraction may be estimated without expensive radiative transfer calculations by the luminosity-weighted average over the simulation, i.e. $f_\text{esc,est}^{\text{Ly}\alpha} \approx \sum L_{\alpha,j} \sigma(f(\bm{x}_j)) / L_\alpha$ where $j$ denotes the cell index. With this simple model we achieve a time-averaged relative error in the estimated escape fractions of only $\langle |1 - f_\text{esc,est}^{\text{Ly}\alpha}/f_\text{esc}^{\text{Ly}\alpha}| \rangle \approx 5\%$, although it is unclear how sensitive this result is to galaxy mass, metallicity, or even the resolution and sub-grid models of particular simulations. Still, Ly$\alpha$ escape based on local conditions could be useful as a correlation tool applied to a large number of similar simulations, or for comparison across simulation frameworks. Either way, we caution against simply applying our model to large-scale simulations without careful recallibration.

\section{Ly\texorpdfstring{$\balpha$}{α} radiation pressure}
\label{sec:Lya_radiation_pressure}
\subsection{Energy density}
\label{sec:energy_density}
We employ Monte Carlo estimators based on the traversed path lengths to calculate the Ly$\alpha$ energy density within each cell \citep[see][]{Smith_RHD_2017}. This allows us to illustrate the internal 3D structure of the resonant scattering process. For comparison, in Fig.~\ref{fig:energy_density} we show the LOS mass-weighted projections of the energy density for ionizing and Ly$\alpha$ radiation. The ionizing photons produce discernible ray-like features whereas the resonant scattering of the Ly$\alpha$ photons leads to a much smoother radiation field. Comparing the energy densities across snapshots shows an increasingly centralized configuration with time. Finally, the Ly$\alpha$ emission originating from recombinations is more concentrated and experiences more trapping than Ly$\alpha$ from collisional excitation.

\subsection{Eddington factor}
We also employ Monte Carlo estimators based on the traversed opacity to calculate the momentum imparted on the gas due to multiple scattering of Ly$\alpha$ photons. The radiation trapping time within a given volume is given by
\begin{equation}
  t_\text{trap} = \frac{\int U_\alpha\,\text{d}V}{\int j_\alpha\,\text{d}V} \, ,
\end{equation}
where $j_\alpha$ denotes the Ly$\alpha$ volume emissivity. Over the entire simulation domain we calculate the time-averaged ratio of trapping to light crossing times to be $\langle t_\text{trap} / t_\text{light}\rangle \approx 1.1$, where $t_\text{light} \approx 2~R_\text{vir}/c$. Therefore, the Ly$\alpha$ photons are well within the optically thin free streaming regime by the time they leave the domain, and all radiative transfer effects due to trapping occur on much smaller scales. However, the trapping ratio increases significantly within the high opacity regions of the galaxy and filaments. Furthermore, such regions also trap radiation from outside sources, thereby producing an artificially strong radiation field when compared to the local intrinsic emission. Thus, a more meaningful measure of the net effect of resonant scattering on the gas is the force multiplier, which is given by
\begin{equation}
  M_\text{F} = \frac{\int \|\bm{a}_{\text{Ly}\alpha}\|\,\rho\,\text{d}V}{\int j_\alpha/c\,\text{d}V} \, ,
\end{equation}
where $\bm{a}_{\text{Ly}\alpha}$ denotes the local acceleration due to Ly$\alpha$ radiation pressure. We calculate a time-averaged force multiplier of $\langle M_\text{F} \rangle \approx 114$ over the entire simulation domain. This is certainly significant, especially as the calculation accounts for vector cancellation of momentum within individual cells.

\begin{figure}
  \centering
  \includegraphics[width=\columnwidth]{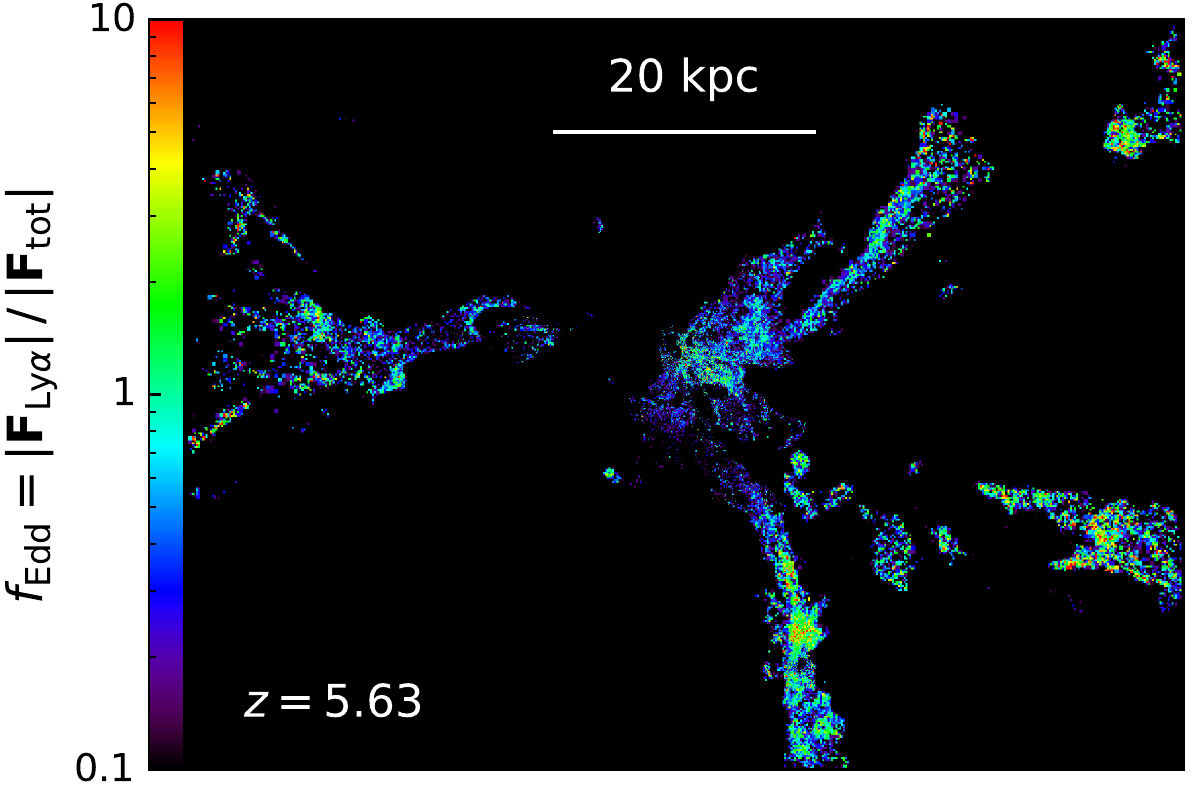}
  \includegraphics[width=\columnwidth]{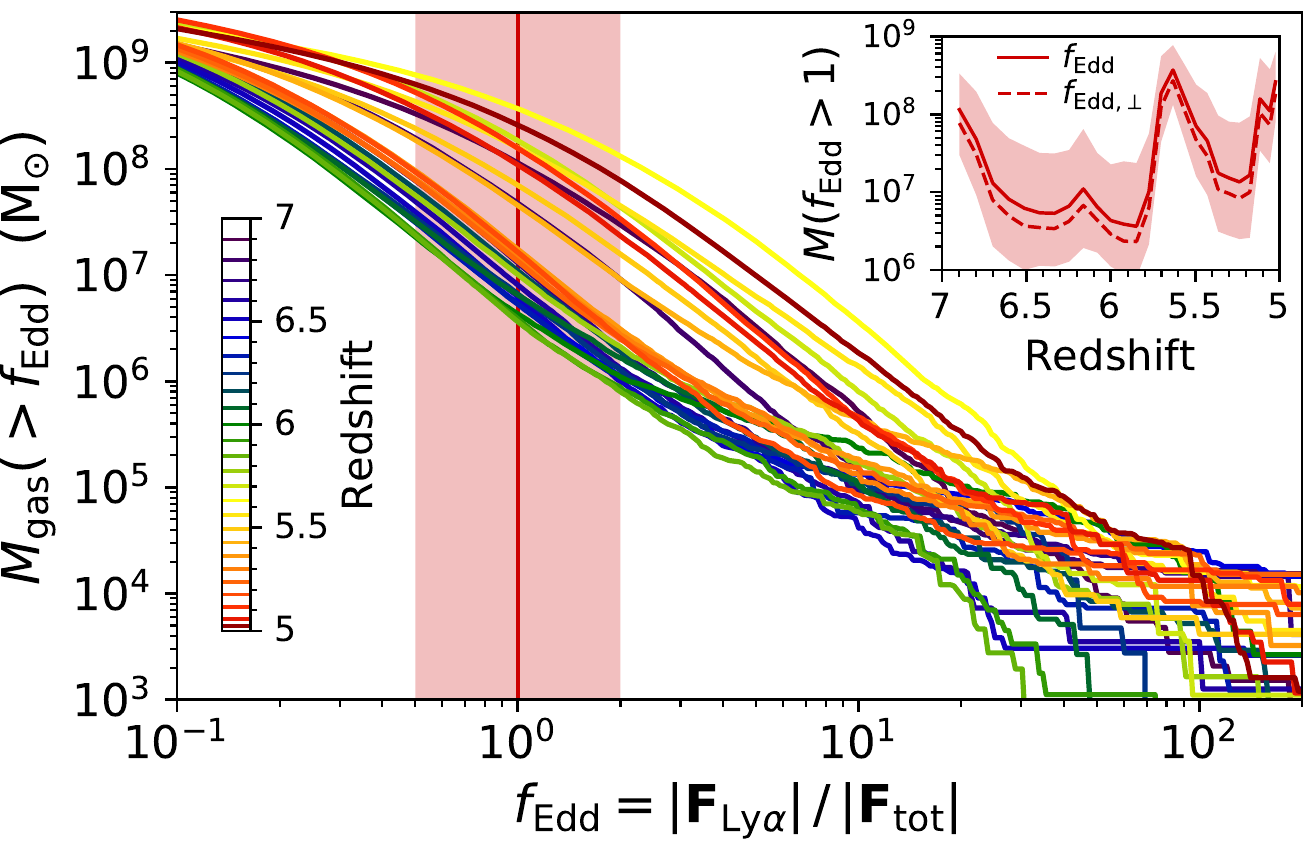}
  \caption{Projected mass-weighted Ly$\alpha$ Eddington factor $f_\text{Edd} = \|\bm{a}_{\text{Ly}\alpha}\| / \|\bm{a}_\text{tot}\|$ shown during the main starburst at $z = 5.63$. The highest factors are along the cosmological filaments feeding the galaxy. The lower panel shows the cumulative mass function of gas above a given Eddington factor for each snapshot, with an inset showing the redshift evolution for $f_\text{Edd} > 1^{+1.0}_{-0.5}$, which is primarily driven by the total luminosity of the galaxy.}
  \label{fig:f_Edd}
\end{figure}

However, determining the potential impact of Ly$\alpha$ radiation pressure requires a direct comparison with other dynamical forces.\footnote{We also note that post-processing estimates can be inaccurate if the trapping becomes comparable to dynamical timescales.} Therefore, we compute the local Eddington factor as the ratio of magnitudes for Ly$\alpha$ radiative and total accelerations experienced by the gas,\footnote{Most regions in the simulation are not self-gravitating, so we consider the total acceleration due to gravity and hydrodynamical (thermal and turbulent) pressure gradients from the simulation.} i.e.
\begin{equation}
  f_\text{Edd} \equiv \frac{\|\bm{a}_{\text{Ly}\alpha}\|}{\|\bm{a}_\text{tot}\|} \, ,
\end{equation}
with (anti)parallel and perpendicular components given by
\begin{equation}
  f_{\text{Edd},\parallel} \equiv -\frac{\bm{a}_{\text{Ly}\alpha} \cdot \bm{a}_\text{tot}}{\|\bm{a}_\text{tot}\|^2}
\end{equation}
and
\begin{equation}
  f_{\text{Edd},\perp} \equiv \sqrt{f_\text{Edd}^2 - f_{\text{Edd},\parallel}^2} \, .
\end{equation}
We note that if the directions of the radiative and gravitational forces are uncorrelated then $\langle |f_{\text{Edd},\parallel}| \rangle / \langle f_\text{Edd} \rangle \approx \int_0^1 \mu\,\text{d}\mu = 1/2$ and $\langle f_{\text{Edd},\perp} \rangle / \langle f_\text{Edd} \rangle \approx \int_0^1 \sqrt{1 - \mu^2}\,\text{d}\mu = \upi/4$, which is more or less the case in our simulations for the mass-weighted and volume-weighted time-averages. In Fig.~\ref{fig:f_Edd} we illustrate the 3D structure of the regions most affected by Ly$\alpha$ trapping with a LOS mass-weighted projection of the Eddington factor, choosing a snapshot corresponding to the main starburst when the Ly$\alpha$ radiation field is strong. It is clear that in this system Ly$\alpha$ radiation pressure is likely to play only a minor role in the overall galactic dynamics. This is a result of the highly inhomogeneous gas distribution, which allows radiation to escape through low-opacity channels. However, the Eddington factor can be quite large along the cosmological filaments feeding the growth of the galaxy. To understand the extent of the high Eddington factor regime, in Fig.~\ref{fig:f_Edd} we also show the cumulative mass function of gas above a given value of $f_\text{Edd}$. We find that $M_\text{gas}(f_\text{Edd}>1)$ fluctuates significantly with redshift in the range of $10^6$--$10^9\,\Msun$, or $0.01$--$10$\% of the total gas mass, primarily driven by the total luminosity of the galaxy. While only a small fraction of the gas has a very high Eddington factor, From Fig.~\ref{fig:f_Edd} it is clear that local trapping can affect the pressure of the infalling filamentary gas, which can have interesting consequences for the nature of filamentary gas infall as a major channel of growth in high-redshift galaxies \citep[e.g.][]{Keres_2005}.

Previous studies have considered the effects of Ly$\alpha$ radiation pressure in highly idealized geometries or with approximate methods for the radiative transfer. In particular, analytic order of magnitude calculations regarding the relative role compared to other feedback mechanisms have been explored by several authors \citep[e.g.][]{Cox_1985,Haehnelt_1995,Oh_Haiman_2002,McKee_Tan_2008,Milosavljevic_2009,Wise_2012}. With the aid of MCRT, \citet{Dijkstra_Loeb_2008,Dijkstra_Loeb_2009} found that multiple scattering within high \HI\ column density shells is capable of significantly enhancing the effective Ly$\alpha$ force. Later \citet{Smith_CR7_2016,Smith_RHD_2017} performed self-consistent MCRT Ly$\alpha$ radiation hydrodynamics (RHD) calculations of galactic winds in the first galaxies, finding that direct collapse black holes (DCBHs), which form in primordial gas, foster an environment that is especially susceptible to Ly$\alpha$ feedback. It has subsequently been found that trapped Ly$\alpha$ cooling radiation potentially affects the initial collapse of these massive black hole seeds through chemical \citep{Johnson_Dijkstra_2017} and thermal \citep{Ge_Wise_2017} feedback. To address the limitations of 1D geometry, \citet{Smith_DCBH_2017} performed a post-processing radiative feedback analysis of a DCBH assembly environment, concluding that fully coupled 3D Ly$\alpha$ RHD will be crucial to consider in future DCBH simulations.

Recently, \citet{Kimm_2018} incorporated a local subgrid model for Ly$\alpha$ momentum transfer into 3D RHD simulations of an isolated metal-poor dwarf galaxy. The authors found that Ly$\alpha$ feedback can regulate the dynamics of star-forming clouds before the onset of supernova explosions, thereby suppressing star formation and indirectly weakening galactic outflows. Our results further demonstrate the need for full 3D Ly$\alpha$ RHD in cosmological simulations to determine the extent to which Ly$\alpha$ trapping in low-metallicity environments is capable of impeding the growth of the first stars and supermassive black holes, disrupting cold gas accretion flows, driving winds in dwarf galaxies, or supplying additional turbulence to regulate star formation. Such simulations based on full solutions of the radiation transport equation will be possible in the near future with acceleration schemes such as the resonant Discrete Diffusion Monte Carlo \citep[rDDMC;][]{Smith_rDDMC_2018}.

\section{Summary and Discussion}
% \section{Summary and Conclusions}
\label{sec:conc}
In this paper, we presented a comprehensive Ly$\alpha$ radiative transfer study of a cosmological zoom-in simulation from the FIRE project \citep{Hopkins_FIRE2_2017,Ma_2018}. We focused on the physics of Ly$\alpha$ escape from an individual galaxy over a redshift range of $z = 5$--$7$. Although our results are subject to cosmic variance and only represent one possible assembly and star formation history, our approach still allows us to treat each LOS as separate observations to quantify the variation due to viewing angle. We also achieve high spatial and angular resolution in the context of the galaxy's redshift evolution, which provides numerous insights about how Ly$\alpha$ observables change in response to starburst activity. Our main conclusions are as follows:
\begin{enumerate}
  \item The dominant driver of the Ly$\alpha$ radiation field is the star formation history, and many properties are susceptible to fluctuations before and after star forming episodes.
  \item High equivalent width sightlines are rare, and
%   naturally arise from the resonant scattering process. %which is unavailable to UV continuum photons.
  are typically associated with outflows with additional coincident UV continuum absorption. The lowest equivalent widths correspond to cosmological filaments.
  \item Individual sources come in and out of visibility during their lifetimes. Thus, multi-object spectroscopic surveys will need to be quite deep to achieve high completeness of all phases of LAE populations for a given mass range. Still, even with $10^4$ second exposure times Ly$\alpha$ detections and spectroscopy of high-$z$ galaxies with the \textit{JWST} is feasible. However, the high (medium) resolution configuration only achieves $\Delta v \approx 100$ ($300$)\,$\text{km\,s}^{-1}$, which is the same order of magnitude as the observed line widths after severe IGM reprocessing. As a consequence, we expect that diagnostics from other lines and cross-correlation studies will be necessary to unravel the detailed properties of high-$z$ LAEs.
  \item The local dust opacity is anti-correlated with Ly$\alpha$ escape, and logistic regression based on the local emission environment can predict the Ly$\alpha$ escape fraction to within 5\% error. Similar models may thus provide efficient alternatives to obtain LAE statistics from hydrodynamical simulations.
  \item Ly$\alpha$ radiation pressure can be dynamically important in dense, neutral, low-metallicity filaments and satellites.
\end{enumerate}

Future Ly$\alpha$ radiative transfer studies should require many of the elements included in this and previous works. For example, post-processing simulations must accurately account for full radiation hydrodynamics of ionizing radiation and other forms of feedback. Results with this approach are increasingly in agreement with observations, although the next major advances may require significantly higher spatial resolution to self-consistently model the multi-phase ISM, CGM, and IGM for statistical samples of simulated high-$z$ galaxies. Capturing the 3D structure of galaxies is vital, and assessing the physics from idealized analytical models is subject to limitations. However, we will continue to benefit from dedicated models and simulations probing radiative transfer effects on small and large scales. Some interesting physics to consider is cosmic ray feedback, which may leave an imprint on the Ly$\alpha$ spectra as the resultant outflows are smoother, colder, and denser than supernova driven winds \citep{Gronke_2018}. As the observational frontier moves to higher redshifts and lower mass galaxies, Ly$\alpha$ radiation hydrodynamics will be increasingly relevant \citep{Smith_RHD_2017,Kimm_2018}, possibly requiring additional acceleration schemes such as the rDDMC method \citep{Smith_rDDMC_2018}. Certainly, an understanding of the intermittency and response to elevated star formation rates is key in interpreting Ly$\alpha$ spectral signatures. Furthermore, the physics of Ly$\alpha$ escape is highly relevant for 21-cm cosmology and reionization studies, and we anticipate the increasing capability of simulations and abundance of data to facilitate the effective confluence of theory and observation at the high-redshift frontier.

\section*{Acknowledgements}
%
% The Acknowledgements section is not numbered. Here you can thank helpful colleagues, acknowledge funding agencies, telescopes and facilities used etc. Try to keep it short.
% Remember to thank the referee.
The authors thank Peter Laursen who kindly provided IGM transmission data and helpful correspondence. AS benefited from numerous discussions with Benny Tsang, Intae Jung, Milo\v{s} Milosavljevi\'{c}, and Yao-Lun Yang. AS also thanks J\'{e}r\'{e}my Blaizot, Max Gronke, Dawn Erb, Anne Verhamme, Andrea Ferrara, Edward Robinson, and Paul Shapiro for insightful conversations. Support for Program number HST-HF2-51421.001-A was provided by NASA through a grant from the Space Telescope Science Institute, which is operated by the Association of Universities for Research in Astronomy, Incorporated, under NASA contract NAS5-26555. VB acknowledges support from NSF grant AST-1413501. CAFG was supported by NSF through grants AST-1412836, AST-1517491, AST-1715216, and CAREER award AST-1652522, by NASA through grant NNX15AB22G, by STScI through grant HST-AR-14562.001, and by a Cottrell Scholar Award from the Research Corporation for Science Advancement. DK was supported by NSF grant AST-1715101 and the Cottrell Scholar Award from the Research Corporation for Science Advancement. The authors acknowledge the Texas Advanced Computing Center (TACC) at the University of Texas at Austin for providing HPC resources.



\bibliographystyle{mnras}
\bibliography{biblio}

%%%%%%%%%%%%%%%%%%%%%%%%%%%%%%%%%%%%%%%%%%%%%%%%%%
%%%%%%%%%%%%%%%%% APPENDICES %%%%%%%%%%%%%%%%%%%%%

% \appendix
% \input{text/A}

%%%%%%%%%%%%%%%%%%%%%%%%%%%%%%%%%%%%%%%%%%%%%%%%%%

% Don't change these lines
\bsp	% typesetting comment
\label{lastpage}
\end{document}